\documentclass[12pt,12pt]{article}
\usepackage{graphicx}
\usepackage{epsfig}
\usepackage{amsmath,amssymb}
\usepackage{slashed}

\catcode`\@=11
\newdimen\ex@
\font\dozeb=cmmib10 scaled \magstep1
\font\dozesyb=cmbsy10 scaled \magstep1
\font\dezb=cmmib10
\def\beq{\begin{equation}}
\def\eeq{\end{equation}}
\def\beqa{\begin{eqnarray}}
\def\eeqa{\end{eqnarray}}
\newcommand{\ba}{\begin{eqnarray}}
\newcommand{\ea}{\end{eqnarray}}
\newcommand\BA{\begin{array}}
\newcommand\EA{\end{array}}

\begin{document}
\def\thefootnote{\fnsymbol{footnote}}

\title{\bf
Thermal resonating Hartree-Bogoliubov theory
based on the projection method\footnotemark[1]
\vspace{-0.1cm}}

\author{Seiya NISHIYAMA$\!$\footnotemark[2]~,~
Jo\~ao da PROVID\^{E}NCIA$\!$\footnotemark[3]~~~and\\ 
Hiromasa OHNISHI\footnotemark[4]\\
\\[-0.3cm]
Centro de F\'\i sica Computacional,
Departamento de F\'\i sica,\\
Universidade de Coimbra,
P-3004-516 Coimbra, Portugal\footnotemark[2]~\footnotemark[3]\\
\\[-0.6cm]
Department of General Science,
Tsuruoka National College of Technology,\\
104 Sawada, Inooka, Tsuruoka, Yamagata, 997-8511, Japan\footnotemark[4]\\
\\[-0.3cm]
{\it Dedicated to the Memory of Hideo Fukutome} }
\maketitle
\vspace{1mm}
\footnotetext[1]{The preliminary work has first been presented by
S. NISHIYAMA at the Workshop on
{\it Thermal Quantum Field Theories and Their Applications}
held at Yukawa Institute for Theoretical Physics,
Kyoto, JAPAN,\\
Seiya Nishiyama and Jo\~ao da Provid\^{e}ncia,
Soryushiron Kenkyu {\bf 114} (2006) C21-C23;
{\bf 116} (2008) B18-B20.
}
\footnotetext[2]{
E-mail address: seikoceu@khe.biglobe.ne.jp}
\footnotetext[3]{
E-mail address: providencia@teor.fis.uc.pt}
\footnotetext[4]{
E-mail address: hohnishi@tsuruoka-nct.ac.jp}

%%%%%%%%%
%                     %
%  0  Abstract  %
%                     % 
%%%%%%%%%

\vspace{-0.9cm}

\begin{abstract}
\vspace{0.2cm}
We propose a rigorous thermal resonating mean-field theory (Res-MFT). 
A state is approximated by superposition of multiple
MF wavefunctions (WFs) composed of
non-orthogonal Hartree-Bogoliubov (HB) WFs.
We adopt a Res-HB subspace spanned by Res-HB
ground and excited states.
A partition function (PF) in a
$SO(2N)$ 
coherent state representation $| g \rangle$
($N$:Number of single-particle states)
is expressed as
$\mbox{Tr}(e^{-\beta H}) 
\!=\!
2^{N\!-\!1} \!\! \int \! \langle g |e^{-\beta H}| g \rangle dg
~\!(\beta \!=\!\! 1/k_B T)$.
Introducing a projection operator $P$ to the Res-HB subspace,
the PF in the Res-HB subspace is given as
$\mbox{Tr}(Pe^{-\beta H})$,
which is calculated within the Res-HB subspace
by using the Laplace transform of 
$e^{-\beta H}$
and the projection method.
The variation of the Res-HB free energy is made,
which leads to
a thermal HB density matrix
$W_{\mbox{{\scriptsize Res}}}
^{\mbox{{\scriptsize thermal}}}$
expressed in terms of
a thermal Res-FB operator
${\cal F}_{\mbox{{\scriptsize Res}}}
^{\mbox{{\scriptsize thermal}}}$
as
$
W_{\mbox{{\scriptsize Res}}}
^{\mbox{{\scriptsize thermal}}}
\!=\!
\{ 1_{\!2N} \!+\! \exp 
( \beta {\cal F} _{\mbox{{\scriptsize Res}}}
^{\mbox{{\scriptsize thermal}}}  )
\}^{\!-1}
$.
A calculation of the PF by an infinite matrix continued fraction
is cumbersome and a procedure of tractable optimization is too complicated.
Instead, we seek for another possible and more practical way of computing
the PF and the Res-HB free energy
within the Res-MFT.  
\vskip0.1cm
$Keywords$: 
Thermal resonating Hartree-Bogoliubov theory; 
Projection method
\vskip0.1cm
PACS numbers; 05.30.$\pm$d, 74.20.Fg
\end{abstract}

\newpage

%%%%%%%%%%%
%                           %
%  1  Introduction  %
%                           %
%%%%%%%%%%%

\def\thesection{\arabic{section}}
\setcounter{equation}{0}
\renewcommand{\theequation}{\arabic{section}.\arabic{equation}}

\section{Introduction}
\vspace{-0.1cm}
~~~
The strongly correlated fermion systems have been attracting much attentions
due to their abundant physical phenomenology such as,
shape transition in nuclei,
metal-insulator transition, 
high-$T_c$ superconductivity, magnetic substances with narrow $d$ bands. 
These latter electron systems show drastic electronic and magnetic changes 
by a slight environment modifications due to
competitions of charge, magnetic and orbital orderings.
All of the above systems are typical examples of
the strongly correlated fermion systems.
In the theoretical studies of such fermion systems, 
it is very important to treat fermion correlations and/or quantum fluctuations rigorously
as far as possible.
For this aim, several quantum many-body theories have been developed,
such as Quantum Monte-Carlo simulation which
provides a general technique to inspect numerically such systems
\cite{QMCM.1993},
Real-Space Density Matrix Renomalization Group method
\cite{White.1992}
and
Exact Diagonalization method
\cite{DagottoMoreo.1985}. 
There has been considerable progress
both in a self-consistent mode-mode coupling theory for weak itinerant magnetism
and a functional integral theory interpolating the extreme regimes
of weak itinerant magnetism and localized spin moments 
\cite{Moriya.1979}.
The nature of electron correlation,
particularly in a 2-D fermion system,
has become an important problem
in connection with high critical-temperature $T_c$ superconductivity.
It is a new and hot unsolved problem.
Twelve years have passed since
Nagamatsu {\it et al.} discovered 
two-gap superconductivity (SC)
of $\mbox{MgB}_2$
with $T_c \!\!=\!\! 39$K
\cite{Akimitsu.01}.
%The $T_c  \!\!=\!\! 39$K in $\mbox{MgB}_2$
%is close to or even above upper theoretical values predicted by
%the BCS theory
%\cite{Parks.69}.
$T_c$ of conventional superconductors 
in the weak coupling regime
has been described by the BCS theory
\cite{BCS.57,Bogo.59} %,BTS.59}
and that  of unconventional ones 
in the strong coupling has been done by the Eliashberg's theory
\cite{Eliashberg.60}. %,Schrieffer.64,GK.82}.
On the contrary, two gaps of
$\mbox{MgB}_2$
have been predicted by Liu {\it et al.}
\cite{LMK.01},
employing a weak-coupling two-gap model with the use of
a ($\sigma$, $\pi$) two-band model.
Remember that the original idea of such a model was proposed long time ago
by Suhl {\it et al.} and Kondo
\cite{SMW.59,Kondo.63}.
On the other hand,
employing the Eliashberg's strong-coupling theory 
\cite{Eliashberg.60},
Choi {\it et al.}
\cite{CRSCL.02}
have also explained such properties of
$\mbox{MgB}_2$.
Further
recent studies of new physics of
high-$T_c$ superconductors have begun as viewed by Tohyama
\cite{Tohyama.2012}
in which
hot topics in cuprate and iron-pnictide superconductors emerge.
In cuprate ones,
two-gap scenario is not necessarily inconsistent with
the Anderson's RVB theory
\cite{Anderson}.
While, according to Tohyama,
it is natural to start with an itinerant model with a weak Coulomb interaction
to describe electric structures of iron-pnictides.
He and Kaneshita could explain a damping in high-energy spin excitation
\cite{KaneToh.2010},
using such an itinerant model and a theory based on
an itinerant five-band Hubbard model and the random phase approximation
\cite{RingSchuk.80}.
These facts emphasize again an inevitable strict manipulation of
the fermion correlations and/or quantum fluctuations mentioned above.

Now we are in a stage to study the above phenomena, particularly
a currently topical high-$T_c$ problems such as $T_c$ itself and multi-gap
standing on the spirit of the resonating mean-field.
There exist many available theories for such problems.
Among them,
the resonating mean-field theory (Res-MFT)
\cite{Fuku.88,NishiFuku.91}
also may stand as a candidate for a possible and effective theory and is considered
to be useful for such a theoretical approach.
This is because the Res-MFT has a following characteristic feature:
Fermion systems with large quantum fluctuations
show serious difficulties in many-body problems at finite temperature.
To approach such problems,
Fukutome has developed
the resonating Hartree-Fock theory (Res-HFT)
\cite{Fuku.88}
and
Fukutome and one of the present authors (S.N.) have extended it
directly to the resonating Hartree-Bogoliubov theory (Res-HBT)
to include pair correlations
\cite{NishiFuku.91}
(referred to as I).
This is our first motivation that we challenge such an exciting physics.

The Res-HBT is equivalent to the coupled Res-HB eigenvalue equations in which
the orbital concept is still surviving though orbitals are resonating.
This means that the band picture has a correspondence to the orbital concept
in the Res-HB approximation
though bands of different structures are resonating.
The superposed HB wave functions (WFs)
are the coherent state representations (CS reps).
The Res-MFT was applied to describe a two-gap SC
\cite{NishiProviOhnishi.10}. 
The Res-HFT has also been applied to the  1-D and 2-D Hubbard models
\cite{Hubbard.63}
and has shown its own effectiveness by Fukutome and Tomita and their collaborators
\cite{Tomita.04}.
The applications of Res-HFT and Res-HBT have been, however,
limited to the ground state property up to the present stage. 
To treat temperature-dependent phenomena such as high-$T_c$ superconductivity, 
an extension of the present theories to theories available for finite temperature case
must be necessarily required.
As the first trial, we have made
an attempt at a Res-MFT description
of thermal behavior of the two-gap SC
\cite{NishiProviOhnishi.13}
(referred to as II).
The thermal variation has been made
in a different way from the usual thermal-BCS theory
\cite{KA.59,AGD.65,RingSchuk.80} %,Abrikosov.88}
and got preliminary results.
To improve solutions exactly,
we construct a rigorous thermal Res-MFT 
basing on the projection-operator method
and give another version of MF approximation 
describable a superconducting fermion system
with $N$ single-particle states.
A partition function in
a $SO(2N)$
CS rep $| g \rangle$
\cite{Perelomov.72}
is expressed as
$\mbox{Tr}(e^{-\beta H}) 
\!\!=\!\!
2^{N\!-\!1} \!\!  \int  \! \langle g |e^{-\beta H}| g \rangle dg~
( \beta \!\!=\!\! 1/k_B T )$ 
where integration is the group integration on the $SO(2N)$.
Introducing a projection operator $P$ to the Res-HB subspace
 proposed in the Res-MFT
\cite{Fuku.88,NishiFuku.91},
the partition function in the Res-HB subspace is given as
$\mbox{Tr}(Pe^{-\beta H})$.
It is calculated within the Res-HB subspace
by using the Laplace transform of 
$e^{-\beta H}$
and the projection operator method
\cite{Naka.58,Zwan.60,Mori.65}
which leads us to an infinite matrix continued fraction (IMCF).
The variation of the Res-HB free energy
is executed parallel to the usual thermal BCS theory
\cite{BCS.57,KA.59,AGD.65}. %,Abrikosov.88}.
It induces
a thermal HB density matrix
$W_{\mbox{{\scriptsize Res}}}
^{\mbox{{\scriptsize thermal}}}$
expressed in terms of
a thermal Res-FB operator
${\cal F}_{\mbox{{\scriptsize Res}}}
^{\mbox{{\scriptsize thermal}}}$
as
$
W_{\mbox{{\scriptsize Res}}}
^{\mbox{{\scriptsize thermal}}}
\!=\!
\{ 1_{\!2N} \!+\! \exp 
( \beta {\cal F} _{\mbox{{\scriptsize Res}}}
^{\mbox{{\scriptsize thermal}}} )
\}^{\!-1}
$.
The Res-HB coupled eigenvalue equation is extended to
the thermal Res-HB coupled eigenvalue equation.  
A calculation of the partition function by the IMCF is very cumbersome
and a procedure of tractable optimization is
too complicated.
Instead, we seek for another possible and more practical way of
computing the Res-HB partition function and the Res-HB free energy
within the framework of the Res-MFT.

In Sec. II we give a brief review of the Res-HBT.   
In Sec. III we derive an IMCF with the use of the projection operator.
In Sec. IV,
we give expressions for the thermal Res-HB density matrix
and the thermal pair density matrix in term of the eigenvalues
of the thermal Res-FB operator.
We also propose tentative expressions for
the thermal Res-HB interstate density matrix and overlap integral.
In Sec. V,
we introduce a quadratic Res-HB Hamiltonian
and extend the HB free energy to the Res-HB free energy.
In the occupation number space in a quasi-particle frame,
we obtain explicit expressions for the approximately calculated
Res-HB partition function and Res-HB free energy.
Finally in Sec. VI,
we give a summary and further perspectives.
In Appendices,
we provite a matrix element of the Laplace transform of 
$e^{-\beta H}$ and various types of matrix elements related to it.
Further we give a proof of the commutability between
thermal Res-FB operator and thermal HB density matrix
and derive some equivalence relation.

\newpage

%%%%%%%%%%%%%%%%%%%%%%%%%%%%%%%%%                                                                                     %                                                                                             %
%  2  Brief review of resonating Hartree-Bogoliubov theory   %                                                                                             %                                                                                             %
%%%%%%%%%%%%%%%%%%%%%%%%%%%%%%%%%

\def\thesection{\arabic{section}}
\setcounter{equation}{0}
\renewcommand{\theequation}{\arabic{section}.\arabic{equation}}

\section{\!\!\!Brief review of resonating Hartree-Bogoliubov theory}

~~First, according to I
let us briefly recapitulate the exact CS rep
on a $SO(2N)$
group of a superconducting fermion system.
We consider a fermion system with $N$ single-particle states.
Let $c_{\alpha }$ and $c^{\dagger }_{\alpha }$,
$\alpha \! = \! 1,\cdots, N$,
be the annihilation and creation operators of the fermion.
Owing to their anticommutation relation,
the pair operators
\begin{eqnarray}
\left.
\begin{array}{rl}
& E_{\;\,\beta }^\alpha
=
{c_\alpha^\dagger }{c_\beta }
-{\frac{1}{2}}\delta_{\alpha \beta },~~
E^{\alpha \dagger }_{\beta }
=
E^{\beta }_\alpha ,\\
\\[-8pt]
& E_{\alpha \beta }
=
{c_\alpha }{c_\beta }, {\quad}{\quad}{\quad}{\quad}\!\!
E^{\alpha \beta }
=
{c_\alpha^\dagger }{c_\beta^\dagger },~~
E^{\alpha \beta }
=
-E_{\alpha \beta }^{\dagger },
\end{array}
\right\}
\label{s02Nalgebra}
\end{eqnarray}
satisfy the commutation relations of the $SO(2N)$ Lie algebra.
They generate a $SO(2N)$ canonical transformation $U(g)$,
which induces the generalized Bogoliubov transformation 
specified by a $SO(2N)$ matrix $g$ as follows:
\begin{eqnarray}
\left.
\begin{array}{rl}
&U(g)[c,c^\dagger ]U^{\dagger }(g)
=
[c,c^\dagger ]g ,\\
& \\[-8pt]
&g
\equiv
\left[ \!\!
\begin{array}{cc}
a&b^\ast\\ 
b&a^\ast 
\end{array} \!\!
\right],{\quad}g^{\dagger }g = gg^{\dagger }
= 1_{2N} ,
\end{array}
\right\}
\label{Bogoliubovtransformation}
\end{eqnarray}
\begin{equation}
U(g)U(g') = U(gg'),{\quad}
U(g^{-1}) = U^{-1}(g) = U^{\dagger }(g),{\quad}
U(1_{2N}) = 1_{2N} ,
\label{productoftransformation}
\end{equation}
where the $(c,c^{\dagger })$ is the $2N$-dimensional row vector
$((c_{\alpha }),(c^{\dagger }_{\alpha }))$
and 
the $a=(a_{\alpha i})$ 
and 
$b = (b_{\alpha i})$ are $N \! \times \! N$ matrices, respectively.
The $1_{2N}$ is the $2N$-dimensional unit matrix.
$U(g)$ is an irreducible representation (irrep) of the $SO(2N)$ group
on the Hilbert space of even fermion numbers.
From this and the orthogonality of irrep matrices,
a state vector is represented as\\[-8pt]
\begin{equation}
|\Psi \rangle 
= 
2^{N-1} \!\! \int U(g)|0\rangle \langle 0|U^\dagger (g)|
\Psi \rangle \;dg
=
2^{N-1} \!\! \int |g \rangle \Psi (g) \;dg ,
\label{CSrep}
\end{equation}\\[-8pt]
where the integration is the group integration on the $SO(2N)$.
The $|0\rangle$ is a vacuum satisfying 
$c_{\alpha }|0\rangle \! = \! 0$.
The $|g\rangle \! = \! U(g)|0\rangle$ is an HB WF and
$\Psi(g) \! = \! \langle 0|U^{\dagger }(g)|\Psi \rangle$.
This is an exact $SO(2N)$ CS rep of a state vector
\cite{Thou.60,Bogo.59}. 
We define the overlap integral $\langle g|g'\rangle$ by\\[-8pt]
\begin{equation}
\langle g|g' \rangle 
= 
\langle 0|U^\dagger (g)U(g')| 0 \rangle = \langle 0|
U(g^\dagger g')|0 \rangle .
\label{overlap integral}
\end{equation}\\[-12pt]
For the Hamiltonian $H$ of the fermion system,
by using Eq.
(\ref{CSrep}),
the Schr\"{o}dinger equation
$(H - E)|\Psi \rangle = 0$
is converted to an integral equation on the $SO(2N)$ group\\[-16pt]
\begin{eqnarray}
\displaystyle \int \{ \langle g|H|g'\rangle
- E \langle g|g'\rangle \} \Psi (g') dg' = 0,~~
\langle g|H|g'\rangle = \langle 0|U^\dagger (g)HU(g')|0 \rangle .
\label{Schroedingerequation}
\end{eqnarray}\\[-14pt]
From $\langle 0|U(g)|0 \rangle = [\det a]^{\frac{1}{2}}$,
where det is determinant, and
(\ref{productoftransformation}),
we have an overlap integral
\begin{eqnarray}
\left.
\begin{array}{rl}
&\langle g|g'\rangle
=
[\det (a^\dagger a' + b^\dagger b')]^{\frac{1}{2}}
= 
[\det z]^{\frac{1}{2}} ,\\
& \\[-8pt]
&z
\equiv
u^\dagger u',{\quad}z = (z_{ij}) ,
\end{array}
\right\}
\label{detformula}
\end{eqnarray}
where we introduce a $2N \! \times \! N$ isometric matrix $u$ by
\begin{eqnarray}
u =
\left[ \!
\begin{array}{c}
b\\a
\end{array} \!
\right],
{\quad}u^\dagger u = 1_N ,
\label{isometricmatrix}
\end{eqnarray}
so that $z$ is an $N \! \times \! N$ matrix.

The matrix elements of the pair operators
(\ref{s02Nalgebra})
and a two-body operator
between two HB WFs are calculated as follows:\\[-6pt]
\begin{equation}
\langle g|{E^{\beta } }_{\alpha }
\!+\!
\frac{1}{2}{\delta }_{\beta \alpha } |g' \rangle
=
R_{\alpha \beta } (g,g')
\cdot 
[\det z]^{\frac{1}{2}} ,~~
R (g,g')
=
b'z^{-1}b^\dagger ,
\label{matrixelement1}
\end{equation}
\begin{equation}
\langle g|E_{\beta \alpha }|g'\rangle
=
K_{\alpha \beta } (g,g')
\cdot
[\det z]^{\frac{1}{2}} ,~~
K (g,g')
=
b'z^{-1}a^\dagger ,
\label{matrixelement2}
\end{equation}
\begin{eqnarray} 
\begin{array}{rl}
\langle g|E^{\alpha \gamma }E_{\delta \beta }
|g' \rangle
=
&\!\!\!
\left\{ R_{\beta \alpha }(g,g')
R_{\delta \gamma }(g,g') 
- R_{\delta \alpha }(g,g')R_{\beta \gamma }(g,g')
\right.\\
\\[-6pt]
&\!\!\!
\left.
-K_{\alpha \gamma }^\ast (g',g) K_{\delta \beta }(g,g')
\right\}
\cdot
[\det z]^{\frac{1}{2}} .
\label{matrixelement3}
\end{array} 
\end{eqnarray}

Let the Hamiltonian $H$ of the system under consideration be\\[-12pt]
\begin{eqnarray}
H
=
h_{\alpha \beta } \!
\left( \!
E_{\;\,\beta }^\alpha 
\!+\!
\frac{1}{2}
{\delta }_{\alpha \beta } \!
\right)
+
\frac{1}{4}
[\alpha \beta |\gamma \delta]E^{\alpha \gamma } E_{\delta \beta } ,
\label{Hamiltonian}
\end{eqnarray}
in which
the matrix
$h_{\alpha \beta }$
related to a single-particle Hamiltonian
includes a chemical potential and 
the antisymmetrized matrix elemeny of the interaction
$[\alpha \beta |\gamma \delta]$
satisfies the relations\\[-12pt]
\begin{eqnarray}
[\alpha \beta |\gamma \delta]
=
-
[\alpha \delta |\gamma \beta]
=
[\gamma \delta | \alpha \beta]
=
[\beta \alpha |\delta \gamma]^\ast .
\label{interactionmatrixelement}
\end{eqnarray}
Here we use the dummy index convention
to take summation over the repeated indices.

Following I,
we briefly recapitulate a new eigenvalue equation called
the Res-HB eigenvalue equation.
in quite a parallel manner to the Res-HF by Fukutome
\cite{Fuku.88}.
We approximate a low energy eigenstate
$|\Psi ^{\mbox{{\scriptsize Res}}} \rangle $
of the Hamiltonian $H$
by a discrete superposition of HB WFs
which are denoted by
$|g_r\rangle ,~|g_s\rangle ~\cdots$.
The $|g_r \rangle $'s are non-orthogonal and represent different
collective correlation states.
Then, the state
$|\Psi ^{\mbox{{\scriptsize Res}}} \rangle $
is given as\\[-12pt]
\beqa
\begin{array}{c}
|\Psi ^{\mbox{{\scriptsize Res}}} \rangle
=
\sum_{s=1}^n |g_s \rangle c_s .
\label{Psi}
\end{array}
\eeqa
The general form of the $N \!\times\! N$ matrix $z$ and
the $2N \!\times\! 2N$ HB interstate density matrix between
$|g_r \rangle$ and $|g_s \rangle$ are, respectively, defined as\\[-12pt]
\beq
z_{rs} 
= 
u_r^\dagger u_s, {\quad}
W_{rs} 
= 
u_s z_{rs}^{-1} u_r^\dagger 
=
W_{sr}^\dagger,
\label{zandW}
\eeq\\[-18pt]
whose matrix form is given as\\[-12pt]
\beqa
\begin{array}{rl}
W_{rs}
=
\left[ \!
\begin{array}{cc}
R_{rs}&K_{rs}\\
\\[-8pt]
-K_{sr}^\ast &1_N - R_{sr}^\ast
\end{array} \!
\right] ,~~
W_{rs}^2
=
W_{rs} ,
\end{array}
\label{generalinterstatedensitymatrix}
\eeqa
where $R_{rr}$ and $K_{rr}$ denote the densities and
the pair densities and
$R_{rs}$ and $K_{rs}$ mean the transition densities.
The mixing coefficients $c_s$ are normalized by the relation
\beqa
\begin{array}{c}
\langle \Psi ^{\mbox{{\scriptsize Res}}} |
\Psi ^{\mbox{{\scriptsize Res}}} \rangle
=
\sum_{r,s=1}^n \langle g_r |g_s \rangle c_r^\ast c_s
=
\sum_{r,s=1}^n [\det z_{rs}]^{\frac{1}{2}} c_r^\ast c_s
=
1 .
\label{generalnormalization}
\end{array}
\eeqa
Using Eq.
(\ref{generalinterstatedensitymatrix}),
the expectation value of the Hamiltonian $H$ by the state
$|\Psi^{\mbox{{\scriptsize Res}}} \rangle$ is given as
\beqa
\begin{array}{c}
\langle \Psi ^{\mbox{{\scriptsize Res}}} |H|
\Psi ^{\mbox{{\scriptsize Res}}} \rangle
=
\sum_{r,s=1}^n \langle g_r |H| g_s \rangle c_r^\ast c_s
=
\sum_{r,s=1}^n H[W_{rs}]
\cdot
[\det z_{rs}]^{\frac{1}{2}}c_r^\ast c_s .
\label{generalexpectationvalue}
\end{array}
\eeqa

For our aim of further discussions,
we adopt the following Lagrangian
$L_{\mbox{\scriptsize Res}}^{\mbox{\scriptsize HB}}$
with the Lagrange multiplier term to secure
the normalization condition (\ref{generalnormalization}):\\[-16pt]
\beqa
\begin{array}{c}
L_{\mbox{\scriptsize Res}}^{\mbox{\scriptsize HB}}
=
\langle \Psi ^{\mbox{{\scriptsize Res}}} |H|
\Psi ^{\mbox{{\scriptsize Res}}} \rangle
-
E
\langle \Psi ^{\mbox{{\scriptsize Res}}}|
\Psi ^{\mbox{{\scriptsize Res}}} \rangle
=
\sum_{r,s=1}^n \{H[W_{rs}] - E \}
\cdot
[\det z_{rs}]^{\frac{1}{2}} c_r^\ast c_s .
\label{Lagrangian}
\end{array}
\eeqa\\[-16pt]
In the case of the temperature $T=0$,
the variation of
$L_{\mbox{\scriptsize Res}}^{\mbox{\scriptsize HB}}$
with respect to $c_r^\ast$ leads to
the Res-HB configuration interaction (CI) equation
to determine the mixing coefficients $c_s$\\[-16pt]
\beqa
\begin{array}{c}
\sum_{s=1}^n \{ H[W_{rs}] - E \}
\cdot
[\det z_{rs}]^{\frac{1}{2}}c_s 
= 
0 .
\label{generalRes-HBCIequation}
\end{array}
\eeqa\\[-14pt]
By the variations with respect to
the HB interstate density matrix $ W $ and
the overlap integral $[\det z]^{\frac{1}{2}}$,
we obtain the Res-HB equation to determine the $u_r$'s as\\[-14pt]
\beqa
\left.
\begin{array}{rl}
&
\sum_{s=1}^n {\cal K}_{rs} c_r^\ast c_s
=
0 , \\
& \\[-10pt]
&
{\cal K}_{rs} 
\equiv
\{
(1_{2N} - W_{rs}) {\cal F} [W_{rs}] + H[W_{rs}] - E
\}
\cdot
W_{rs}
\cdot
[\det z_{rs}]^{\frac{1}{2}} ,
\end{array}
\right\}
\label{generalRes-HBequation}
\eeqa\\[-8pt]
in which the Fock-Bogoliubov (FB) operator is introduced as\\[-14pt]
\beqa
\begin{array}{rl}
{\cal F}[W_{rs}]
=
\left[ \!
\begin{array}{cc}
F_{rs}&D_{rs} \\
\\[-10pt]
-D_{sr}^\ast &-F_{sr}^\ast
\end{array} \!
\right] ,
\end{array}
\label{generalFBoperator}
\eeqa\\[-12pt]
where the $N \!\times\! N$ matrices $F_{rs}$ and $D_{rs}$ are
the functional derivatives of the $H[W_{rs}]$
with respect to the $W^{\mbox{{\scriptsize T}}}_{rs}$
in which $\mbox{{\scriptsize T}}$ stands for
the transposition of a matrix.
They are defined as\\[-16pt]
\begin{eqnarray}
\left.
\begin{array}{rl}
&F_{rs;\alpha \beta }
\equiv
\displaystyle \frac{\delta H[W_{rs}]}{\delta R_{rs;\beta \alpha }}
= h_{\alpha \beta }
+
[\alpha \beta |\gamma \delta]R_{rs;\delta \gamma } ,\\
& \\[-10pt]
&D_{rs;\alpha \beta }
\equiv
\displaystyle \frac{\delta H[W_{rs}]}{\delta K_{sr;\alpha \beta }^\ast }
=
-
\frac{1}{2}[\alpha \gamma |\beta \delta]K_{rs;\delta \gamma } .
\end{array}
\right\}
\label{FandDoperators}
\end{eqnarray}\\[-10pt]
It has already been shown in I that
equation (\ref{generalRes-HBequation}) is equivalent to
the following Res-HB coupled eigenvalue equations:\\[-14pt]
\beqa
\left.
\begin{array}{rl}
[{\cal F}_r u_r]_i
=
\epsilon_{ri}u_{ri},~~
\epsilon_{ri}
\equiv
\widetilde{\epsilon_r}_i
- \{ H[W_{rr}]-E\} |c_r|^2 , \\
\\[-8pt]
{\cal F}_r
\equiv
{\cal F}[W_{rr}]|c_r|^2
+
\sum _{s =1}^{\prime~n} ({\cal K}_{rs} c_r^\ast c_s
+
{\cal K}_{rs}^\dagger c_r c_s^\ast )
=
{\cal F}_r ^\dagger ,
\end{array}
\right \}
\label{generalRes-HBeigenvalueequation}
\eeqa\\[-10pt]
where the primed summation is made
under the restriction $s \! \ne \! r$
and the quasi-particle energy
$ \widetilde{\epsilon }_r
\! = \!
u_r ^\dagger {\cal F}[W_{rr}]|c_r|^2 u_r
\! = \!
\delta_{ij} \widetilde{\epsilon }_{ri} $
satisfies the property of the usual HB orbital energy:\\[-14pt]
\beqa
\begin{array}{rl}
&\widetilde{\epsilon }_r
=
\left[
\begin{array}{cccccc}
\widetilde{\epsilon }_{r1}&&&&&\\[-8pt]
&\ddots&&&0&\\[-8pt]
&&\widetilde{\epsilon }_{rN}&&&\\
&&&- \widetilde{\epsilon }_{r1}&&\\[-8pt]
&0&&&\ddots&\\[-8pt]
&&&&&- \widetilde{\epsilon }_{rN}
\end{array}
\right] .
\end{array}
\label{quasi-particleenergy}
\eeqa\\[-8pt]
The $2N \!\!\times\!\! 2N $ matrix ${\cal F}_r$ is called
the Res-Fock-Bogoliubov (Res-FB) operator.
The Res-HB eigenvalue equation
(\ref{generalRes-HBeigenvalueequation}) means that
every HB eigenfunction in a HB resonating state has
its own orbital energies $\epsilon _r$,
which declares
{\it the survival of the orbital concept}
in the Res-HB theory,
though orbitals of different structures are resonating.
The Res-HB CI equation (\ref{generalRes-HBCIequation}) and
the Res-HB eigenvalue equation
(\ref{generalRes-HBeigenvalueequation})
is solved iteratively if we start from
suitable trial $u_r$'s.
Once the HB WFs $|g_r \rangle $
in the Res-HB ground state are determined,
then the other solutions of the Res-HB CI equation give
a series of the Res-HB excited states
called $resonon$ excitations.
To show the usefulness of the Res-HBT
without unnecessary complications,
we have first applied it to a problem of describing the coexistence phenomenon of 
two deformed-shapes occurred 
in a simple and schematic model of nuclei
%identical nucleons
\cite{NishiFuku.92},
using the Bogoliubov-Valatin transformation
\cite{Vala.58,Bogo.59}.
The Res-HFT with the usual S-det
has been applied to a simple LMG model of nuclei
\cite{LMG.65,Nishi.94},
a NJL model of hadron
\cite{NJL.61,NishiProvi.99,NishiProviOhno.01}
and that with spin-projected S-det
is applied to a 1-D half-filled Hubbard model
\cite{GuilAlvarez.96,Hubbard.63,Tomita.04}.

\newpage

%%%%%%%%%%%%%%%%%%%%%%%%%%%%%%                                                                                 
%                                                                                    %
%  3    Derivation o infinite matrix continued fraction    %
%                                                                                    %
%               with the use of  projection operator              %                                                                                 %                                                                                    %
%%%%%%%%%%%%%%%%%%%%%%%%%%%%%%

\def\thesection{\arabic{section}}
\setcounter{equation}{0}
\renewcommand{\theequation}{\arabic{section}.\arabic{equation}}

\section{Derivation of infinite matrix continued fraction
with the use of projection operator}

~~~~As proved in II,
the explicit form of partition function in the Res-HB subspace
is given as
\\[-20pt]
\beqa
\begin{array}{c}
\mbox{Tr} (Pe^{- \beta H})
=
\sum _{r,s=1}^n
\langle g _r |
e^{- \beta H}
|g _s \rangle (S^{-1})_{sr} ,~
(P
\!\equiv\!
\sum _{r,s=1}^n
|g _r \rangle | (S^{-1})_{rs} \langle g _s| ) ,
\end{array}
\label{traceformP}
\eeqa
the R.H.S. of which
%(\ref{traceformP})
is brought into a form suitable for
applying the projection method by Fulde
\cite{Fulde.93}.
For this purpose we introduce the Laplace transform
of $f(\beta ) \!=\! e^{-\beta H}$ and $Q \!=\! 1 - P$,\\[-16pt]
\beqa
{\cal L}\{f(\beta )\}
\!=\!
{\widetilde{f}(\tau )}
\!\equiv\!
\int_0 ^\infty \!
e^{-\tau \beta } e^{- \beta H}
d\beta 
\!=\!
{\displaystyle \frac{1}{\tau \!+\! H} }
\!=\!
{\displaystyle \frac{1}{\tau \!+\! HQ}
\!-\! 
\frac{1}{\tau \!+\! HQ}HP\frac{1}{\tau \!+\! H}}
, ~(\mbox{Re}\{\tau \} \!>\! 0) ,
\label{Laplacetransformof-bH}
\eeqa
We aim at evaluating the matrix
${\widetilde{\mathbb{R}}(\tau )}$ whose matrix elements
are given by\\[-10pt]
\beq
{\widetilde{\mathbb{R}}_{uv}(\tau )}
=
\langle g _u |
\frac{1}{\tau + H}
|g _v \rangle 
\equiv
{\widetilde{\mathbb{R}}_{uv}^{(0)}(\tau )} .
~~(u,v = 1, \cdots, n)
\label{Rmatrix} 
\eeq
Let us denote
$H,~|g _v \rangle,~P$, $Q$ and $S$
as
$H^{(0)},~|g _v^{(0)} \rangle,~P^{(0)}$, $Q^{(0)}$ and $S^{(0)}$,
respectively.
Using 
(\ref{Laplacetransformof-bH})
we can express
$
{\widetilde{\mathbb{R}}_{uv}^{(0)}(\tau )}
$
in the following form:\\[-10pt]
\beq
{\widetilde{\mathbb{R}}_{uv}^{(0)}(\tau )}
=
\langle g _u^{(0)} |
\frac{1}{\tau + H^{(0)}Q^{(0)}}
|g _v^{(0)} \rangle 
-
\langle g _u^{(0)} |
\frac{1}{\tau + H^{(0)}Q^{(0)}}H^{(0)}P^{(0)}
\frac{1}{\tau + H^{(0)}}
|g _v^{(0)} \rangle.
\label{Rmatrix2}
\eeq
The first term is simply calculated as\\[-20pt]
\beqa
\begin{array}{c}
{\displaystyle 
\langle g _u^{(0)} |
\frac{1}{\tau + H^{(0)}Q^{(0)}}
|g _v^{(0)} \rangle 
=
\frac{1}{\tau }}
\sum _{k = 0}^\infty
\langle g _u^{(0)}|
\left( 
{\displaystyle - \frac{H^{(0)}Q^{(0)}}{\tau }}
\right)^k
|g _v^{(0)} \rangle
=
{\displaystyle \frac{1}{\tau }}{\widetilde{\mathbb{S}}_{uv}^{(0)}},
\end{array}
\label{S0}
\eeqa\\[-12pt]
where we have used
$Q^{(0)}|g _v^{(0)} \rangle 
= 
(1 - P^{(0)})|g _v^{(0)} \rangle = 0$
which is easily proved.
Inserting the projection operator $P^{(0)}$ into
the second term in the second line of
(\ref{Rmatrix2}),
it is rewritten as\\[-16pt]
\beqa
\begin{array}{rl}
&\!\!\!
- 
\langle g _u^{(0)} |
{\displaystyle \frac{1}{\tau + H^{(0)}Q^{(0)}}}
H^{(0)}P^{(0)}
{\displaystyle \frac{1}{\tau + H^{(0)}}}
|g _v^{(0)} \rangle \\
\\[-10pt]
=
&\!\!\!
-
{\displaystyle \frac{1}{\tau }}
\sum _{r,s=1}^n
\langle g _u^{(0)} |
{\displaystyle 
\frac{\tau + H^{(0)}Q^{(0)} - H^{(0)}Q^{(0)}}{\tau + H^{(0)}Q^{(0)}}
}
H^{(0)}
|g _r^{(0)} \rangle 
(S^{(0)-1})_{rs} 
\widetilde{\mathbb{R}}_{sv}^{(0)}(\tau ) \\
\\[-12pt]
=
&\!\!\!
{\displaystyle \frac{1}{\tau }}
\sum _{r,s=1}^n
\left[ 
\mathbb{L}_{ur}^{(0)} + \mathbb{M}_{ur}^{(0)}(\tau ) 
\right]
\left(
\mathbb{S}^{(0)-1}
\right)_{rs} 
\widetilde{\mathbb{R}}_{sv}^{(0)}(\tau ) ,
\label{L(0)+M(0)} 
\end{array}
\eeqa
where
$\mathbb{L}_{ur}^{(0)}$ and 
$\mathbb{M}_{ur}^{(0)}(\tau )$
are defined as\\[-10pt]
\beq
\mathbb{L}_{ur}^{(0)}
\equiv
-
\langle g _u^{(0)} |
H^{(0)}
|g _r^{(0)} \rangle , ~~
\mathbb{M}_{ur}^{(0)}(\tau )
\equiv
\langle g _u^{(0)} |
H^{(0)}Q^{(0)}
{\displaystyle \frac{1}{\tau + Q^{(0)}H^{(0)}Q^{(0)}}}
Q^{(0)}H^{(0)}
|g _r^{(0)} \rangle .
\label{L0andM0}
\eeq
In expressing the
$\mathbb{M}_{ur}^{(0)}(\tau )$,
second equation of
(\ref{L0andM0}),
we have used the idempotency relation
$Q^{(0)2} = Q^{(0)}$.
Finally, substituting
(\ref{S0})
and
(\ref{L(0)+M(0)})
into
(\ref{Rmatrix2})
and solving
${\widetilde{\mathbb{R}}^{(0)}(\tau )}$
in equation
(\ref{Rmatrix2})
in a matrix notation, we have\\[-10pt]
\beq
{\widetilde{\mathbb{R}}^{(0)}(\tau )}
=
\frac{1}
{\tau \mathbb{I}
-
\left[
\mathbb{L}^{(0)} + \mathbb{M}^{(0)}(\tau )
\right] 
\mathbb{S}^{(0)-1}}\mathbb{S}^{(0)}.
\label{R0matrix}
\eeq
In the above,
different from the projection method by Fulde
\cite{Fulde.93},
we do not introduce the Liouville operator 
$LA \equiv [H,~A]$
for any operator $A$
but use the original Hamiltonian $H$.
Further,
to calculate the elements of the so-called {\it memory matrix}
$\mathbb{M}_{ur}^{(0)}(\tau )$,
we introduce the state
$
|g _r^{(1)} \rangle
\equiv
Q^{(0)}H^{(0)}|g _r^{(0)} \rangle
$
and the projection operator as

\beqa
\left.
\begin{array}{rl}
P^{(1)}
\equiv
&\!\!\!
\sum _{r,s=1}^n
|g _r^{(1)} \rangle ({\mathbb{S}^{(1)-1}})_{rs} \langle g _s^{(1)} |
=
P^{(1)\dagger }, ~~Q^{(1)} = 1 - P^{(1)} , \\
\\[-8pt]
P^{(1)2}
=
&\!\!\!
P^{(1)} ,~~Q^{(1)2} = Q^{(1)} ,~~
P^{(1)}Q^{(1)} = Q^{(1)}P^{(1)} = 0 ,
\end{array}
\right\}
\label{projectionoperator1}
\eeqa
where
$
{\mathbb{S}^{(1)}}
\!=\!
({\mathbb{S}_{rs}^{(1)}})
\left(
\!=\!
\langle g _r^{(1)} | g _s^{(1)} \rangle
\!=\!
\langle g _r^{(0)} |
H^{(0)}Q^{(0)}H^{(0)}
|g _s^{(0)} \rangle
\right)
$
is an $n \times n$ matrix
and ${\mathbb{S}}^{(1)\dagger }
\!=\!
{\mathbb{S}}^{(1)}$.
Using 
$H^{(1)} \equiv Q^{(0)}H^{(0)}Q^{(0)}$
and
the second equation of
(\ref{L0andM0})
we can express
$\mathbb{M}_{uv}^{(0)}(\tau )$
as\\[-12pt]
\beqa
\begin{array}{rl}
\mathbb{M}_{uv}^{(0)}(\tau )
=
&\!\!\!
\langle g _u^{(1)} |
{\displaystyle
\frac{1}{\tau + H^{(1)}(Q^{(1)} + P^{(1)})}
}
|g _v^{(1)} \rangle
=
{\widetilde{\mathbb{R}}_{uv}^{(1)}(\tau )} \\
\\[-12pt]
=
&\!\!\!
\langle g _u^{(1)} |
{\displaystyle 
\frac{1}{\tau + H^{(1)}Q^{(1)}}
}
|g _v^{(1)} \rangle
-
\langle g _u^{(1)} |
{\displaystyle
\frac{1}{\tau + H^{(1)}Q^{(1)}}
}
H^{(1)}P^{(1)}
{\displaystyle
\frac{1}{\tau + H^{(1)}}
}
|g _v^{(1)} \rangle.
\label{M0matrix}
\end{array}
\eeqa
The first term is simply calculated 
in the same way as 
(\ref{S0}).
Inserting 
the projection operator $P^{(1)}$,
first equation of
(\ref{projectionoperator1}),
into
the second term in second line of
(\ref{M0matrix}),
we have\\[-12pt]
\beqa
\begin{array}{rl}
&\!\!\!
-
\langle g _u^{(1)} |
{\displaystyle \frac{1}{\tau + H^{(1)}Q^{(1)}}}
H^{(1)}P^{(1)}
{\displaystyle \frac{1}{\tau + H^{(1)}}}
|g _v^{(1)} \rangle\\
\\[-10pt]
=
&\!\!\!
-
{\displaystyle \frac{1}{\tau }}
\sum _{r,s=1}^n
\langle g _u^{(1)} |
{\displaystyle
\frac{\tau + H^{(1)}Q^{(1)} - H^{(1)}Q^{(1)}}{\tau + H^{(1)}Q^{(1)}}
}
H^{(1)}
|g _r^{(1)} \rangle
(S^{(1)-1})_{rs}
{\widetilde{\mathbb{R}}_{sv}^{(1)}(\tau )} \\
\\[-10pt]
=
&\!\!\!
{\displaystyle \frac{1}{\tau }}
\sum _{r,s=1}^n
\left[
\mathbb{L}_{ur}^{(1)} + \mathbb{M}_{ur}^{(1)}(\tau )
\right]
(\mathbb{S}^{(1)-1})_{rs} 
{\widetilde{\mathbb{R}}_{sv}^{(1)}(\tau )} ,
\end{array}
\eeqa
\beq
\mathbb{L}_{ur}^{(1)}
\equiv
-
\langle g _u^{(1)} |
H^{(1)}
|g _r^{(1)} \rangle , ~~
\mathbb{M}_{ur}^{(1)}(\tau )
\equiv
\langle g _u^{(1)} |
H^{(1)}Q^{(1)}
{\displaystyle \frac{1}{\tau + Q^{(1)}H^{(1)}Q^{(1)}}}
Q^{(1)}H^{(1)}
|g _r^{(1)} \rangle .
\label{L1andM1}
\eeq
Solving 
${\widetilde{\mathbb{R}}^{(1)}(\tau )}$
of
(\ref{M0matrix})
in a matrix form, we have\\[-8pt]
\beq
{\widetilde{\mathbb{R}}^{(1)}(\tau )}
=
\frac{1}
{\tau \mathbb{I}
-
\left[
\mathbb{L}^{(1)} + \mathbb{M}^{(1)}(\tau )
\right] 
\mathbb{S}^{(1)-1}}\mathbb{S}^{(1)} ,~~
\mathbb{M}^{(1)}(\tau )
=
{\widetilde{\mathbb{R}}^{(2)}(\tau )} .
\label{R1matrix}
\eeq
Substituting
(\ref{R1matrix})
into
(\ref{R0matrix})
and repeating the above procedure,
ultimately we can get the final expression for
${\widetilde{\mathbb{R}}^{(0)}(\tau )}$
in terms of IMCFs
\cite{Akhiezer.65,Moricfrac.65,Dupuis.67}
and
\cite{Wall.48,Jones.Thron.80,BJJK.09,Forster.75,Risken.89,
Fick.Sauermann.90}:\\[-12pt]
\beqa
\begin{array}{c}
\widetilde{\mathbb{R}}^{(0)}(\tau )
=
{\displaystyle
\frac{1}
{
\tau \mathbb{I}
- 
\left[
\mathbb{L}^{(0)} +
{\displaystyle
\frac{1}
{
\tau \mathbb{I}
-
\left[
\mathbb{L}^{(1)} +
{\displaystyle
\frac{1}
{
\tau \mathbb{I}
-
\left[
\mathbb{L}^{(2)} 
+
\cdots
\right]
\mathbb{S}^{(2)-1}
}
}
\mathbb{S}^{(2)}
\right]
\mathbb{S}^{(1)-1}
}
}
\mathbb{S}^{(1)}
\right]
\mathbb{S}^{(0)-1}
}
}
\mathbb{S}^{(0)},
\end{array}
\label{fractalR0matrix}
\eeqa
where we have introduced the following state and
projection operator
\beqa
\left.
\!\!\!
\begin{array}{rl}
&\!\!\!
|g _r^{(l)} \rangle
\equiv
Q^{(l-1)}H^{(l-1)}|g _r^{(l-1)} \rangle
=
Q^{(l-1)}H^{(l-1)}Q^{(l-2)}H^{(l-2)}
\cdots
Q^{(0)}H^{(0)}
|g _r^{(0)} \rangle,\\
\\[-8pt]
&\!\!\!
Q^{(l)}
\equiv
1 
- 
\sum _{r,s=1}^n 
|g _r^{(l)} \rangle (S^{(l)-1})_{rs} \langle g _s^{(l)} | ,~
Q^{(l)2} = Q^{(l)} , ~
\mathbb{S}_{rs}^{(l)}
\equiv
\langle g _r^{(l)} | g _s^{(l)} \rangle,~(l \!\ge\! 2)
\end{array}
\right\}
\eeqa
and $\tau$-independent matrix element
$
\mathbb{L}_{ur}^{(l)}
\!\equiv\!
-
\langle g _u^{(l)} |
H^{(l)}
|g _r^{(l)} \rangle
$
where
$
H^{(l)}
\!\equiv\!
Q^{(l-1)}H^{(l-1)}Q^{(l-1)} 
$.
The first few matrices
$\mathbb{S}_{rs}^{(l)}$
and 
$\mathbb{L}_{ur}^{(l)}$
are given in Appendix A.
Finally, going back to the original expression
${\widetilde{\mathbb{R}}_{rs}}
(\!=\! {\widetilde{\mathbb{R}}_{rs}^{(0)}})$
and making the inverse Laplace transform
${\cal L}^{-1}$,
we can reach the desired partition function
in the Res-HB subspace as
\beqa
\begin{array}{c}
\mbox{Tr} (Pe^{- \beta H})
=
\sum _{r,s=1}^n
{\cal L}^{-1}
\left\{
{\widetilde{\mathbb{R}}_{rs}(\tau )}
\right\}
(S^{-1})_{sr}
=
\sum _{r,s=1}^n
{\mathbb{R}_{rs}(\beta )}
(S^{-1})_{sr} .
\end{array}
\label{invLaptraceformP}
\eeqa

The form of IMCF
(\ref{fractalR0matrix})
is also derived from the so-called
tridiagonal vector recurrence relation for
a first-order time derivative of
${\bf c}_{m}(t)$,
\beq
\dot{\bf c}_{m}(t)
=
\mathbb{Q}_{m}^{-}{\bf c}_{m-1}(t)
+
\mathbb{Q}_{m}{\bf c}_{m}(t)
+
\mathbb{Q}_{m}^{+}{\bf c}_{m+1}(t)  .
\label{tridiagonal}
\eeq
${\bf c}_{m}(t)$
is a time-dependent $n$-dimensional columun vector
and
$\mathbb{Q}_{m}^{\pm },~\mathbb{Q}_{m}$
are time-independent  and constant $n  \!\times\! n$ matrices.
$\!\!$To solve
(\ref{tridiagonal}),
following exactly the method developed by Risken
\cite{Risken.89}
(p.p. 217-219),
let us introduce a Green's function matrix
$\mathbb{G}_{m,m^\prime }(t)$.
The general solution of
(\ref{tridiagonal})
is expressed in terms of the Green's function matrix
with an initial condition as
\beqa
\begin{array}{c}
{\bf c}_{m}(t)
=
\sum_{m^\prime = 0 }^{\infty }
\mathbb{G}_{m,m^\prime }(t) {\bf c}_{m}(0) ,~~
\mathbb{G}_{m,m^\prime }(0)
=
\mathbb{I} \delta_{m,m^\prime } .
\end{array}
\label{generalsol}
\eeqa
Substituting
(\ref{generalsol}) into (\ref{tridiagonal})
and making a Laplace transform
$
\widetilde{\mathbb{G}}_{m,m^\prime }(\tau)
\!\!=\!\!
\int_0^{\infty } \!\! e^{-\tau t} \mathbb{G}_{m,m^\prime }(t)dt
$,
we obtain the following relation as a sufficient condition
for the solution of
(\ref{tridiagonal}):
\beq
\mathbb{Q}_{m}^{-}\widetilde{\mathbb{G}}_{m-1,m^\prime }(\tau)
+
\left(\mathbb{Q}_{m} - \tau \mathbb{I}\right)
\widetilde{\mathbb{G}}_{m,m^\prime }(\tau)
+
\mathbb{Q}_{m}^{+}\widetilde{\mathbb{G}}_{m+1,m^\prime }(\tau)
=
- \mathbb{I} \delta_{m,m^\prime } .
\label{tridiagonaleq}
\eeq
Further introduce a matrix
$\widetilde{\mathbb{S}}_m^+(\tau) $
which raises the number of the left index of
the Laplace transformed Green's function matrix
through the relation
\beq
\widetilde{\mathbb{G}}_{m+1,m^\prime }(\tau)
=
\widetilde{\mathbb{S}}_m^+(\tau) 
\widetilde{\mathbb{G}}_{m,m^\prime }(\tau) .
\label{raisingSmatrix}
\eeq
For the moment neglecting the inhomogeneous term in
(\ref{tridiagonaleq}),
we have
\beq
\mathbb{Q}_{m}^{-}\widetilde{\mathbb{G}}_{m-1,m^\prime }(\tau)
\!+\!
\left[ \!
\mathbb{Q}_{m} - \tau \mathbb{I}
+
\mathbb{Q}_{m}^{+}\widetilde{\mathbb{S}}_m^+(\tau) \!
\right] \!
\widetilde{\mathbb{G}}_{m,m^\prime }(\tau)
=
0 .
\label{tridiagonalSeq}
\eeq
By multiplying
(\ref{tridiagonalSeq})
with the inverse of the matrix in the parenthesis
and by comparing the result with the relation
$
\widetilde{\mathbb{G}}_{m,m^\prime }(\tau)
\!=\!
\widetilde{\mathbb{S}}_{m-1}^+(\tau) 
\widetilde{\mathbb{G}}_{m-1,m^\prime }(\tau)
$,
which is derived from
(\ref{raisingSmatrix}),
we can immediately obtain
$
\widetilde{\mathbb{S}}_{m-1}^+(\tau) 
\!\!=\!\!
\left[ \!
\tau \mathbb{I} \!-\! \mathbb{Q}_{m}
\!-\!
\mathbb{Q}_{m}^{+}\widetilde{\mathbb{S}}_m^+(\tau) \!
\right]^{\!-1} \!
\mathbb{Q}_{m}^{-}
$.
By the iteration we get 
the following relation in terms of IMCFs:\\[-10pt]
\beq
\widetilde{\mathbb{S}}_m^+(\tau) 
\!=\!
\left[ \!
\tau \mathbb{I} \!-\! \mathbb{Q}_{m+1}
\!-\!
\mathbb{Q}_{m+1}^{+} \!
\left[ \!
\tau \mathbb{I} \!-\! \mathbb{Q}_{m+2}
\!-\!
\mathbb{Q}_{m+2}^{+} \!
\left[ \!
\tau \mathbb{I} \!-\! \mathbb{Q}_{m+3}
\!-\!
\cdots \!
\right]^{-1}
\mathbb{Q}_{m+3}^{-} \!
\right]^{-1}
\mathbb{Q}_{m+2}^{-} \!
\right]^{-1} \!\!
\mathbb{Q}_{m+1}^{-} .
\label{infinitematrixcontinuedfraction}
\eeq
Putting $m \!=\! 0$,
we can obtain the final expression for
$\widetilde{\mathbb{S}}_{0}^{+}(\tau )$
as\\[-14pt]
\beqa
\begin{array}{c}
\widetilde{\mathbb{S}}_{0}^{+}(\tau )
=
{\displaystyle
\frac{1}
{
\tau \mathbb{I}
-
\mathbb{Q}_{1} - \mathbb{Q}_{1}^{+}
{\displaystyle
\frac{1}
{
\tau \mathbb{I}
-
\mathbb{Q}_{2} - \mathbb{Q}_{2}^{+}
{\displaystyle
\frac{1}
{
\tau \mathbb{I} 
-
\mathbb{Q}_{3} 
-
\cdots
}
}
\mathbb{Q}_{3}^{-}
}
}
\mathbb{Q}_{2}^{-}
}
}
\mathbb{Q}_{1}^{-} ,
\end{array}
\label{fractalmatrix}
\eeqa
which is
quite similar to the form of
(\ref{fractalR0matrix}).
In the above
if we choose
$
\widetilde{\mathbb{S}}_{0}^{+}(\tau ),~
\mathbb{Q}_{m},~\mathbb{Q}_{m}^{+}
$
and
$\mathbb{Q}_{m}^{-}$
as\\[-16pt]
\beqa
\left.
\begin{array}{llll}
&
\widetilde{\mathbb{S}}_{0}^{+}(\tau )
=
\widetilde{\mathbb{R}}_{0}(\tau ) , &
\mathbb{S}^{(-1)} \!=\! \mathbb{I} , & \\
\\
&
\mathbb{Q}_{m}
=
\mathbb{L}^{(m-1)}
\mathbb{S}^{(m-1)-1},&
\mathbb{Q}_{m}^{+}
=
\mathbb{I} ,
&
\mathbb{Q}_{m}^{-}
=
\mathbb{S}^{(m-1)}
\mathbb{S}^{(m-2)-1} ,
\end{array}
\right\}
\label{coincidence}
\eeqa
for $m \!=\! 1,~2, \cdots$,
respectively,
the IMCF
(\ref{fractalmatrix})
just coincides with
the previous IMCF
(\ref{fractalR0matrix}).

\newpage

%%%%%%%%%%%%%%%%%%%%%%%%%%%%%%%%%
%                                                                                             %
%   4          Expression for thermal HB density matrix             %
%                                                                                             %
%      in terms of eigenvalues of thermal Res-FB operator       %
%                                                                                             %
%%%%%%%%%%%%%%%%%%%%%%%%%%%%%%%%%

\def\thesection{\arabic{section}}
\setcounter{equation}{0}
\renewcommand{\theequation}{\arabic{section}.\arabic{equation}}
\section{Expression for thermal HB density matrix
in terms of eigenvalues of thermal Res-FB operator} 

~~~~First we give
the commutability relation
and
the equivalence relation
\beqa
[{\cal F}_{\mbox{{\scriptsize Res}}:r}
^{\mbox{{\scriptsize thermal}}} ,
W_{\mbox{{\scriptsize Res}}:rr}^{\mbox{{\scriptsize thermal}}}] = 0 ,
\label{modificationofWFW1}
\eeqa
\vspace{-0.8cm}
\beqa
\begin{array}{c}
\sum_{k =1}^n \sum_{s = 1} ^n
{\cal K}_{\mbox{{\scriptsize Res}}:rs}
^{\mbox{{\scriptsize thermal}}(k)}
c_r ^{(k)*} c_s^{(k)}
\equiv
{\cal F}_{\mbox{{\scriptsize Res}}:r}^{\mbox{{\scriptsize thermal}}}
W_{\mbox{{\scriptsize Res}}:rr}^{\mbox{{\scriptsize thermal}}}
\!-\!
W_{\mbox{{\scriptsize Res}}:rr}^{\mbox{{\scriptsize thermal}}}
{\cal F}_{\mbox{{\scriptsize Res}}:r}^{\mbox{{\scriptsize thermal}}}
W_{\mbox{{\scriptsize Res}}:rr}^{\mbox{{\scriptsize thermal}}}.
\end{array}
\label{modificationofWFW2}
\eeqa
where
${\cal K}_{\mbox{{\scriptsize Res}}:rs}^{\mbox{{\scriptsize thermal}}(k)}$
is defined as
\beqa
{\cal K}_{\mbox{{\scriptsize Res}}:rs}
^{\mbox{{\scriptsize thermal}}(k)}
\!\!\equiv\!\!
\left\{ \!
(1_{2N} \!\!-\!\! W_{\mbox{{\scriptsize Res}}:rs}
^{\mbox{{\scriptsize thermal}}})
{\cal F} [W_{\mbox{{\scriptsize Res}}:rs}
^{\mbox{{\scriptsize thermal}}}]
\!\!+\!\! 
H[W_{\mbox{{\scriptsize Res}}:rs}
^{\mbox{{\scriptsize thermal}}}]
\!\!-\!\!
E^{(k)} \!
\right\}
\!\cdot\! 
W_{\mbox{{\scriptsize Res}}:rs}
^{\mbox{{\scriptsize thermal}}}
\!\cdot\! 
[\det z^{\mbox{{\scriptsize thermal}}}_{rs}]^{\frac{1}{2}} .
\label{thermalRes-HBequation}
\eeqa
Both the relations
(\ref{modificationofWFW1}) and (\ref{modificationofWFW2})
are proved in Appendix B.
As shown in II,
using
(\ref{modificationofWFW2}),
the direct variation of the Res-HB free energy leads us to
the $r$th thermal HB density matrix
$W_{\mbox{{\scriptsize Res}}:rr}
^{\mbox{{\scriptsize thermal}}}$,
which is expressed in terms of
the $r$th thermal Res-FB operator
${\cal F}_{\mbox{{\scriptsize Res}}:r}
^{\mbox{{\scriptsize thermal}}}$,
as
\beqa
W_{\mbox{{\scriptsize Res}}:rr}
^{\mbox{{\scriptsize thermal}}}
= 
\frac{1}
{1_{2N} + \exp 
\{ \beta {\cal F} _{\mbox{{\scriptsize Res}}:r}
^{\mbox{{\scriptsize thermal}}}
\}
} ,~(r \!=\! 1, \cdots, n) ,
\label{solutionWrr}
\eeqa
where we have used again the relation
$[{\cal F}_{\mbox{{\scriptsize Res}}:r}
^{\mbox{{\scriptsize thermal}}} , 
W_{\mbox{{\scriptsize Res}}:rr}
^{\mbox{{\scriptsize thermal}}}] \!=\! 0$.
We rewrite
$
W_{\mbox{{\scriptsize Res}}:rr}
^{\mbox{{\scriptsize thermal}}}
$
as 
$
W_{\mbox{{\scriptsize Res}}:rr}
^{\mbox{{\scriptsize thermal}}\ast }
$
for later convenience.
By using a Bogoliubov transformation $g_r$,
the
$
W_{\mbox{{\scriptsize Res}}:rr}
^{\mbox{{\scriptsize thermal}}\ast }
$ 
is diagonalized as follows:
\beqa
\begin{array}{rl}
\widetilde{W}_r
&\!\!\!\!
= 
g_r ^{\dagger }
W_{\mbox{{\scriptsize Res}}:rr}
^{\mbox{{\scriptsize thermal}}\ast }
g_r \\
\\
& \!\!\!\!
=
\left[
\begin{array}{cccccc}
\widetilde{w}_{r1}&&&&&\\
&\ddots&&&0&\\
&&\widetilde{w}_{rN}&&&\\
&&&1 - \widetilde{w}_{r1}&&\\
&0&&&\ddots&\\
&&&&&1 - \widetilde{w}_{rN}
\end{array}
\right]
=
\left[
\begin{array}{cc}
\widetilde{w}_r&0\\
&\\
0&1_N - \widetilde{w}_r
\end{array}
\right] ,
\end{array}
\label{tildeCalFrmatrix}
\eeqa
where
\beqa
\widetilde{w}_{ri}
=
\frac{1}
{1
+
\exp 
\left\{
\beta \widetilde{\epsilon }_{ri}^{\mbox{{\scriptsize thermal}}}
\right\}
},~~
1 - \widetilde{w}_{ri}
=
\frac{1}
{1
+
\exp 
\left\{
-\beta \widetilde{\epsilon }_{ri}^{\mbox{{\scriptsize thermal}}}
\right\}
}.~~
(r = 1, \cdots, n)
\eeqa
The diagonalization of
the $r$th thermal Res-FB operator
${\cal F}_{\mbox{{\scriptsize Res}}:r}
^{\mbox{{\scriptsize thermal}}}$
by the same Bogoliubov transformation $g_r$
leads to the eigenvalue
$\epsilon_{ri}^{\mbox{{\scriptsize thermal}}}$.
To this eigenvalue by adding a term
$\sum_{k=1} ^n
\left\{ 
H[W_{\mbox{{\scriptsize Res}}:rr}
^{\mbox{{\scriptsize thermal}}}]
\! - \! E^{(k)}
\right\}
|c_r ^{(k)} |^2
\cdot 1_{2N}$,
the usual HB type of the eigenvalue
$\widetilde{\epsilon }_{ri}^{\mbox{{\scriptsize thermal}}}$
is realized.
Using 
(\ref{tildeCalFrmatrix}) 
we can derive
the inverse transformation of
(\ref{tildeCalFrmatrix})
in the form
\beqa
\!\!\!\!
\begin{array}{rl}
g_r ^{\dagger } W_{\mbox{{\scriptsize Res}}:rr}
^{\mbox{{\scriptsize thermal}}\ast }
[{\cal F}_{\mbox{{\scriptsize Res}}:r}
^{\mbox{{\scriptsize thermal}}}] g_r
&\!\!\!\!
\! = \!
g_r ^{\dagger }
{\displaystyle
\frac{1}
{1_{\! 2N}
\!\! + \!\!
\exp \!
\left[ \!
\beta \!
\left( \!
{\cal F}_{\mbox{{\scriptsize Res}}:r}
^{\mbox{{\scriptsize thermal}}}
\!\!+ \!\!
\sum_{k=1} ^n \!
\left\{ \!
H[W_{\mbox{{\scriptsize Res}}:rr}
^{\mbox{{\scriptsize thermal}}}]
\!\!-\!\!
E^{(k)} \!
\right\} \!
|c_r ^{(k)}|^2
\!\cdot\!
1_{\! 2N} \!
\right) \!
\right]
}  g_r \!
} \\
\\
&\!\!\!\!
=
\widetilde{W}_r
= \!
\left[ \!
\begin{array}{cc}
\widetilde{w}_r&0\\
&\\
0&1_N - \widetilde{w}_r \!
\end{array}
\right] .
\end{array}
\label{CalFrmatrix}
\eeqa
Inversely transforming again,
the desired $r$th thermal HB density matrix is obtained as
\beqa
\begin{array}{rl}
W_{\mbox{{\scriptsize Res}}:rr}
^{\mbox{{\scriptsize thermal}}}
&\!\!\!
=
g_r ^\ast \widetilde{W}_r g_r ^{\mbox{{\scriptsize T}}}  
=
\left[ \!
\begin{array}{cc}
a_r ^\ast&b_r\\
&\\
b_r ^\ast&a_r
\end{array} \!
\right] \!\!
\left[
\begin{array}{cc}
\widetilde{w}_r&0\\
&\\
0&1_N - \widetilde{w}_r
\end{array}
\right] \!\!
\left[
\begin{array}{cc}
a_r ^{\mbox{{\scriptsize T}}}&b_r ^{\mbox{{\scriptsize T}}}\\
&\\
b_r ^{\dagger }&a_r ^{\dagger }
\end{array}
\right] \\
\\
&\!\!\!
= \!
\left[ \!
\begin{array}{cc}
a_r ^\ast \widetilde{w}_r a_r ^{\mbox{{\scriptsize T}}} 
+
b_r (1_N - \widetilde{w}_r ) b_r ^{\dagger }
&a_r^\ast \widetilde{w}_r b_r ^{\mbox{{\scriptsize T}}} 
+
b_r (1_N - \widetilde{w}_r ) a_r ^{\dagger }
\\
&\\
b_r ^\ast \widetilde{w}_r a_r ^{\mbox{{\scriptsize T}}}
+
a_r (1_N - \widetilde{w}_r ) b_r ^{\dagger }
&b_r ^\ast \widetilde{w}_r b_r ^{\mbox{{\scriptsize T}}}
+
a_r (1_N - \widetilde{w}_r ) a_r ^{\dagger }
\\
\end{array} \!
\right] ,
\end{array}
\label{Wrrmatrix}
\eeqa
from which and the thermal density matrix
$W_{\mbox{{\scriptsize Res}}:rs}
^{\mbox{{\scriptsize thermal}}}$
defined by the second equation of
(2.6) in II
but with $s = r$,
we have\\[-16pt]
\beqa
\left.
\begin{array}{rl}
R_{\mbox{{\scriptsize Res}}:rr}
^{\mbox{{\scriptsize thermal}}}
=
a_r ^\ast \widetilde{w}_r a_r ^{\mbox{{\scriptsize T}}}
+
b_r (1_N - \widetilde{w}_r ) b_r ^{\dagger },\\
\\
K_{\mbox{{\scriptsize Res}}:rr}
^{\mbox{{\scriptsize thermal}}}
=
a_r ^\ast \widetilde{w}_r b_r ^{\mbox{{\scriptsize T}}}
+
b_r (1_N - \widetilde{w}_r ) a_r ^{\dagger }.
\end{array} \!
\right\}
\label{temperaturedependentWrrblockmatrix}
\eeqa
Further substituting 
(\ref{temperaturedependentWrrblockmatrix})
into $F$ and $D$ matrices given by
(\ref{FandDoperators}),
we can obtain\\[-10pt]
\beqa
\left.
\begin{array}{lr}
F_{\mbox{{\scriptsize Res}}:rr;\alpha \beta }
^{\mbox{{\scriptsize thermal}}}
=
h_{\alpha \beta }
\!+\!
[\alpha \beta |\gamma \delta]
R_{\mbox{{\scriptsize Res}}:rr;\delta \gamma }
^{\mbox{{\scriptsize thermal}}} , \\
\\[-6pt]
D_{\mbox{{\scriptsize Res}}:rr;\alpha \beta }
^{\mbox{{\scriptsize thermal}}}
=
-
\displaystyle{\frac{1}{2}}
[\alpha \gamma |\beta \delta]
K_{\mbox{{\scriptsize Res}}:rr;\delta \gamma}
^{\mbox{{\scriptsize thermal}}} .
\end{array} \!
\right\}
\label{temperaturedependentFandD}
\eeqa
The above formulas are the direct extension of the result
and the condition for the self-consistent field (SCF) given
by the single HB WF
\cite{Ozaki.85,Goodman.81} %,AKPY.93}
to those by the multiple HB WFs.
The explicit temperature dependence of 
the thermal HB interstate density matrix
$
W_{\mbox{{\scriptsize Res}}:rs}
^{\mbox{{\scriptsize thermal}}}
$,
however,
can not be determined directly
within the framework of the present theory.

The whole Res-HB subspace and
the thermal Res-FB operator are represented
in the forms of the direct sum, respectively, as\\[-20pt]
\beqa
\begin{array}{rl}
&g
=
\left[
\begin{array}{ccccc}
g_1&&&&\\
&\ddots&&0&\\
&&g_r&&\\
&0&&\ddots&\\
&&&&g_n
\end{array}
\right]
= 
\sum_{r=1} ^n \oplus g_r,~~
g_r
=
\left[
\begin{array}{cc}
a_r&b_r ^\ast\\
&\\
b_r&a_r ^\ast
\end{array}
\right] ,
\end{array}
\label{gmatrix}
\eeqa
\vspace{-0.5cm}
\beqa
\begin{array}{rl}
&{\cal F}_{\mbox{{\scriptsize Res}}}
^{\mbox{{\scriptsize thermal}}}
=
\left[
\begin{array}{ccccc}
{\cal F}_{\mbox{{\scriptsize Res}}:1}
^{\mbox{{\scriptsize thermal}}}&&&&\\
&\ddots&&0&\\
&&{\cal F}_{\mbox{{\scriptsize Res}}:r}
^{\mbox{{\scriptsize thermal}}}&&\\
&0&&\ddots&\\
&&&&{\cal F}_{\mbox{{\scriptsize Res}}:n}
^{\mbox{{\scriptsize thermal}}}
\end{array}
\right]
=
\sum_{r=1} ^n \oplus {\cal F}_{\mbox{{\scriptsize Res}}:r}
^{\mbox{{\scriptsize thermal}}} .
\end{array}
\label{CalFmatrix}
\eeqa
The HB density matrix in the whole Res-HB subspace
$W_{\mbox{{\scriptsize Res}}}$
is also given in the form of the direct sum
$\sum_{r=1} ^n \oplus W_{\mbox{{\scriptsize Res}}:rr}$.
On the other hand,
the thermal HB density matrix in the whole Res-HB subspace
is also given in the form of the direct sum\\[-20pt]
\beqa
\begin{array}{l}
W_{\mbox{{\scriptsize Res}}}
^{\mbox{{\scriptsize thermal}}}
=
g ^\ast \widetilde{W}g ^{\mbox{{\scriptsize T}}}
= 
\sum_{r=1} ^n \oplus W_{\mbox{{\scriptsize Res}}:rr}
^{\mbox{{\scriptsize thermal}}}
,~~
W_{\mbox{{\scriptsize Res}}:rr}
^{\mbox{{\scriptsize thermal}}}
=
g_r ^\ast \widetilde{W}_r g_r ^{\mbox{{\scriptsize T}}},
\end{array}
\label{WrrCalFrmatrix}
\eeqa
however,
in which
the idempotency relation
$
W_{\mbox{{\scriptsize Res}}:rr}
^{\mbox{{\scriptsize thermal}}~2}
\!=\!
W_{\mbox{{\scriptsize Res}}:rr}
^{\mbox{{\scriptsize thermal}}}
$
is no longer approved.

Up to the present stage,
we have not any knowledge concerning
the thermal HB interstate density matrix.
Although it is not a strict manner,
for the time being,
the $rs$-component of the thermal HB interstate density matrix
is reasonably approximated as\\[-16pt]
\beqa
\!\!\!\!\!\!
\begin{array}{rl}
&\!\!\!
W_{\mbox{{\scriptsize Res}}:rs}
^{\mbox{{\scriptsize thermal}}}
\!\approx\!
g_r ^\ast \sqrt{\widetilde{W}_r} \sqrt{\widetilde{W}_s}
g_s^{\mbox{{\scriptsize T}}}
\!=\!
\left[ \!\!
\begin{array}{cc}
a_r ^\ast &b_r \\
&\\[-4pt]
b_r ^\ast &a_r
\end{array} \!\!
\right] \!\!
\left[ \!\!\!
\begin{array}{cc}
\sqrt{\widetilde{w}_r}&0\\
&\\[-4pt]
0&\sqrt{1_N - \widetilde{w}_r}
\end{array} \!\!
\right] \!\!
\left[ \!\!\!
\begin{array}{cc}
\sqrt{\widetilde{w}_s}&0\\
&\\[-4pt]
0&\sqrt{1_N - \widetilde{w}_s}
\end{array} \!\!
\right] \!\!
\left[ \!\!
\begin{array}{cc}
a_s ^{\mbox{{\scriptsize T}}}&b_s ^{\mbox{{\scriptsize T}}}\\
&\\[-4pt]
b_s ^{\dagger }&a_s ^{\dagger }
\end{array} \!\!
\right] \\
\\[-8pt]
&\!\!\!
= \!
\left[ \!\!
\begin{array}{cc}
a_r ^\ast \sqrt{\widetilde{w}_r}\sqrt{\widetilde{w}_s}
a_s ^{\mbox{{\scriptsize T}}}
\!+\!
b_r \sqrt{1_N \!-\! \widetilde{w}_r}
\sqrt{1_N \!-\! \widetilde{w}_s} b_s ^{\dagger }
&a_r ^\ast \sqrt{\widetilde{w}_r}\sqrt{\widetilde{w}_s}
b_s ^{\mbox{{\scriptsize T}}}
\!+\!
b_r \sqrt{1_N \!-\! \widetilde{w}_r}
\sqrt{1_N \!-\! \widetilde{w}_s} a_s ^{\dagger } \\
&\\[-4pt]
b_r ^\ast \sqrt{\widetilde{w}_r}\sqrt{\widetilde{w}_s}
a_s ^{\mbox{{\scriptsize T}}}
\!+\!
a_r \sqrt{1_N \!-\! \widetilde{w}_r}
\sqrt{1_N \!-\! \widetilde{w}_s} b_s ^{\dagger }
&b_r ^\ast \sqrt{\widetilde{w}_r}\sqrt{\widetilde{w}_s}
b_s^{\mbox{{\scriptsize T}}}
\!+\!
a_r \sqrt{1_N \!-\! \widetilde{w}_r}
\sqrt{1_N \!-\! \widetilde{w}_s} a_s ^{\dagger }
\end{array} \!\!
\right] .
\end{array}
\label{Wrsmatrix}
\eeqa\\[-10pt]
Needless to say,
the idempotency relation
$
W_{\mbox{{\scriptsize Res}}:rs}
^{\mbox{{\scriptsize thermal}}~2}
\!=\! 
W_{\mbox{{\scriptsize Res}}:rs}
^{\mbox{{\scriptsize thermal}}}
$
is also no longer approved
though the property
$
W_{\mbox{{\scriptsize Res}}:rs}
^{\mbox{{\scriptsize thermal}}~\dag }
\!=\! 
W_{\mbox{{\scriptsize Res}}:sr}
^{\mbox{{\scriptsize thermal}}}
$
is satisfied.
From
(\ref{Wrsmatrix})
and the thermal density matrix
$
W_{\mbox{{\scriptsize Res}}:rs}
^{\mbox{{\scriptsize thermal}}}
$
defined by the same form as that given by
(\ref{generalinterstatedensitymatrix}),
we have\\[-16pt]
\beqa
\left.
\begin{array}{rl}
R_{\mbox{{\scriptsize Res}}:rs}
^{\mbox{{\scriptsize thermal}}}
=
a_r ^\ast \sqrt{\widetilde{w}_r}\sqrt{\widetilde{w}_s}
a_s ^{\mbox{{\scriptsize T}}}
+
b_r\sqrt{1_N - \widetilde{w}_r}
\sqrt{1_N - \widetilde{w}_s} b_s ^{\dagger },\\
\\[-6pt]
K_{\mbox{{\scriptsize Res}}:rs}
^{\mbox{{\scriptsize thermal}}}
=
a_r ^\ast \sqrt{\widetilde{w}_r}\sqrt{\widetilde{w}_s}
b_s ^{\mbox{{\scriptsize T}}}
+
b_r \sqrt{1_N - \widetilde{w}_r}
\sqrt{1_N - \widetilde{w}_s} a_s ^{\dagger }.
\end{array}
\right\}
\label{temperaturedependentWrsblockmatrix}
\eeqa\\[-12pt]
Further substituting 
(\ref{temperaturedependentWrsblockmatrix}) 
into $F$ and $D$ matrices given by
(\ref{FandDoperators}),
we can obtain the matrices
$
F_{\mbox{{\scriptsize Res}}:rs;\alpha \beta }
^{\mbox{{\scriptsize thermal}}}
~\mbox{and}~
D_{\mbox{{\scriptsize Res}}:rs;\alpha \beta }
^{\mbox{{\scriptsize thermal}}}
$
which have the same forms as those given by
(\ref{temperaturedependentFandD}).
Of course, the temperature-dependent overlap integral
$
z_{\mbox{{\scriptsize Res}}:rs}
^{\mbox{{\scriptsize thermal}}}
$
can not been sought in the same form as that
given by the second equation of
(\ref{zandW}).
Therefore, we are forced to take it in a suitable way
as far as possible.
For the moment, we assume it to be the form as\\[-18pt]
\beqa
\begin{array}{rl}
z_{\mbox{{\scriptsize Res}}:rs}
^{\mbox{{\scriptsize thermal}}}
\approx &\!\!\!\!
\begin{array}{cc}
[\sqrt{\widetilde{w}_r}b_r ^\dag ,&\! \sqrt{\widetilde{w}_r}a_r ^\dag ]
\end{array} \!\!\!
\left[ \!
\begin{array}{c}
b_s \sqrt{\widetilde{w}_s}\\
\\[-4pt]
a_s \sqrt{\widetilde{w}_s}
\end{array} \!
\right]
\!\!+\!\!
\begin{array}{cc}
[\sqrt{1_N \!-\! \widetilde{w}_r}b_r ^\dag ,
&\! \sqrt{1_N \!-\! \widetilde{w}_r}a_r ^\dag ]
\end{array} \!\!\!
\left[ \!
\begin{array}{c}
b_s \sqrt{1_N \!-\! \widetilde{w}_s}\\
\\[-4pt]
a_s \sqrt{1_N \!-\! \widetilde{w}_s}
\end{array} \!
\right] \\
\\[-8pt]
=&\!\!\!\!
\sqrt{\widetilde{w}_r}
(a_r ^\dag a_s \!+\! b_r ^\dag b_s )
\sqrt{\widetilde{w}_s}
\!+\!
\sqrt{1_N \!-\! \widetilde{w}_r}
(a_r ^\dag a_s \!+\! b_r ^\dag b_s )
\sqrt{1_N \!-\! \widetilde{w}_s} .
\end{array}
\label{zthermalapprxrs}
\eeqa\\[-8pt]
Equations
(\ref{temperaturedependentWrrblockmatrix}),
(\ref{temperaturedependentFandD})
and
(\ref{temperaturedependentWrsblockmatrix})
yield the  temperature dependence of
the HB interstate density matrix,
Hamiltonian matrix element
and
FB operator and
equation
(\ref{zthermalapprxrs})
gives approximately a temperature-dependent overlap integral.
Then all of these determine the temperature dependence of the
$
{\cal K}_{\mbox{{\scriptsize Res}}:rs}
^{\mbox{{\scriptsize thermal}}(k)}
$\!
(\ref{thermalRes-HBequation}).
Thus, 
the thermal Res-HB coupled eigenvalue equations
together with the thermal Res-FB operator
(\ref{thermalRes-HBeigenvalueequation0}),
which are given later,
are naturally set out
within the framework of the present formal theory.

However, a calculation of the partition function by the IMCF
is very cumbersome and a method of solving
the above equations is too complicated to execute.
As suggested by Fukutome
\cite{Fuku.88}, 
we pay attention mainly to a contribution from a diagonal part in the formula
$\mbox{Tr}(Pe^{-\beta H})$
(\ref{traceformP}).
Then the partition function in the Res-HB subspace
is approximately calculated as\\[-24pt]
\beqa
\begin{array}{c}
\mbox{Tr} (Pe^{- \beta H})
\!\approx\!
\sum _{r=1}^n
\langle g _r |
e^{- \beta H[W_{rr}]}e^{- \beta <H>_{g_r}}
|g _r \rangle (S^{-1})_{rr} ,
\end{array}
\label{ApproxTraceformP}
\eeqa\\[-18pt]
where
$H[W_{rr}]$
corresponds to the $r$-th Res-HB energy functional and
$<H>_{g_r}$
is the quasi-particle Hamiltonian in the $g_{r}$ quasi-particle frame
whose explicit expression is given later.
The formula
(\ref{ApproxTraceformP})
shows that if the Res-HB energy functional
has multiple low energy local minima
then they give a large contribution to the partition function
and the $resonon$ spectrum recovers the energy levels of the system
in the previous section.
Based on such a useful observation, 
we will seek for another possible and more practical way of
approximating a partition function and a free energy
within the Res-MFT.
This is made in the next section.

\newpage

%%%%%%%%%%%%%%%%%%%%%%%%%%%%%%%%%
%                                                                                             %
%  5  Approximation of partition function and free energy     %
%                                                                                             %
%%%%%%%%%%%%%%%%%%%%%%%%%%%%%%%%%

\def\thesection{\arabic{section}}
\setcounter{equation}{0}
\renewcommand{\theequation}{\arabic{section}.\arabic{equation}}
\section{Approximation of partition function and
free energy}
 
~~
Here,
a more practical way to the Res-HB theory is given.
To calculate approximately
a partition function in the Res-HB subspace,
consider  
a matrix-valued variable ${\cal Z}$
with the same form as the one of
(\ref{generalFBoperator})
and
introduce a quadratic HB Hamiltonian\\[-18pt]
\beqa
H[{\cal Z}]
\!\equiv\!
\frac{1}{2}
[c^{\dagger },~c]{\cal Z} \!
\left[ \!\!
\begin{array}{c}
c ,\\
\\[-12pt]
c^{\dagger }
\end{array} \!\!
\right]
,~
{\cal Z}^{\dagger } \!=\! {\cal Z}
\!=\!
\left[ \!\!
\begin{array}{cc}
F&D \\
\\[-10pt]
-D^\ast &-F^\ast
\end{array} \!\!
\right] ,
\begin{array}{c}
F
= (F_{\alpha \beta }) , \\
\\[-8pt]
D
= (D_{\alpha \beta }) ,
\end{array} \!
\label{quadraticHBHamiltonian}
\eeqa\\[-16pt]
though we use the same symbols,
$F$ and $D$ are not identical with the $F$ and $D$ used in
(\ref{generalFBoperator}).
In the statistical density matrix
$\stackrel{\circ }{W} \! (\!=\! e^{-\beta H} \! /\mbox{Tr}(e^{-\beta H}))$,
instead of the original Hamiltonian $H$,
we adopt the above quadratic Hamiltonian
$H[{\cal Z}] $
(\ref{quadraticHBHamiltonian}).
Then,  the approximate free energy, i.e.,
the usual HB free energy
$F[{\cal Z}] $
is given in the following form
\cite{Ozaki.85}:\\[-20pt]
\beqa
\!\!\!\!
F \! [{\cal Z}] 
\!=\! 
\langle H \!\!-\!\! H \! [{\cal Z}] \rangle _{\!{\cal Z}}
\!-\!
{\displaystyle \frac{1}{\beta }} 
\ln \! \mbox{Tr} (e^{- \beta H[{\cal Z}]}) , 
\langle H \rangle _{\!{\cal Z}} 
\!\equiv\!
{\displaystyle \frac{\mbox{Tr}(e^{- \beta H[{\cal Z}]} \! H)}
{\mbox{Tr} (e^{- \beta H[{\cal Z}]})}}, 
\langle H[{\cal Z}] \rangle _{\!{\cal Z}}
\!\equiv\!
{\displaystyle 
\frac{\mbox{Tr}(e^{- \beta H[{\cal Z}]} \! H \! [{\cal Z}])}
{\mbox{Tr} (e^{- \beta H[{\cal Z}]})}
}.
\label{usualHBfreeenergyandtraceform}
\eeqa\\[-12pt]
A natural extension of the HB free energy to
the Res-HB free one is easily made.
We strongly assume 
${\cal Z} \!=\! \sum_{r=1} ^n \! \oplus {\cal F}_{\! r}$
where ${\cal F}_{\! r}$
is the already known Res-FB operator
(\ref{generalRes-HBeigenvalueequation}).
It is a crucial and essential point that instead of
(\ref{quadraticHBHamiltonian})
we introduce a quadratic Res-HB Hamiltonian\\[-18pt]
\beqa
\begin{array}{cc}
H[{\cal Z}]_{\mbox{{\scriptsize Res}}}
\!\equiv\!&\!
{\displaystyle
\frac{1}{2}}
[c^{\dagger },~c,
\cdots,
c^{\dagger },~c,
\cdots,
c^{\dagger },~c] \!
\left[ \!
\begin{array}{ccccc}
{\cal F}_1&&&&\\
\\[-16pt]
&\!\!\!\!\ddots&&~~~0&\\
&&\!\!\!\!\!\!{\cal F}_r\!\!\!\!&&\\
&\!\!\!\!\!\!\!\!\!\!\!\!0&\!\!\!\!&\!\!\!\!\!\!\!\!\vspace{-0.3cm}\ddots&\\
\\[-6pt]
&&&&\!\!\!\!\!\!\!\!{\cal F}_n
\end{array} \!
\right] \!
\left[ \!
\begin{array}{c}
     c \\[-2pt]
     c^{\dagger }, \\[-8pt]
     \vdots \\[-10pt]
     c \\[-2pt]
     c^{\dagger }, \\[-8pt]
     \vdots \\[-10pt]
     c \\[-2pt]
     c^{\dagger }
\end{array} \!
\right] . 
\end{array}
\label{quadraticHBHamiltonian2}
\eeqa\\[-14pt]
We can extend the HB free energy
to the Res-HB free energy in the form\\[-10pt]
\beq
F_{\mbox{{\scriptsize Res}}}
\!=\!
\mbox{Tr}(\stackrel{\circ }{W}_{\mbox{{\scriptsize Res}}} H)  
\!+\!
\frac{1}{\beta }
\mbox{Tr}
\left\{ \!
\stackrel{\circ }{W}_{\mbox{{\scriptsize Res}}}
\ln \!
\stackrel{\circ }{W}_{\mbox{{\scriptsize Res}}} \!
\right\},~
\stackrel{\circ }{W}_{\mbox{{\scriptsize Res}}}
\equiv
\frac{Pe^{-\beta H}P}
{\mbox{Tr}(Pe^{-\beta H})} .
\label{Res-HBfreeenergy}
\eeq\\[-12pt] 
According to
(\ref{Res-HBfreeenergy}), 
with use of
(\ref{quadraticHBHamiltonian2})
the Res-HB free energy is modified as follows:\\[-10pt]
\beq
F[{\cal Z}]_{\mbox{{\scriptsize Res}}}
\! = \!
\mbox{Tr}
(\stackrel{\circ }{W}\![{\cal Z}]_{\mbox{{\scriptsize Res}}} H)  
\! + \!
\frac{1}{\beta }
\mbox{Tr}
\left\{ \!
\stackrel{\circ }{W}\![{\cal Z}]_{\mbox{{\scriptsize Res}}} \ln
(\stackrel{\circ }{W}\![{\cal Z}]_{\mbox{{\scriptsize Res}}}) \!
\right\},
\stackrel{\circ }{W}\![{\cal Z}]_{\mbox{{\scriptsize Res}}} 
\! \equiv \!
\frac{Pe^{-\beta H[{\cal Z}]_{\mbox{{\scriptsize Res}}}}P}
{\mbox{Tr}(Pe^{-\beta H[{\cal Z}]_{\mbox{{\scriptsize Res}}}})} ,
\label{Res-HB free energy 3} 
\eeq\\[-10pt]
in which,
by making the Taylor expansion of
$\ln \! P \!=\! \ln \{1 \!-\! (1 \!-\! P) \}$
and using $P^2 \!=\! P$,
we obtain\\[-8pt]
\beq
\frac{1}{\beta }
\mbox{Tr}
\left\{
\stackrel{\circ}{W}\![{\cal Z}]_{\mbox{{\scriptsize Res}}} \ln
(\stackrel{\circ}{W}\![{\cal Z}]_{\mbox{{\scriptsize Res}}})
\right\} 
\!=\!
- \langle H[{\cal Z}]_{\mbox{{\scriptsize Res}}} \rangle
_{{\cal Z};\mbox{{\scriptsize Res}}}
\!-\!
\frac{1}{\beta } \ln \mbox{Tr} 
(Pe^{- \beta H[{\cal Z}]_{\mbox{{\scriptsize Res}}}}).
\label{traceform2}
\eeq\\[-10pt]
Using
(\ref{traceform2}),
the Res-HB free energy is also rewritten as\\[-12pt]
\beq
F[{\cal Z}]_{\mbox{{\scriptsize Res}}}
\!=\!
\langle H - H[{\cal Z}]_{\mbox{{\scriptsize Res}}} \rangle
_{{\cal Z};\mbox{{\scriptsize Res}}}
\!-\!
\frac{1}{\beta } \ln \mbox{Tr}
(Pe^{- \beta H[{\cal Z}]_{\mbox{{\scriptsize Res}}}}),
\label{Res-HB free energy 2}
\eeq
\vspace{-0.7cm}
\beqa
\langle H[{\cal Z}]_{\mbox{{\scriptsize Res}}} \rangle
_{{\cal Z};\mbox{{\scriptsize Res}}} 
\!\equiv\!
{\displaystyle
\frac{\mbox{Tr}(Pe^{- \beta H[{\cal Z}]_{\mbox{{\scriptsize Res}}}}P
H[{\cal Z}]_{\mbox{{\scriptsize Res}}})}
{\mbox{Tr} (Pe^{- \beta H[{\cal Z}]_{\mbox{{\scriptsize Res}}}})}
}, ~~
\langle H \rangle
_{{\cal Z};\mbox{{\scriptsize Res}}}
\!\equiv\!
{\displaystyle
\frac{\mbox{Tr}(Pe^{- \beta H[{\cal Z}]_{\mbox{{\scriptsize Res}}}}PH)}
{\mbox{Tr} (Pe^{- \beta H[{\cal Z}]_{\mbox{{\scriptsize Res}}}})}
} .
\label{traceform3}
\eeqa\\[-28pt]

With the use of the solved set of the Res-HB eigenvalue equations
$[{\cal F}_r u_r]_i \!=\! \epsilon _{ri} u_{ri}$
(\ref{generalRes-HBeigenvalueequation}),
the quadratic Hamiltonian $H[{\cal Z}]_{\mbox{{\scriptsize Res}}} $
(\ref{quadraticHBHamiltonian2})
is diagonalized in the $g_t$ quasi-particle frame as\\[-20pt]
\beqa
\begin{array}{c}
H[{\cal Z}]_{\mbox{{\scriptsize Res}}}
=
- {\displaystyle \frac{1}{2}}
\epsilon 
+
\sum _{t=1}^n \!
\sum _{i_t =1}^N \!  \epsilon _{t i_t} \!
d_{{g_t}i_t} ^\dagger d_{{g_t}i_t},~\!
[d_{g_t},d_{g_t} ^\dagger ]
\!\equiv\!
[c, c^\dagger ]g_t, ~\!
(\epsilon
\!=\!
\sum _{t=1}^n \!
\sum _{i_t =1}^N \!
\epsilon _{t i_t}) .
\end{array}
\label{diagonalizationingquasipartcleframe}
\eeqa\\[-18pt]
We introduce the trace manipulation {\bf tr} 
in the occupation number space
and
consider eigenvalues of the occupation number operator 
$n_{{g_t}i_t} (= d_{{g_t}i_t} ^\dagger d_{{g_t}i_t})$ 
are either 0 or 1.
From
(\ref{diagonalizationingquasipartcleframe}),
using the formula
(\ref{traceformP}),
we can compute approximately the second of the R.H.S. of
(\ref{traceform2})
as
\newpage
\beqa
\begin{array}{c}
\mbox{Tr} \!
\left( \! Pe^{- \beta H[{\cal Z}]_{\mbox{{\scriptsize Res}}}} \! \right)
=
e^{\frac{\beta \epsilon }{2}} \!
\sum _{r,s=1}^n
\langle g _r |
\mbox{{\bf tr}} \!
\left( \!
e^
{
- \beta \!
\sum _{t=1}^n
\sum _{i_t =1}^N
{\displaystyle
\epsilon _{t i_t}
n_{{g_t}i_t}
}
} \!
\right) \!
|g _s \rangle (S^{-1})_{sr} \\
\\[-12pt]
\!\!\!\!
\!=\!
e^{\frac{\beta \epsilon }{2}} \!
\sum _{r,s=1}^n
\langle g _r |
\mbox{{\bf tr}} \!
\left( \!
e^
{
- \beta \!
\sum _{i_r =1}^N \!
{\displaystyle
\epsilon _{ri_r} \!
n_{{g_r}i_r}
}
}
e^
{
- \beta \!
\sum _{i_s =1}^N 
{\displaystyle
\epsilon _{si_s} \!
n_{{g_s}i_s}
}
} 
e^
{
- \beta \!
\sum _{t \neq r, s=1}^n \!
\sum _{i_t =1}^N 
{\displaystyle
\epsilon _{t i_t} \!
n_{{g_t}i_t}
}
} \!\!
\right) \!
|g _s \rangle (S^{-1})_{sr} \\
\\[-12pt]
\!\!
=
e^{\frac{\beta \epsilon }{2}}
\sum _{r=1}^n \!
\Pi _{i_r =1}^N \!
\left( 
1 \!+\!
e^{- \beta \epsilon _{r i_r}}
\right) \!
f_r ,~
f_r
\!\equiv\!
\sum _{s=1}^n \!
S _{rs}
\Pi _{i_s =1}^N \!
\left( 
1 \!+\!
e^{- \beta \epsilon _{s i_s}} 
\right) \!
(S^{-1})_{sr} ,
\end{array}
\label{explicittraceform}
\eeqa\\[-10pt]
then, the logarithm of which divided by $\beta$
is approximately calculated as\\[-18pt]
\beqa
\!\!\!\!\!\!\!\!
\begin{array}{cc}
{\displaystyle 
\frac{1}{\beta }}
\ln
\mbox{Tr} \!
\left( Pe^{- \beta H[{\cal Z}]_{\mbox{{\scriptsize Res}}}} \right)
\!\simeq\!
{\displaystyle \frac{1}{2}}\epsilon
\!-\!
\sum _{r=1}^n \!\!
\sum _{i_r =1}^N \! 
\epsilon _{r i_r} \!
\!-\!
{\displaystyle 
\frac{1}{\beta }} \!
\sum _{r=1}^n \!
\sum _{i_r =1}^N
\ln 
{\displaystyle \frac{w _{r i_r}}{f_r}}, ~
w _{r i_r}
\!\equiv\!
{\displaystyle \frac{1}{1 \!+\! e^{\beta \epsilon _{r i_r}}}}  .
\end{array}
\label{logarithmoftraceform}
\eeqa\\[-30pt]

Further,
the quantity
$
\langle H[{\cal Z}]_{\mbox{{\scriptsize Res}}} \rangle
_{{\cal Z};\mbox{{\scriptsize Res}}}
$
defined by the first equation of
(\ref{traceform3})
is computed as\\[-18pt]
\beqa
\begin{array}{l}
\langle H[{\cal Z}]_{\mbox{{\scriptsize Res}}} \rangle
_{{\cal Z};\mbox{{\scriptsize Res}}}
=
- {\displaystyle \frac{1}{2}}
\epsilon
+
{\displaystyle
\frac{e^{\frac{\beta \epsilon }{2}}
}
{\mbox{Tr}
\left( \! Pe^{-\beta H[{\cal Z}]_{\mbox{{\scriptsize Res}}}} \! \right)}
} \!
\sum _{t=1}^n
\sum _{i_t =1}^N
\epsilon _{t i_t} \\
\\[-12pt]
\!\times
\left\{
\sum _{r, s=1}^n
\sum _{r^\prime,s^\prime=1}^n
\langle g _r |
\mbox{{\bf tr}} \!
\left( \!
e^
{
- 
\beta 
\sum _{t^\prime =1}^n
\sum _{i^\prime _{t^\prime } =1}^N
\epsilon _{t^\prime \! i^\prime _{t^\prime }}
n_{{g_{t^\prime } }i^\prime _{t^\prime }}
} \!
|g _{r^\prime } \rangle 
(\! S^{- \! 1} \!)_{r^\prime s^\prime }
\langle g _{s^\prime } |
n_{{g_t}i_t} \!
\right) \!
|g _s \rangle \!
( S^{- \! 1} )_{sr} \!
\right\} \\
\\[-16pt]
\simeq
- {\displaystyle \frac{1}{2}}
\epsilon 
\!+\!
{\displaystyle
\frac{1}
{
\sum _{t=1}^n
\Pi _{i_t =1}^N \!
\left( 
1
\!+\! 
e^{- \beta \epsilon _{t i_t}}
\right) \!
f_t}
} \\
\\[-12pt]
\!\times
{\displaystyle \frac{1}{2}}
\left\{ \!
\sum _{r,s=1}^n 
\sum _{i_s =1}^N 
\epsilon _{s i_s} \!
\sum _{r^\prime,s^\prime=1}^n \!
\Pi _{i^\prime_s \neq i_s =1}^N \!
\left(  
1 
\!+\!
e^{-\beta \epsilon _{s i^\prime _s }} \!
\right) \!
e^{-\beta \epsilon _{s i_s}} \!
S_{rr^\prime }
(S^{-1})_{r^\prime s^\prime }
S_{s^\prime s}
(S^{-1})_{sr}
\right. \\
\\[-12pt]
\!\!\!\!
+\!
\left.
\sum _{r^\prime \! , s^\prime  \! =1}^n \!
\sum _{i_{s^\prime } =1}^N \!
\epsilon _{s^\prime i_{s^\prime }} \!
\sum _{r, s=1}^n \!
\Pi _{i^\prime _{s^\prime } \neq i_{s^\prime } \! =1}^N \!
\left(
1 
\!+\!
e^{-\beta \epsilon _{s^\prime \! i^\prime _{s^\prime }}} \!
\right) \!
e^{-\beta \epsilon _{s^\prime \! i _{s^\prime }}} \!
S_{r r^\prime }
(S^{-1})_{r^\prime s^\prime }
(S^{-1})_{s^\prime s }
S_{s r} \!
\right\} \! .
\end{array}
\label{explicittraceform2}
\eeqa\\[-10pt]
Substituting the last equation of
(\ref{explicittraceform})
into the denominator, i.e.,
$\mbox{Tr} (\! Pe^{- \beta H[{\cal Z}]_{\mbox{{\scriptsize Res}}}} \!)$
and taking only the term with $t \!=\! s$ or $t \!=\! s^\prime$,
the
$\langle H[{\cal Z}]_{\mbox{{\scriptsize Res}}} \rangle
_{{\cal Z};\mbox{{\scriptsize Res}}}$
can be changed to a more compact form\\[-20pt]
\beqa
\!\!\!\!\!\!\!\!
\begin{array}{c}
\langle H[{\cal Z}]_{\mbox{{\scriptsize Res}}} \rangle
_{{\cal Z};\mbox{{\scriptsize Res}}}
\!\simeq\!
- {\displaystyle \frac{1}{2}}
\epsilon
\!+\!
\sum _{r=1}^n \!
\sum _{i_r =1}^N \! 
\epsilon _{r i_r} \!
{\displaystyle \frac{w _{r i_r}}{f_r}} .
\end{array}
\label{expliciHZexpectationvalue}
\eeqa\\[-16pt]
The addition of 
(\ref{logarithmoftraceform}) to
(\ref{expliciHZexpectationvalue})
and the uses of
(\ref{tildeCalFrmatrix})
and
$
\epsilon _{r i_r}
\!\simeq\!
{\displaystyle \frac{1}{\beta}} \!
\ln \!
\left\{ \!
\left( \!
\! 1
\!-\! 
\frac{w _{r i_r}}{f_r} \!
\right) \!\!
/ 
\frac{w _{r i_r}}{f_r} \!
\right\}
$
lead to\\[-12pt]
\beqa
\!\!\!\!\!
\begin{array}{c}
-\langle H[{\cal Z}]_{\mbox{{\scriptsize Res}}} \rangle
_{\!{\cal Z};\mbox{{\scriptsize Res}}}
\!-\!\!
{\displaystyle 
\frac{1}{\beta }} \!
\ln \!
\mbox{Tr} \!
\left( \! Pe^{- \beta H[{\cal Z}]_{\mbox{{\scriptsize Res}}}} \! \right)
\!\!\simeq\!\!
{\displaystyle \frac{1}{\beta }} \!\!
\sum _{r\!=\!1}^n \!\!
\sum _{i_r \!=\!1}^N \!\! 
\left[ \!
\left( \!
\! 1
\!\!-\! 
{\displaystyle \frac{w _{r i_r}}{f_r}} \!\!
\right) \!
\ln \!
\left\{ \!\!
\left( \!
\! 1
\!-\! 
{\displaystyle \frac{w _{r i_r}}{f_r}} \!\!
\right) \!\!
/ 
{\displaystyle \frac{w _{r i_r}}{f_r}} \!\!
\right\}
\!+\!
\ln \!
{\displaystyle 
\frac{w _{r i_r}}{f_r}} \!
\right] \\
\\[-12pt]
\!=\!
{\displaystyle \frac{1}{2}
\frac{1}{\beta }} \!
\sum_{r =1} ^n \!
\mbox{Tr} \!
\left\{ \!
\widetilde{W}_r
\ln \! \widetilde{W}_r
\!+\!
(1_{2N} \!-\! \widetilde{W}_r)
\ln (1_{2N} \!-\! \widetilde{W}_r) \!
\right\} ,~
\widetilde{W}_r
\!\equiv\!
g_r ^\dagger W_{\mbox{{\scriptsize Res}}:rr}^{\mbox{{\scriptsize thermal}}\ast} g_r .
\end{array}
\label{explicittraceform4}
\eeqa\\[-14pt]
This is identical with the entropy
$S_{\mbox{{\scriptsize Res}}}^{\mbox{{\scriptsize thermalHB}}}$
in II
except the modification
$
\widetilde{w} _{r i_r}
\!\!\rightarrow\!\!
w _{r i_r}
\!\!\rightarrow\!\!
\frac{w _{r i_r}}{f_r}
$
in
(\ref{tildeCalFrmatrix}).
Taking only a single HB WF,
(\ref{explicittraceform4})
reduces to
the usual statistical expectation value
by HB $\!$WF.
The summation with respect to $r$ in R.H.S.
in
(\ref{explicittraceform4})
and multiplication by 
$\frac{1}{f_r}$
arise due to the superposition of HB WFs.

Finally,
a computation of the quantity
$
\langle H \rangle
_{{\cal Z};\mbox{{\scriptsize Res}}}
$
defined by the second of
(\ref{traceform3})
is made through the following procedures:
Let us define new fermion $SO(2N)$ Lie operators as\\[-20pt]
\beqa
E^{{{g_t}i_t}}_{~~{{g_t}j_t}}
\!\equiv\!
d_{{g_t}i_t} ^\dagger d_{{g_t}j_t}
\!-\!
{\displaystyle \frac{1}{2} \delta_{i_t j_t}} , ~
E^{{{g_t}i_t} {{g_t}j_t}}
\!\equiv\!
d_{{g_t}i_t} ^\dagger d_{{g_t}j_t}^\dagger , ~
E_{{{g_t}i_t} {{g_t}j_t}}
\!\equiv\!
d_{{g_t}i_t} d_{{g_t}j_t} , ~
[d_{g_t},d_{g_t} ^\dagger ] \!\equiv\! [c, c^\dagger ]g_t .
\label{newfermionLieop}
\eeqa\\[-18pt]
\def\erw#1{{<\!\!#1\!\!>_{g_t}}}
The original fermion $\!SO(2N)\!$ Lie ones are expressed
in terms of the quasi-particle expectation values
$\erw{E^{\alpha }_{~\beta } }, \erw{E_{\alpha \beta } }$ and
$\erw{E^{\alpha \beta } }$
($c$-numbers) and of the new $SO(2N)$ Lie ones
(\ref{newfermionLieop}) (quantum mechanical fluctuations)
as follows:\\[-20pt]
\beqa
\!\!\!\!\!\!\!\!\!\!\!\!\!\!
\left.
\BA{ll}
&E^{\alpha }_{~\beta }
\!=\!
\erw{E^{\alpha }_{~\beta } }
\!+\!
\left( \!
a^{\alpha }_{~i_t}
a^{\beta \star }_{~j_t}
\!-\!
b_{\beta i_t}
b^\star_{\alpha j_t} \!
\right) \!\!
\left( \!\!
E^{{{g_t}i_t}}_{~~{{g_t}j_t}}
\!+\!
{\displaystyle \frac{1}{2} \delta_{i_t j_t}} \!\!
\right)
\!+
a^{\alpha }_{~i_t}b_{\beta j_t} \!
E^{{{g_t}i_t} {{g_t}j_t}}
\!+
b^\star_{\alpha i_t}a^{\beta \star }_{~j_t} \!
E_{{{g_t}i_t} {{g_t}j_t}} , \\
\\[-12pt]
&E_{\alpha \beta }
\!=\!
\erw{E_{\alpha \beta } }
\!+\!
\left( \!
b_{\alpha i_t}
a^{\beta \star }_{~j_t}
\!-\!
b_{\beta i_t}
a^{\alpha \star }_{~j_t} \!
\right) \!\!
\left( \!\!
E^{{{g_t}i_t}}_{~~{{g_t}j_t}}
\!+\!
{\displaystyle \frac{1}{2} \delta_{i_t j_t}} \!\!
\right)
\!+
b_{\alpha i_t} \! b_{\beta j_t} \!
E^{{{g_t}i_t} {{g_t}j_t}}
\!+
a^{\alpha \star }_{~i_t} \! a^{\beta \star }_{~j_t} \!
E_{{{g_t}i_t} {{g_t}j_t}} , \\
\\[-8pt]
&E^{\alpha \beta }
\!=\!
-E_{\alpha \beta }^{\dag } ,~
\erw{E^{\alpha \beta } }
\!=\!
- \erw{E_{\alpha \beta } }^{\!\!\!\!\star } .
\EA \!\!\!\!
\right\}
\label{bosonrotation2}
\eeqa\\[-14pt]
On the $g_t$ quasi-particle frame,
using
(\ref{bosonrotation2}),
the original Hamiltonian
$H$
(\ref{Hamiltonian})
is transformed into $H_{g_t}$ expressed as\\[-16pt]
\beqa
\!\!\!\!\!\!\!\!\!\!\!\!\!\!\!\!\!\!\!
\BA{ll}
&H_{g_t}
\!=\!
\erw{H}\! \\
\\[-10pt]
&
\!\!+\!\!
\left\{ \!\!
\left( \!\!
a^{\alpha }_{~i_t}
a^{\beta \star }_{~j_t}
\!\!-\!\!
b_{\beta i_t}
b^\star_{\alpha j_t} \!
\right) \!\!
F_{\alpha \beta }^{g_t}
\!+\!
{\displaystyle \frac{1}{2}} \!
\left( \!
b_{\alpha i_t}
a^{\beta \star }_{~j_t}
\!\!-\!\!
b_{\beta i_t}
a^{\alpha \star }_{~j_t} \!
\right) \!\!
D^{g_t \star }_{\beta \alpha }
\!+\!
{\displaystyle \frac{1}{2}} \!
\left( \!
a^{\alpha }_{~i_t}
b^\star _{\beta j_t}
\!\!-\!\!
a^{\beta }_{~i_t}
b^\star _{\alpha j_t} \!
\right) \!\!
D_{\!\alpha \beta }^{g_t} \!
\right\} \!\!
\left( \!\!
E^{{{g_t}i_t}}_{~~{{g_t}j_t}}
\!\!+\!\!
{\displaystyle \frac{1}{2} \delta_{i_t j_t}} \!\!
\right) \\
\\[-12pt]
&
\!\!+\!\!
\left\{ \!\!
a^{\alpha }_{~i_t} \! b_{\beta j_t} \!
F_{\!\alpha \beta }^{g_t}
\!\!+\!\!
{\displaystyle \frac{1}{2}} 
b_{\alpha i_t} \! b_{\beta j_t} \!
D^{g_t \star }_{\!\beta \alpha }
\!\!+\!\!
{\displaystyle \frac{1}{2}}
a^{\alpha }_{~i_t} \! a^{\beta }_{\!~j_t} \!
D_{\!\alpha \beta }^{g_t} \!\!
\right\} \!\!
E^{{{g_t}i_t} {{g_t}j_t}} 
\!\!+\!\!
\left\{ \!\!
b^\star_{\alpha i_t} \! a^{\beta \star }_{\!~j_t} \!
F_{\!\alpha \beta }^{g_t}
\!\!+\!\!
{\displaystyle \frac{1}{2}}
a^{\alpha \star }_{~i_t} \! a^{\beta \star }_{\!~j_t} \!
D^{g_t \star }_{\!\beta \alpha } 
\!\!+\!\!
{\displaystyle \frac{1}{2}}
b^\star _{\alpha i_t} \! b^\star _{\beta j_t} \!
D_{\!\alpha \beta }^{g_t} \!\!
\right\} \!\!
E_{{{g_t}i_t} {{g_t}j_t}} \\
\\[-12pt]
&
\!\!+\!
{\displaystyle \frac{1}{16}}
[\alpha \beta|\gamma \delta] \!
\left[ \!
\left( ^{^{^{^{.}}}} \!\!\!
a^{\alpha }_{~i_t} \! a^{\gamma }_{~j_t}
\!\!-\!\!
a^{\alpha }_{~j_t} \! a^{\gamma }_{~i_t}
{ }^{^{^{^{.}}}} \!\!\!
\right) \!\!
\left( \!
a^{\delta \star }_{~i^\prime _t} \! a^{\beta \star }_{~j^\prime _t}
\!\!-\!\!
a^{\delta \star }_{~j^\prime _t} \! a^{\beta \star }_{~i^\prime _t} \!
\right)
\!\!+\!\!
\left( \!
b^\star _{\alpha i^\prime_t}
b^\star _{\gamma j^\prime_t}
\!\!-\!\!
b^\star _{\alpha j^\prime_t}
b^\star _{\gamma i^\prime_t} \!
\right) \!\!
\left( ^{^{^{^{.}}}} \!\!\!
b _{\delta i_t}
b _{\beta j_t}
\!\!-\!\!
b _{\delta j_t}
b _{\beta i_t}
{ }^{^{^{^{.}}}} \!\!\!
{ }^{^{^{^{.}}}} \!\!\!\right)
\right. \\
\\[-12pt]
&
\left.
~~~~~~~~~~
\!\!-
4 \!
\left( \!
a^{\alpha }_{~i_t} 
b^\star _{\gamma i^\prime_t}
\!\!-\!\!
a^{\gamma }_{~i_t}
b^\star _{\alpha i^\prime_t} \!
\right) \!\!
\left( \!
b_{\delta j_t}
a^{\beta \star }_{~j^\prime _t}
\!\!-\!\!
b_{\beta j_t}
a^{\delta \star }_{~j^\prime _t} \!
\right) \!
\right] \!\!
E^{{{g_t}i_t} {{g_t}j_t}} \!
E_{{{g_t}i^\prime _t} {{g_t}j^\prime _t}} \\
\\[-12pt]
&
\!\!+\!
{\displaystyle \frac{1}{8}}
[\alpha \beta|\gamma \delta] \!\!
\left[ \!\!
\left( ^{^{^{^{.}}}} \!\!\!
a^{\alpha }_{~i_t} \! a^{\gamma }_{~j_t}
\!\!-\!\!
a^{\alpha }_{~j_t} \! a^{\gamma }_{~i_t}
{ }^{^{^{^{.}}}} \!\!\!
\right) \!\!
\left( ^{^{^{^{.}}}} \!\!\!
b_{\delta i^\prime _t}
a^{\beta \star }_{~j^\prime _t}
\!\!-\!\!
b_{\beta i^\prime _t}
a^{\delta \star }_{~j^\prime _t}
{ }^{^{^{^{.}}}} \!\!\!
\right) \!
\!+\!
2 \!\!
\left( \!
a^{\alpha }_{~i_t}
b^\star _{\gamma j^\prime _t}
\!\!-\!\!
a^{\gamma }_{~i_t}
b^\star _{\alpha j^\prime _t} \!
\right) \!\!
b_{\delta j _t} b_{\beta i^\prime _t} \!
\right] \!\!
E^{{{g_t}i_t} {{g_t}j_t}} \!\!
\left( \!\!
E^{{{g_t}i^\prime _t}}_{~~{{g_t}j^\prime _t}}
\!\!+\!\!
{\displaystyle \frac{1}{2} \delta_{i^\prime _t j^\prime _t}} \!\!
\right)  \\
\\[-12pt]
& 
\!\!+\!
{\displaystyle \frac{1}{8}}
[\alpha \beta|\gamma \delta] \!\!
\left[ \!\!
\left( ^{^{^{^{.}}}} \!\!\!
a^{\alpha }_{~i_t} \! b^\star _{\gamma j_t}
\!\!-\!\!
a^{\gamma }_{~i_t} \! b^\star _{\alpha j_t}
{ }^{^{^{^{.}}}} \!\!\!
\right) \!\!
\left( \!
a^{\delta \star }_{~i^\prime _t} \! a^{\beta \star }_{~j^\prime _t}
\!\!-\!\!
a^{\delta \star }_{~j^\prime _t} \! a^{\beta \star }_{~i^\prime _t} \!
\right) \!
\!+\!
2
b^\star _{\alpha j_t} \! b^\star _{\gamma i^\prime _t} \!\!
\left( \!
b_{\delta i_t} \! a^{\beta \star }_{~j^\prime_t} \!
\!\!-\!\!
b_{\beta i_t} \! a^{\delta \star }_{~j^\prime_t } \!
\right) \!
\right] \!\!\!
\left( \!\!
E^{{{g_t}i_t}}_{~~{{g_t}j_t}}
\!\!+\!\!
{\displaystyle \frac{1}{2} \delta_{i_t j_t}} \!\!
\right) \!\!
E_{{{g_t}i^\prime _t} {{g_t}j^\prime _t}} \\
\\[-12pt]
&
\!\!+\!
{\displaystyle \frac{1}{16}}
[\alpha \beta|\gamma \delta] \!\!
\left( ^{^{^{^{.}}}} \!\!\!
a^{\alpha }_{~i_t} \! a^{\gamma }_{~j_t}
\!\!-\!\!
a^{\alpha }_{~j_t} \! a^{\gamma }_{~i_t}
{ }^{^{^{^{.}}}} \!\!\!
\right) \!\!
\left( ^{^{^{^{.}}}} \!\!\!
b_{\delta i^\prime _t} \! b_{\beta j^\prime _t}
\!\!-\!\!
b_{\delta j^\prime _t} \! b_{\beta i^\prime _t}
{ }^{^{^{^{.}}}} \!\!\!
\right) \!\!
E^{{{g_t}i_t} {{g_t}j_t}} \!
E^{{{g_t}i^\prime_t} {{g_t}j^\prime_t}} \\
\\[-12pt]
&
\!\!+\!
{\displaystyle \frac{1}{16}}
[\alpha \beta|\gamma \delta] \!\!
\left( ^{^{^{^{.}}}} \!\!\!
b^\star _{\alpha i_t} \! b^\star _{\gamma j_t}
\!\!-\!\!
b^\star _{\alpha j_t} \! b^\star _{\gamma i_t}
{ }^{^{^{^{.}}}} \!\!\!
\right) \!\!
\left( ^{^{^{^{.}}}} \!\!\!
a^{\delta \star }_{~i^\prime _t} \! a^{\beta \star }_{~j^\prime _t}
\!\!-\!\!
a^{\delta \star }_{~j^\prime _t} \! a^{\beta \star }_{~i^\prime _t}
{ }^{^{^{^{.}}}} \!\!\!
\right) \!\!
E_{{{g_t}i_t} {{g_t}j_t}} \!
E_{{{g_t}i^\prime_t} {{g_t}j^\prime _t}} ,
\EA \!\!
\label{Hamiltonianimage2}
\eeqa\\[-6pt]
where\\[-18pt]
\beqa
\BA{ll}
\!\!\!\!\!\!\!\!
&\!\erw{H}\!
\equiv
h_{\alpha \beta }
\erw{E^{\alpha }_{~\beta }
\!+\!
{\displaystyle \frac{1}{2}}\delta_{\alpha \beta }
} \\
\\[-12pt]
&~~~~~~~~~~+
{\displaystyle \frac{1}{2}}
[\alpha \beta|\gamma \delta]
\erw{E^{\alpha }_{~\beta }
\!+\!
{\displaystyle \frac{1}{2}}\delta_{\alpha \beta }
} \!
\erw{E^{\gamma }_{~\delta }
\!+\!
{\displaystyle \frac{1}{2}}\delta_{\gamma \delta }
}
+
{\displaystyle \frac{1}{4}}
[\alpha \beta|\gamma \delta]
\erw{E^{\alpha \gamma } } \!
\erw{E_{\delta \beta } } .
\EA
\label{Hamiltonianimage3}
\eeqa\\[-8pt]
Here the
$E^{\alpha }_{~\beta }, E_{\alpha \beta }$ and
$E^{\alpha \beta }$
are generators of rotation in the $2N$-dimensional Euclidian space.
The
$\erw{H}\!$
means
an energy of classical motion of 
$SO(2N)$ fermion top.
Under a
{\it quasi anti-commutation-relation approximation},
a new aspect of the top is described in our recent work
\cite{NishiProviijmpa2012}.
Using the expression for
$H\!$
(\ref{Hamiltonianimage2}),
the
$
\!\langle H \rangle _{{\cal Z};\mbox{{\scriptsize Res}}}
$
defined by second of
(\ref{traceform3})
is computed as
\\[-14pt]
\beqa
\!\!\!\!\!\!\!
\begin{array}{l}
\langle H \rangle _{{\cal Z};\mbox{{\scriptsize Res}}}
\!=\!
{\displaystyle
\frac{e^{\frac{\beta \epsilon }{2}}
}
{\mbox{Tr} \!
\left( \! Pe^{-\beta H[{\cal Z}]_{\mbox{{\scriptsize Res}}}} \! \right)}
} \!
\sum _{t=1}^n \!
\sum _{r, s=1}^{ ~n}
\langle g _r |
\mbox{{\bf tr}} \!
\left( \!\!
e^
{
- 
\beta 
\sum _{t^\prime =1}^n \!
\sum _{i^\prime _{t^\prime } =1}^N \!
\epsilon _{t^\prime i^\prime _{t^\prime }}
n_{{g_{t^\prime } }i^\prime _{t^\prime }}
} \!
PH_{g_t} \!\!
\right) \!
|g _s \rangle
( S^{- 1} )_{sr} \\
\\[-12pt]
\simeq
\sum _{t=1}^n
\erw{H}
\!+\!
{\displaystyle
\frac{e^{\frac{\beta \epsilon }{2}}
}
{\mbox{Tr} \!
\left( \! Pe^{-\beta H[{\cal Z}]_{\mbox{{\scriptsize Res}}}} \! \right)}
} \!
\sum _{t=1}^n \!
\sum _{i_t =1}^N \\
\\[-12pt]
\times
\left\{ \!\!
\left( \!
a^{\alpha }_{~i_t}
a^{\beta \star }_{~i_t}
\!\!-\!\!
b_{\beta i_t}
b^\star_{\alpha i_t} \!
\right) \!\!
F_{\alpha \beta }^{g_t}
\!+\!
{\displaystyle \frac{1}{2}} \!
\left( \!
b_{\alpha i_t}
a^{\beta \star }_{~i_t}
\!\!-\!\!
b_{\beta i_t}
a^{\alpha \star }_{~i_t} \!
\right) \!\!
D^{g_t \star }_{\beta \alpha }
\!+\!
{\displaystyle \frac{1}{2}} \!
\left( \!
a^{\alpha }_{~i_t}
b^\star _{\beta i_t}
\!\!-\!\!
a^{\beta }_{~i_t}
b^\star _{\alpha i_t} \!
\right) \!\!
D_{\alpha \beta }^{g_t} \!
\right\} \\
\\[-10pt]
\times
\left\{ \!
\sum _{r, s=1}^{n} \!\!
\sum _{r^\prime,s^\prime=1}^n \!
\langle g _r |
\mbox{{\bf tr}} \!
\left( \!\!
e^
{
- 
 \beta 
\sum _{t^\prime =1}^n \!
\sum _{i^\prime _{t^\prime } =1}^N 
\epsilon _{t^\prime i^\prime _{t^\prime }}
n_{{g_{t^\prime } }i^\prime _{t^\prime }}
} 
|g _{r^\prime } \rangle 
(\! S^{- \! 1} \!)_{r^\prime s^\prime }
\langle g _{s^\prime } |
n_{{g_t} {i_t}} \!\!
\right) \!
|g _s \rangle 
( S^{- \! 1} )_{sr} \!
\right\} \\
\\[-16pt]
+
{\displaystyle
\frac{e^{\frac{\beta \epsilon }{2}}
}
{\mbox{Tr} \!
\left( \! Pe^{-\beta H[{\cal Z}]_{\mbox{{\scriptsize Res}}}} \! \right)}
} \!
\sum _{t=1}^n \!
\sum _{i_t =1}^N \!
\sum _{j_t =1}^N \\
\\[-10pt]
\times
{\displaystyle \frac{1}{8}}
[\alpha \beta|\gamma \delta] \!
\left[ 
\left( ^{^{^{^{.}}}} \!\!\!
a^{\alpha }_{~i_t} a^{\gamma }_{~j_t}
\!-\!
a^{\alpha }_{~j_t} a^{\gamma }_{~i_t}
{ }^{^{^{^{.}}}} \!\!\!
\right) \!\!
\left( ^{^{^{^{.}}}} \!\!\!
a^{\delta \star }_{~j_t} a^{\beta \star }_{~i_t}
\!-\!
a^{\delta \star }_{~i_t} a^{\beta \star }_{~j_t}
{ }^{^{^{^{.}}}} \!\!\!
\right)
\!+\!
\left( ^{^{^{^{.}}}} \!\!\!
b^\star _{\alpha j_t}
b^\star _{\gamma i_t}
\!\!-\!\!
b^\star _{\alpha i_t}
b^\star _{\gamma j_t}
{ }^{^{^{^{.}}}} \!\!\!
\right) \!\!
\left( ^{^{^{^{.}}}} \!\!\!
b _{\delta i_t}
b _{\beta j_t}
\!\!-\!\!
b _{\delta j_t}
b _{\beta i_t}
{ }^{^{^{^{.}}}} \!\!\!
\right)
\right. \\
\\[-8pt]
\left.
-
2 \!
\left\{ \!
\left( ^{^{^{^{.}}}} \!\!\!
a^{\alpha }_{~i_t}
b^\star _{\gamma j_t}
\!-\!
a^{\gamma }_{~i_t}
b^\star _{\alpha j_t}
{ }^{^{^{^{.}}}} \!\!\! 
\right) \!\!
\left( ^{^{^{^{.}}}} \!\!\!
b_{\delta j_t}
a^{\beta \star }_{~i_t}
\!-\!
b_{\beta j_t}
a^{\delta \star }_{~i_t} \!
{}^{^{^{^{.}}}} \!\!\!
\right)
\!-\!
\left( ^{^{^{^{.}}}} \!\!\!
a^{\alpha }_{~i_t}
b^\star _{\gamma i_t}
\!-\!
a^{\gamma }_{~i_t}
b^\star _{\alpha i_t}
 { }^{^{^{^{.}}}} \!\!\! 
\right) \!\!
\left( ^{^{^{^{.}}}} \!\!\!
b_{\delta j_t}
a^{\beta \star }_{~j_t}
\!-\!
b_{\beta j_t}
a^{\delta \star }_{~j_t}
{}^{^{^{^{.}}}} \!\!\!
\right) \!
\right\} 
\right] \\
\\[-8pt]
\times
\left\{ \!
\sum _{r, s=1}^{n} \!
\sum _{r^\prime,s^\prime=1}^n
\langle g _r |
\mbox{{\bf tr}} \!
\left( \!\!
e^
{
- 
\beta 
\sum _{t^\prime =1}^n \!
\sum _{i^\prime _{t^\prime } =1}^N 
\epsilon _{t^\prime i^\prime _{t^\prime }}
n_{{g_{t^\prime } }i^\prime _{t^\prime }}
} \!
|g _{r^\prime } \rangle
(\! S^{- \! 1} \!)_{r^\prime s^\prime }
\langle g _{s^\prime } |
n_{{g_t} {i_t}} \!
n_{{g_t} {j_t}} \!\!
\right) \!
|g _s \rangle
( S^{- \! 1} )_{sr} \!\!
\right\} \! ,
\end{array}
\label{explicittraceform3}
\eeqa\\[-12pt]
where we have used the definitions
(\ref{FandDoperators})
and the following ensemble averages of the new fermion $SO(2N)$ Lie operators
in the $g_t$ quasi-particle frame:\\[-16pt]
\beqa
\!\!\!\!
\left.
\begin{array}{l}
\mbox{Tr} \!
\left\{ \!\!
Pe^{- \beta H[{\cal Z}]_{\mbox{{\scriptsize Res}}}}P \!
\left( \!\!
E^{\!~{{g_t}i_t}}_{~~{{g_t}j_t}}
\!\!+\!\!
{\displaystyle \frac{1}{2} \delta_{i_t j_t}} \!\!
\right) \!\!
\right\}
\!=\!
\delta_{i_t j_t}
\mbox{Tr} \!
\left( \!
Pe^{- \beta H[{\cal Z}]_{\mbox{{\scriptsize Res}}}}P
n_{{g_t}i_t}
\right) , \\
\\[-10pt]
\mbox{Tr} \!
\left( \!
Pe^{- \beta H[{\cal Z}]_{\mbox{{\scriptsize Res}}}}P
E^{{{g_t}i_t} {{g_t}j_t}}
\right)
\!=\!
0 , ~
\mbox{Tr} \!
\left( \!
Pe^{- \beta H[{\cal Z}]_{\mbox{{\scriptsize Res}}}}P
E_{{{g_t}i_t} {{g_t}j_t}}
\right)
\!=\!
0 , \\
\\[-6pt]
\mbox{Tr} \!
\left( \!
Pe^{- \beta H[{\cal Z}]_{\mbox{{\scriptsize Res}}}}P
E^{{{g_t}i_t} {{g_t}j_t}} \!
E_{{{g_t}i^\prime _t} {{g_t}j^\prime _t}} \!
\right)
\!=\!
\left(
\delta_{i_t j^\prime _t}
\delta_{j_t i^\prime _t}
\!-\!
\delta_{i_t i^\prime _t}
\delta_{j_t j^\prime _t}
\right) \!
\mbox{Tr} \!
\left( \!
Pe^{- \beta H[{\cal Z}]_{\mbox{{\scriptsize Res}}}}P
n_{{g_t}i_t}
n_{{g_t}j_t}
\right) .
\end{array} \!
\right\}
\label{ensembleaverages}
\eeqa\\[-10pt]
The trace manipulation {\bf tr} 
is made as was done previously
and
the eigenvalue of $n_{{g_t}i_t}$
is taken to be either 0 or 1.
Using this,
further computation of
(\ref{explicittraceform3})
can be made approximately as\\[-18pt]
\def\egr#1{{<\!\!#1\!\!>_{\!g_r}}}
\def\egsp#1{{<\!\!#1\!\!>_{g_{s^\prime }}}}
\def\egs#1{{<\!\!#1\!\!>_{g_s}}}
\beqa
\!\!\!\!
\begin{array}{l}
\langle H \rangle _{{\cal Z};\mbox{{\scriptsize Res}}}
\simeq
\sum _{t=1}^n
\erw{H}
+
{\displaystyle
\frac{1}
{
\sum _{t=1}^n \!
\Pi _{i_t =1}^N \!
\left( \!
1
\!+\! 
e^{- \beta \epsilon _{t i_t}} \!
\right) \!
f_t
} 
} \\
\\[-10pt]
\!\times\!
\left[ 
^{^{^{^{^{.}}}}} \!\!\!
\sum _{s=1}^{n} \!
\sum _{i_s =1}^N \!\!
\left\{ \!
\left( \!
a^{\alpha }_{~i_s}
a^{\beta \star }_{~i_s}
\!\!-\!\!
b_{\beta i_s}
b^\star_{\alpha i_s} \!
\right) \!\!
F_{\alpha \beta }^{g_s}
\!\!+\!\!
{\displaystyle \frac{1}{2}} \!
\left( \!
b_{\alpha i_s}
a^{\beta \star }_{~i_s}
\!\!-\!\!
b_{\beta i_s}
a^{\alpha \star }_{~i_s} \!
\right) \!\!
D^{g_s \star }_{\beta \alpha }
\!\!+\!\!
{\displaystyle \frac{1}{2}} \!
\left( \!
a^{\alpha }_{~i_s}
b^\star _{\beta i_s}
\!\!-\!\!
a^{\beta }_{~i_s}
b^\star _{\alpha i_s} \!
\right) \!\!
D_{\alpha \beta }^{g_s} \!
\right\}
\right. \\
\\[-12pt]
\!\times
{\displaystyle \frac{1}{2}} \!
\left\{ \!
^{^{^{^{^{.}}}}} \!\!\!
\sum _{r^\prime,s^\prime=1}^n \!
\Pi _{i^\prime_s \neq i_s =1}^N \!
\left( \!
1 
\!+\!
e^{-\beta \epsilon _{s i^\prime_s }} \!
\right) \!
e^{- \beta  \epsilon _{s i_s }} \!\!
\sum _{r=1}^n
S_{rr^\prime }
(S^{-1})_{r^\prime s^\prime }
S_{s^\prime s}
(S^{-1})_{sr} \!
\right. \\
\\[-12pt]
\left.
\left.
\!+
\sum _{r^\prime, s^\prime=1}^n \!
\Pi _{i^\prime _{s } \neq i_{s } =1}^N \!
\left( \!
1 
\!+\!
e^{-\beta \epsilon _{s i^\prime _{s }}} \!
\right) \!
e^{- \beta \epsilon _{s i_{s }}} \!\!
\sum _{r=1}^n \! 
S_{rr^\prime }
(S^{-1})_{r^\prime s^\prime }
S_{s^\prime s }
(S^{-1})_{s r} \!\!\!
^{^{^{^{^{.}}}}} \!
\right\} 
\right]  \\
\\[-10pt]
+
{\displaystyle
\frac{1}
{\sum _{t=1}^n \!
\Pi _{i_t =1}^N \!
\left( \!
1 \!+\! 
e^{- \beta \epsilon _{t i_t}} \!
\right) \!
f_t
}
} \!
\sum _{s=1}^n
{\displaystyle \frac{1}{4}}
[\alpha \beta|\gamma \delta] \\
\\[-8pt]
\!\times\!
\left[
2
\left( \!
b^\star_{\alpha i_s} \!
b_{\beta i_s}
\!-\!
a^{\alpha }_{~i_s} \!
a^{\beta \star }_{~i_s} \!
\right) \!\!
\left( ^{^{^{^{.}}}} \!\!\!
b^\star_{\gamma j_s} \!
b_{\delta j_s} \!
\!-\!
a^{\gamma }_{~j_s} \!
a^{\delta \star }_{~j_s} \!\!
\right)
\!+\!
\left( \!\! 
^{^{^{^{.}}}} \!
a^{\alpha }_{~i_s} \!
b^\star_{\gamma i_s}
\!-\!
a^{\gamma }_{~i_s} \!
b^\star_{\alpha i_s} \!\!
\right) \!\!
\left( \!
b_{\delta j_s}
a^{\beta \star }_{~j_s}
\!-\!
b_{\beta j_s}
a^{\delta \star }_{~j_s} \!
\right) 
\right] \\
\\[-12pt]
\!\times
{\displaystyle \frac{1}{2}}
\left\{ \!
^{^{^{^{^{.}}}}} \!\!\!
\sum _{r^\prime,s^\prime=1}^n  \!
\Pi _{i^\prime_s \neq i_s \neq j_s =1}^N \!
\left( \!
1
\!+\!
e^{-\beta \epsilon _{s i^\prime_s }} \!
\right) \!
e^{- \beta \epsilon _{s i_s }}
e^{- \beta \epsilon _{s j_s }} \!\!
\sum _{r=1}^{n} \!
S_{r r^\prime }
( S^{- 1} )_{r^\prime s^\prime }
S_{s^\prime s}
( S^{- 1} )_{sr}
\right. \\
\\[-12pt]
\left.
\!+
\sum _{r^\prime,s^\prime =1}^n \!
\Pi _{i^\prime _{s } \neq i_{s } \neq j_{s } =1}^N \!
\left( \!
1
\!+\!
e^{-
\beta \epsilon _{s j^\prime _{s }}} \!
\right) \!
e^{-  \beta \epsilon _{s i_{s }}}
e^{- \beta  \epsilon _{s j_{s }}} \!\!
\sum _{r=1}^n \!
S_{r r^\prime }
( S^{- 1})_{r^\prime s^\prime }
S_{s^\prime s }
( S^{- 1} )_{s r} \!\!
^{^{^{^{^{.}}}}} \!
\right\} \! ,
\end{array}
\label{explicittraceform5}
\eeqa\\[-8pt]
where we have used
(\ref{Hamiltonianimage3})
and the equation in the last line of
(\ref{explicittraceform}).
Taking only the term with $t \!=\! s$
in the denominator,
equation
(\ref{explicittraceform5})
is converted into\\[-12pt]
\beqa
\!\!\!\!
\begin{array}{l}
\langle H \rangle _{{\cal Z};\mbox{{\scriptsize Res}}}
\simeq
\sum _{t=1}^n
\erw{H} \\
\\[-6pt]
+
^{^{^{^{^{.}}}}} \!\!\!
\sum _{s=1}^{n} \!
\sum _{i_s =1}^N 
{\displaystyle \frac{1}{2}} \!\!
\left[ ^{^{^{^{^{.}}}}} \!\!\!\!
\left\{ \!
a^{\alpha \star}_{~i_s} \!
{\displaystyle \frac{w_{s i_s}}{f_s}} \!
a^{\beta }_{~i_s} \!\!
\!+\!
b_{\alpha i_s} \!\!
\left( \!\! 1 \!\!-\!\! {\displaystyle \frac{w_{s i_s}}{f_s}} \!\! \right) \!
b^\star_{\beta i_s} \!
\right\} \!\!
F_{\alpha \beta }^{g_s \star }
\!\!+\!\!
\left\{ \!
a^{\beta \star }_{~i_s}
{\displaystyle \frac{w_{s i_s}}{f_s}}
a^{\alpha }_{~i_s}
\!\!+\!\!
b_{\beta i_s} \!\!
\left( \!\! 1 \!\!-\!\! {\displaystyle \frac{w_{s i_s}}{f_s}} \!\! \right) \!
b^\star_{\alpha i_s} \!\!
\right\} \!\!
F_{\alpha \beta }^{g_s }
\right. \\
\\[-8pt]
~~~~~~~~~~~~~~~~~~~
\!+\!\!
\left\{ \!\!
a^{\alpha \star }_{~i_s}
{\displaystyle \frac{w_{s i_s}}{f_s}}
b_{\beta i_s} \!
\!\!+\!
b_{\alpha i_s} \!\!
\left( \!\! 1 \!\!-\!\! {\displaystyle \frac{w_{s i_s}}{f_s}} \!\! \right) \!
a^{\beta \star }_{~i_s} \!\!
\right\} \!\!
D^{g_s \star }_{\alpha \beta  }
\!\!-\!\!
\left\{ \!
a^{\beta }_{~i_s} \!
{\displaystyle \frac{w_{s i_s}}{f_s}} \!
b^\star _{\alpha i_s}
\!\!+\!\!
b^\star _{\beta i_s} \!\!
\left( \!\! 1 \!\!-\!\! {\displaystyle \frac{w_{s i_s}}{f_s}} \!\! \right) \!
a^{\alpha }_{~i_s} \!
\right\} \!\!
D_{\alpha \beta }^{g_s} \\
\\[-8pt]
\left.
~~~~~~~~~~~~~~~~~~~
\!\!-
b_{\alpha i_s} \!\!
\left( \!
F_{\alpha \beta }^{g_s \star}
b^\star_{\beta i_s}
\!+\!
D_{\alpha \beta }^{g_s \star}
a^{\beta \star}_{~i_s} \!
\right) \!
\!-\!
b^\star_{\alpha  i_s} \!\!
\left( ^{^{^{^{.}}}} \!\!\!
F_{\alpha \beta }^{g_s}
b_{\beta i_s}
\!+\!
D_{\alpha \beta }^{g_s}
a^{\beta}_{~i_s} \!\!\!\!
^{^{^{^{.}}}} 
\right) \!
^{^{^{^{^{.}}}}} \!\!\!
\right]  \\
\\[-8pt]
+
\sum _{s=1}^{n}
{\displaystyle \frac{1}{4}}
[\alpha \beta|\gamma \delta] \\
\\[-6pt]
\!\times\!
\left[
2
\left\{ \!
-
a^{\alpha }_{~i_s} \!\!
{\displaystyle \frac{w_{s i_s}}{f_s}} \!
a^{\beta \star }_{~i_s}
\!-\!
b^\star_{\alpha i_s} \!\!
\left( \!\! 1 \!\!-\!\! {\displaystyle \frac{w_{s i_s}}{f_s}} \!\! \right) \!
b_{\beta i_s} \!
\!+\!
b^\star _{\alpha i_s}
b_{\beta i_s} \!\!
\right\} \!\!
\left\{ ^{^{^{^{.}}}} \!\!\!
\!-\!
a^{\delta \star }_{~j_s} \!
{\displaystyle \frac{w_{s j_s}}{f_s}} \!
a^{\gamma }_{~j_s} \!\!
\!-\!
b_{\delta j_s} \!\!
\left( \!\! 1 \!\!-\!\! {\displaystyle \frac{w_{s j_s}}{f_s}} \!\! \right) \!
b^\star_{\gamma j_s} \!
\!+\!
b_{\delta j_s} \!
b^\star_{\gamma j_s}  \! 
{}^{^{^{^{.}}}} \!\!\!
\right\}
\right. \\
\\[-6pt]
\left.
\!+
\left\{  ^{^{^{^{.}}}} \!\!\!
a^{\alpha }_{~i_s} \!
{\displaystyle \frac{w_{s i_s}}{f_s}}
b^\star_{\gamma i_s}
\!+\!
b^\star_{\alpha i_s} \!\!
\left( \!\! 1 \!\!-\!\! {\displaystyle \frac{w_{s i_s}}{f_s}} \!\! \right) \!
a^{\gamma }_{~i_s} \!\!
\!-\!
a^{\gamma }_{~i_s}
b^\star_{\alpha i_s}
{ }^{^{^{^{.}}}} \!\!\!\! \right\} \!\!
\left\{ \!
a^{\beta \star }_{~j_s}
{\displaystyle \frac{w_{s j_s}}{f_s}}
b_{\delta j_s}
\!+\!
b_{\beta j_s} \!\!
\left( \!\! 1 \!\!-\!\! {\displaystyle \frac{w_{s j_s}}{f_s}} \!\! \right) \!
a^{\delta \star }_{~j_s} \!
\!-\!
b_{\beta j_s}
a^{\delta \star }_{~j_s} \!\!
\right\}
\right]  .
\end{array}
\label{explicittraceform6}
\eeqa
Using
the usual HB orbital energy
$\widetilde{\epsilon }_{r i_r}$
(\ref{generalRes-HBeigenvalueequation}),
(\ref{temperaturedependentWrrblockmatrix})
and
(\ref{temperaturedependentFandD})
but with the modification
$
\widetilde{w} _{r i_r}
\!\rightarrow\!
w _{r i_r}
\!\rightarrow\!
\frac{w _{r i_r}}{f_r}
$,
equation
(\ref{explicittraceform6})
is rewritten into a more compact form\\[-14pt]
\beqa
\!\!\!\!\!\!\!
\begin{array}{l}
\langle H \rangle _{{\cal Z};\mbox{{\scriptsize Res}}}
\simeq
\def\erw#1{{<\!\!#1\!\!>_{\!g_r}}}
\sum _{r=1}^n \!
\erw{H} \\
\\[-10pt]
\!+\!
\sum _{r=1}^n \!
{\displaystyle \frac{1}{2}} \!
\left\{ ^{^{^{^{.}}}} \!\!\!
R_{\mbox{{\scriptsize Res}}:rr;\beta \alpha }
^{\mbox{{\scriptsize thermal}}}
F_{\alpha \beta }^{g_r }
\!-\!
\left(
\delta_{\beta \alpha } \!
\!-\!
R_{\mbox{{\scriptsize Res}}:rr;\beta \alpha }
^{\mbox{{\scriptsize thermal}} \star }
\right) \!
F_{\alpha \beta }^{g_r \star }
\!-\!
K_{\mbox{{\scriptsize Res}}:rr;\beta \alpha }
^{\mbox{{\scriptsize thermal}}}
D^{g_r \star }_{\alpha \beta }
\!-\!
K_{\mbox{{\scriptsize Res}}:rr;\beta \alpha }
^{\mbox{{\scriptsize thermal}} \star }
D_{\alpha \beta }^{g_r} \!
\right\} \\
\\[-8pt]
\!+\!
\sum _{r=1}^n \! 
\left\{ \!
{\displaystyle \frac{1}{2}}
F_{\alpha \alpha }^{g_r \star } \!
+
\sum _{i_r=1}^N \!
\widetilde{\epsilon }_{r i_r}
b^\star_{\alpha i_r}
b_{\alpha i_r} \!\!
\right\} \\
\\[-10pt]
\!+
\sum _{r=1}^n
{\displaystyle \frac{1}{4}}
[\alpha \beta|\gamma \delta] \\
\\[-10pt]
\!\times\!\!
\left[
2 \!
\left( ^{^{^{^{.}}}} \!\!\!\!
\!-\!
R_{\mbox{{\scriptsize Res}}:rr;\alpha \beta }
^{\mbox{{\scriptsize thermal}} \star } \!
\!+\!\!
b^\star_{\alpha i_r} \!
b_{\beta i_r}  \!
\right) \!\!
\left( ^{^{^{^{.}}}} \!\!\!\!
\!-\!
R_{\mbox{{\scriptsize Res}}:rr; \delta \gamma}
^{\mbox{{\scriptsize thermal}} } \!
\!+\!\!
b_{\delta j_r} \!
b^\star_{\gamma j_r} \!
\right)
\!\!+\!\!
\left( ^{^{^{^{.}}}} \!\!\!\!
\!-\!
K_{\mbox{{\scriptsize Res}}:rr;\gamma \alpha }
^{\mbox{{\scriptsize thermal}} \star }
\!-\!
a^{\gamma }_{~i_r}
b^\star_{\alpha i_r} \!
\right) \!\!
\left( ^{^{^{^{.}}}} \!\!\!\!
K_{\mbox{{\scriptsize Res}}:rr;\beta \delta }
^{\mbox{{\scriptsize thermal}}}
\!-\!
b_{\beta j_r}
a^{\delta \star }_{~j_r} \!
\right) \!
^{^{^{^{.}}}} \!\!\!
\right] \\
\\[-10pt]
=
\def\erw#1{{<\!\!#1\!\!>_{g_r}}}
\sum _{r=1}^n
\erw{H}
\!+\!
{\displaystyle \frac{1}{2}}
\sum _{r=1}^n \!
\mbox{Tr}
\left\{ \!
\left[ \!\!
\begin{array}{cc}
R_{\mbox{{\scriptsize Res}}:rr }
^{\mbox{{\scriptsize thermal}}} &
K_{\mbox{{\scriptsize Res}}:rr }
^{\mbox{{\scriptsize thermal}}} \\
\\[-10pt]
-K_{\mbox{{\scriptsize Res}}:rr }
^{\mbox{{\scriptsize thermal}} \ast } &
1_N - R_{\mbox{{\scriptsize Res}}:rr }
^{\mbox{{\scriptsize thermal}} \ast }
\end{array} \!\!
\right] \!
\left[ \!\!
\begin{array}{cc}
F^{g_r}&D^{g_r} \\
\\[-10pt]
-D^{g_r \ast } &-F^{g_r \ast }
\end{array} \!\!
\right] \!
\right\} \\
\\[-12pt]
\!+\!
\sum _{r=1}^n \! 
\left\{ \!
{\displaystyle \frac{1}{2}}
\mbox{Tr} \left\{ F^{g_r \star }\right\} \!
+
\sum _{i_r=1}^N
\widetilde{\epsilon }_{r i_r}
b^\star_{\alpha i_r}
b_{\alpha i_r} \!\!
\right\} \\
\\[-10pt]
+
\sum _{r=1}^n
{\displaystyle \frac{1}{4}}
[\alpha \beta|\gamma \delta] \\
\\[-12pt]
\!\!\times\!\!
\left[ \!
^{^{^{^{^{.}}}}} \!\!\!
2 \!
R_{\mbox{{\scriptsize Res}}:rr;\alpha \beta }
^{\mbox{{\scriptsize thermal}} \star } 
R_{\mbox{{\scriptsize Res}}:rr;\delta \gamma }
^{\mbox{{\scriptsize thermal}} }
\!\!-\!\!
K_{\mbox{{\scriptsize Res}}:rr;\gamma \alpha }
^{\mbox{{\scriptsize thermal}} \star } 
K_{\mbox{{\scriptsize Res}}:rr;\beta \delta }
^{\mbox{{\scriptsize thermal}} } 
\def\erw#1{{<\!\!#1\!\!>_{\!g_r}}}
\!\!-\!\!
2 \!
R_{\mbox{{\scriptsize Res}}:rr;\alpha \beta }
^{\mbox{{\scriptsize thermal}} \star }
\erw{E^{\gamma }_{~\delta }
\!\!+\!\!
{\displaystyle \frac{1}{2}} \delta_{\gamma \delta } } \!
\!\!-\!\!
2 \!
\erw{E^{\alpha }_{~\beta }
\!\!+\!\!
{\displaystyle \frac{1}{2}} \delta_{\alpha \beta } } \!
R_{\mbox{{\scriptsize Res}}:rr;\delta \gamma }
^{\mbox{{\scriptsize thermal}}} 
\right. \\
\\[-14pt]
\left.
\def\erw#1{{<\!\!#1\!\!>_{\!g_r}}}
\!+\!
K_{\mbox{{\scriptsize Res}}:rr;\gamma \alpha }
^{\mbox{{\scriptsize thermal}} \star } \!
\erw{E_{\delta \beta } }
\!-\!
\erw{E^{\alpha \gamma } } \!
K_{\mbox{{\scriptsize Res}}:rr;\beta \delta }
^{\mbox{{\scriptsize thermal}} } 
\def\erw#1{{<\!\!#1\!\!>_{\!g_r}}}
\!+\!
2 \!
\erw{E^{\alpha }_{~\beta }
\!\!+\!\!
{\displaystyle \frac{1}{2}} \delta_{\alpha \beta } } \!
\erw{E^{\gamma }_{~\delta }
\!\!+\!\!
{\displaystyle \frac{1}{2}} \delta_{\gamma \delta }}
\!+\!
\erw{E^{\alpha \gamma } } \!
\erw{E_{\delta \beta } } \!
\right] \! .
\end{array}
\label{explicittraceform7}
\eeqa\\[-8pt]
Owing to  the relations
(\ref{FandDoperators}), i.e.,
$\!
F^{g_r}_{\alpha \beta }
\!\!=\!\!
h_{\alpha \beta }
\!+\!
\def\erw#1{{<\!\!#1\!\!>_{\!g_r}}}
[\alpha \beta|\gamma \delta] 
\erw{E^{\gamma }_{~\delta }
\!\!+\!\!
{\displaystyle \frac{1}{2}} \delta_{\gamma \delta }} 
~\!\mbox{and}\!~
D^{g_r}_{\alpha \gamma }
\!\!=\!\!
{\displaystyle \frac{1}{2}}
[\alpha \beta|\gamma \delta]
\erw{E_{\delta \beta }} 
$,
the final expression for equation
(\ref{explicittraceform7})
leads to the following simple form:\\[-16pt]
\beqa
\!\!\!\!\!
\begin{array}{l}
\langle H \rangle _{{\cal Z};\mbox{{\scriptsize Res}}} 
\!\!=\!\!
\def\erw#1{{<\!\!#1\!\!>_{\!g_r}}}
\sum _{r=1}^n \!
\erw{H}
\def\erw#1{{<\!\!#1\!\!>_{\!g_r}}}
\!\!+\!\!
\sum _{r=1}^n \!
{\displaystyle \frac{1}{2}}
[\alpha \beta|\gamma \delta] \!\!
\left[ \!
\erw{E^{\alpha }_{~\beta }
\!\!+\!\!
{\displaystyle \frac{1}{2}} \delta_{\alpha \beta }} \!
\erw{E^{\gamma }_{~\delta }
\!\!+\!\!
{\displaystyle \frac{1}{2}} \delta_{\gamma \delta }}
\!\!+\!\!
{\displaystyle \frac{1}{2}}
\erw{E^{\alpha \gamma } } \!
\erw{E_{\delta \beta }} \!
\right] \\
\\[-8pt]
+
{\displaystyle \frac{1}{2}} \!
\sum _{r=1}^n \!
\mbox{Tr}
\left\{ \!
\left[ \!\!
\begin{array}{cc}
R_{\mbox{{\scriptsize Res}}:rr }
^{\mbox{{\scriptsize thermal}}} &
K_{\mbox{{\scriptsize Res}}:rr }
^{\mbox{{\scriptsize thermal}}} \\
\\[-10pt]
-K_{\mbox{{\scriptsize Res}}:rr }
^{\mbox{{\scriptsize thermal}} \ast } &
1_N - R_{\mbox{{\scriptsize Res}}:rr }
^{\mbox{{\scriptsize thermal}} \ast }
\end{array} \!\!
\right] \!
\left[ \!\!
\begin{array}{cc}
F^{g_r}&D^{g_r} \\
\\[-10pt]
-D^{g_r \ast } &-F^{g_r \ast }
\end{array} \!\!
\right] \!
\right\}
\!+\!
\sum _{r=1}^n \!\!
\sum _{i_r=1}^N \!
\widetilde{\epsilon }_{r i_r}
b^\star_{\alpha i_r}
b_{\alpha i_r} \\
\\[-10pt]
-
\sum _{r=1}^n \!
{\displaystyle \frac{1}{2}} \!
\left\{ ^{^{^{^{.}}}} \!\!\!
R_{\mbox{{\scriptsize Res}}:rr;\beta \alpha }
^{\mbox{{\scriptsize thermal}} }
F^{g_r}_{\alpha \beta }
-\!
\left(
\delta_{\beta \alpha }
\!-\!
R_{\mbox{{\scriptsize Res}}:rr;\beta \alpha }
^{\mbox{{\scriptsize thermal}} \star }
\right) \!
F^{g_r \star }_{\alpha \beta }
\!-\!
K_{\mbox{{\scriptsize Res}}:rr;\beta \alpha }
^{\mbox{{\scriptsize thermal}} }
D^{g_r \ast }_{\alpha \beta }
\!-\!
K_{\mbox{{\scriptsize Res}}:rr;\beta \alpha }
^{\mbox{{\scriptsize thermal}} \star }
D^{g_r}_{\alpha \beta }
\right\}  \\
\\[-8pt]
+
\sum _{r=1}^n \!
{\displaystyle \frac{1}{2}} \!
\left\{ ^{^{^{^{.}}}} \!\!\!\!
-
2 \!
K_{\mbox{{\scriptsize Res}}:rr;\beta \alpha }
^{\mbox{{\scriptsize thermal}} }
D^{g_r \ast }_{\alpha \beta }
\right\} \\
\\[-10pt]
+
\sum _{r=1}^n \!
\left[
h_{\alpha \beta }
R_{\mbox{{\scriptsize Res}}:rr;\beta \alpha }
^{\mbox{{\scriptsize thermal}} } \!
+
{\displaystyle \frac{1}{2}}
[\alpha \beta|\gamma \delta] \!
\left\{ \!
R_{\mbox{{\scriptsize Res}}:rr;\beta \alpha }
^{\mbox{{\scriptsize thermal}} }
R_{\mbox{{\scriptsize Res}}:rr;\delta \gamma }
^{\mbox{{\scriptsize thermal}} }
\!-\!
{\displaystyle \frac{1}{2}}
K_{\mbox{{\scriptsize Res}}:rr;\alpha \gamma }
^{\mbox{{\scriptsize thermal}} \star }
K_{\mbox{{\scriptsize Res}}:rr;\delta \beta }
^{\mbox{{\scriptsize thermal}} } \!
\right\} \!
\right] \\
\\[-12pt]
=
\def\erw#1{{<\!\!#1\!\!>_{\!g_r}}}
2 \! \sum _{r=1}^n \!
\erw{H}
\!-\!
\sum _{r=1}^n \!
h_{\alpha \beta }
\erw{E^{\alpha }_{~\beta }
\!\!+\!\!
{\displaystyle \frac{1}{2}} \delta_{\alpha \beta }}
\!+\!
\sum _{r=1}^n \!\!
\sum _{i_r=1}^N \!
\widetilde{\epsilon }_{r i_r}
b^\star_{\alpha i_r}
b_{\alpha i_r} 
\!+\! 
\def\erw#1{{<\!\!#1\!\!>_{\!g_r}^{\mbox{{\scriptsize thermal}}}}}
\sum _{r=1}^n \!
\erw{H} .
\end{array}
\label{explicittraceform8}
\eeqa\\[-8pt]
The first three terms in the last line of the above equation
(\ref{explicittraceform8})
are temperature independent while the last term is temperature dependent and is defined as\\[-16pt]
\beqa
\begin{array}{l}
\def\erw#1{{<\!\!#1\!\!>_{\!g_r}^{\mbox{{\scriptsize thermal}}}}}
\erw{H}
\!\equiv\!
h_{\alpha \beta }
R_{\mbox{{\scriptsize Res}}:rr;\beta \alpha }
^{\mbox{{\scriptsize thermal}} } \!
+
{\displaystyle \frac{1}{2}}
[\alpha \beta|\gamma \delta] \!
\left\{ \!
R_{\mbox{{\scriptsize Res}}:rr;\beta \alpha }
^{\mbox{{\scriptsize thermal}} }
R_{\mbox{{\scriptsize Res}}:rr;\delta \gamma }
^{\mbox{{\scriptsize thermal}} }
\!-\!
{\displaystyle \frac{3}{2}}
K_{\mbox{{\scriptsize Res}}:rr;\alpha \gamma }
^{\mbox{{\scriptsize thermal}} \star }
K_{\mbox{{\scriptsize Res}}:rr;\delta \beta }
^{\mbox{{\scriptsize thermal}} } \!
\right\} ,
\end{array}
\label{explicittraceform9}
\eeqa\\[-12pt]
which is a thermal
expectation value of the Hamiltonian
\cite{Ozaki.85,Goodman.81}
in the $g_r$ quasi-particle frame
but the numerical factor in the curl blackets is modified
from $-\frac{1}{2}$ to $-\frac{3}{2}$.
Such a modification can take place
since the HB WFs with different structures are resonating.

\newpage

Let us remind the variable
$w_{r i_r} \!
\left(
\!\equiv\! 
[ 1 \!+\! e^{ \beta \epsilon _{r i_r }}]^{-1} 
\right)
$.
Then the variation of the Res-HB free energy
$F[{\cal Z}]_{\mbox{{\scriptsize Res}}}$
(\ref{Res-HB free energy 2})
with respect to the variable
${\displaystyle \frac{w_{r i_r}}{f_r}}$
is calculated as\\[-14pt]
\beqa
\begin{array}{l}
\delta F[{\cal Z}]_{\mbox{{\scriptsize Res}}}
\!=\!
\delta \langle H  \rangle
_{{\cal Z};\mbox{{\scriptsize Res}}}
\!-
\delta \!
\left\{ \!
\langle H[{\cal Z}]_{\mbox{{\scriptsize Res}}} \rangle
_{{\cal Z};\mbox{{\scriptsize Res}}}
\!+\!
{\displaystyle \frac{1}{\beta }} \ln \! \mbox{Tr} \!
\left( Pe^{- \beta H[{\cal Z}]_{\mbox{{\scriptsize Res}}}} \right) \!
\right\} .
\end{array}
\label{variationRes-HBfreeenergy}
\eeqa\\[-8pt]
First we give the variational formula for the variable
${\displaystyle \frac{w_{r i_r}}{f_r}}$
as follows:\\[-12pt]
\beqa
\begin{array}{ll}
\delta \! 
\left( \! {\displaystyle \frac{w_{r i_r}}{f_r}}  \! 
\right)
\!=\!
&\!\!
{\displaystyle \frac{\delta w_{r i_r}}{f_r}} \!
-
{\displaystyle \frac{w_{r i_r}}{f_r}}
{\displaystyle \frac{\delta f_r}{f_r}} \\
\\[-10pt]
&\!\!\!\!\!\!\!\!
\!=
-
{\displaystyle \frac{w_{r i_r}}{f_r}}
\left( 1 - w_{r i_r} \right) \!
\delta \! 
\left( \beta \epsilon _{r i_r }\right)
-
{\displaystyle \frac{w_{r i_r}}{f_r}}
{\displaystyle \frac{1}{f_r}}
S _{rr}
{\displaystyle \frac{1}{1 -w_{r i_r}}}
{\displaystyle \frac{\delta w_{r i_r}}{1 -w_{r i_r}}}
(S^{-1})_{rr}  \\
\\[-10pt]
&\!\!\!\!\!\!\!\!
\!=
-
{\displaystyle \frac{w_{r i_r}}{f_r}}
\left( 1 - w_{r i_r} \right) \!
\delta \! 
\left( \beta \epsilon _{r i_r }  \right)
+
{\displaystyle \frac{w_{r i_r}}{f_r}}
{\displaystyle \frac{1}{f_r}}
S _{rr}
{\displaystyle \frac{w_{r i_r}}{1 -w_{r i_r}}}
\delta \! \left( \beta \epsilon _{r i_r }  \right) \!
(S^{-1})_{rr}  \\
\\[-10pt]
&\!\!\!\!\!\!\!\!
\!=
-
{\displaystyle \frac{w_{r i_r}}{f_r}}
\left\{ \!
1 - w_{r i_r}
\!-\!
{\displaystyle \frac{1}{f_r}}
S _{rr}
{\displaystyle \frac{w_{r i_r}}{1 -w_{r i_r}}} \!
(S^{-1})_{rr}
\right\} \!
\delta \! 
\left( \beta \epsilon _{r i_r } \right)   \\
\\[-6pt]
&\!\!\!\!\!\!\!\!
\!\equiv 
\Delta w_{r i_r}
\delta \! 
\left( \beta \epsilon _{r i_r } \right)   .
\end{array}
\label{variationalformula}
\eeqa
Using
(\ref{variationalformula}),
the variational formula for the
$R_{\mbox{{\scriptsize Res}}:rr}^{\mbox{{\scriptsize thermal}}}$
and
$K_{\mbox{{\scriptsize Res}}:rr}^{\mbox{{\scriptsize thermal}}}$
matrices are given as
\beqa
\left.
\begin{array}{rl}
\delta R_{\mbox{{\scriptsize Res}}:rr;\alpha \beta}
^{\mbox{{\scriptsize thermal}}}
=
a^{\alpha \star }_{~i_r}
\delta \!
\left( \! {\displaystyle \frac{w_{r i_r}}{f_r}} \! \right) \!
a^{\beta }_{~i_r}
\!-\!
b_{\alpha i_r}
\delta \!
\left( \! {\displaystyle \frac{w_{r i_r}}{f_r}} \! \right) \!
b^\star_{\beta i_r} , \\
\\[-6pt]
\delta K_{\mbox{{\scriptsize Res}}:rr;\alpha \beta}
^{\mbox{{\scriptsize thermal}}}
=
a^{\alpha \star}_{~i_r}
\delta \!
\left( \! {\displaystyle \frac{w_{r i_r}}{f_r}} \! \right) \!
b_{\beta i_r}
\!-\!
b_{\alpha i_r}
\delta \!
\left( \! {\displaystyle \frac{w_{r i_r}}{f_r}} \! \right) \!
a^{\beta \star}_{~i_r} .
\end{array} \!
\right\}
\label{deltaRandK}
\eeqa
The second part of the variation in
(\ref{variationRes-HBfreeenergy})
is computed as\\[-14pt]
\beqa
\begin{array}{c}
-~
\delta
\left\{ \!
\langle H[{\cal Z}]_{\mbox{{\scriptsize Res}}} \rangle
_{\!{\cal Z};\mbox{{\scriptsize Res}}}
\!+\!
{\displaystyle 
\frac{1}{\beta }} \!
\ln \!
\mbox{Tr} \!
\left( \! Pe^{- \beta H[{\cal Z}]_{\mbox{{\scriptsize Res}}}} \! \right) \!
\right\} \\
\\[-10pt]
\!=\!
{\displaystyle \frac{1}{\beta }}
\sum _{r\!=\!1}^n \!
\sum _{i_r \!=\!1}^N \!
\delta \!
\left[ \!
{\displaystyle \frac{w _{r i_r}}{f_r}}
\ln \!
{\displaystyle \frac{w _{r i_r}}{f_r}}
\!+\!
\left( \!
\! 1
\!-\! 
{\displaystyle \frac{w _{r i_r}}{f_r}} \!\!
\right) \!
\ln \!
\left( \!\!
1
\!-\! 
{\displaystyle \frac{w _{r i_r}}{f_r}} \!
\right)
\right]  \\
\\[-4pt]
\!=\!
- {\displaystyle \frac{1}{\beta }}
\sum _{r\!=\!1}^n \!
\sum _{i_r \!=\!1}^N
\ln 
\left[
{\displaystyle
\frac{
f_r
\!-\!
w _{r i_r}
}{
w _{r i_r}
}
}
\right] \! 
\Delta w_{r i_r}
\delta \! 
\left( \beta \epsilon _{r i_r } \right) .
\end{array}
\label{deltaexplicittraceform4}
\eeqa
The variational formula for the thermal expectation value
$
\def\erw#1{{<\!\!#1\!\!>_{g_r}^{\mbox{{\scriptsize thermal}}}}}
\erw{H}
$
(\ref{explicittraceform9})
is calculated as\\[-12pt]
\beqa
\begin{array}{c}
\delta
\def\erw#1{{<\!\!#1\!\!>_{g_r}^{\mbox{{\scriptsize thermal}}}}}
\erw{H}
=
{\displaystyle \frac{1}{2}} \!
\left\{
F_{\mbox{{\scriptsize Res}}:rr;\beta \alpha }
^{\mbox{{\scriptsize thermal}} }
\delta R_{\mbox{{\scriptsize Res}}:rr;\alpha \beta }
^{\mbox{{\scriptsize thermal}} }
+
F_{\mbox{{\scriptsize Res}}:rr;\beta \alpha }
^{\mbox{{\scriptsize thermal}} \star }
\delta R_{\mbox{{\scriptsize Res}}:rr;\alpha \beta }
^{\mbox{{\scriptsize thermal}} \star  }
\right. \\
\\[-6pt]
\left.
-
3 D_{\mbox{{\scriptsize Res}}:rr;\beta \alpha }
^{\mbox{{\scriptsize thermal}} }
\delta K_{\mbox{{\scriptsize Res}}:rr;\alpha \beta }
^{\mbox{{\scriptsize thermal}}  \star}
-
3 D_{\mbox{{\scriptsize Res}}:rr;\beta \alpha }
^{\mbox{{\scriptsize thermal}} \star }
\delta K_{\mbox{{\scriptsize Res}}:rr;\alpha \beta }
^{\mbox{{\scriptsize thermal}} }
\right\}  \\
\\[-4pt]
\!=\!
{\displaystyle \frac{1}{2}} \!
\left[ \!
F_{\mbox{{\scriptsize Res}}:rr;\beta \alpha }
^{\mbox{{\scriptsize thermal}} } \!\!
\left\{ \!\!
a^{\alpha \star }_{~i_r} \!
\delta \!\!
\left( \!\! {\displaystyle \frac{w_{r i_r}}{f_r}} \!\! \right) \!\!
a^{\beta }_{~i_r}
\!\!-\!\!
b_{\alpha i_r}
\delta \!\!
\left( \!\! {\displaystyle \frac{w_{r i_r}}{f_r}} \!\! \right) \!\!
b^\star_{\beta i_r} \!\!
\right\}
\!\!+\!\!
F_{\mbox{{\scriptsize Res}}:rr;\beta \alpha }
^{\mbox{{\scriptsize thermal}} \star } \!\!
\left\{ \!\!
a^{\alpha }_{~i_r} \!
\delta \!\!
\left( \!\! {\displaystyle \frac{w_{r i_r}}{f_r}} \!\! \right) \!\!
a^{\beta \star }_{~i_r}
\!\!-\!\!
b^\star_{\alpha i_r}
\delta \!\!
\left( \!\! {\displaystyle \frac{w_{r i_r}}{f_r}} \!\! \right) \!\!
b_{\beta i_r} \!\!
\right\}
\right. \\
\\[-6pt]
\left.
-
3 D_{\mbox{{\scriptsize Res}}:rr;\beta \alpha }
^{\mbox{{\scriptsize thermal}} } \!\!
\left\{ \!\!
a^{\alpha }_{~i_r} \!
\delta \!\!
\left( \!\! {\displaystyle \frac{w_{r i_r}}{f_r}} \!\! \right) \!\!
b^\star _{\beta i_r}
\!\!-\!\!
b^\star_{\alpha i_r}
\delta \!\!
\left( \!\! {\displaystyle \frac{w_{r i_r}}{f_r}} \!\! \right) \!\!
a^{\beta }_{~i_r} \!
\right\}
\!\!-\!\!
3 D_{\mbox{{\scriptsize Res}}:rr;\beta \alpha }
^{\mbox{{\scriptsize thermal}} \star } \!\!
\left\{ \!\!
a^{\alpha \star }_{~i_r} \!
\delta \!\!
\left( \!\! {\displaystyle \frac{w_{r i_r}}{f_r}} \!\! \right) \!\!
b_{\beta i_r}
\!\!-\!\!
b_{\alpha i_r}
\delta \!\!
\left( \!\! {\displaystyle \frac{w_{r i_r}}{f_r}} \!\! \right) \!\!
a^{\beta \star }_{~i_r} \!\!
\right\} \!
\right] ,
\end{array}
\label{deltaexplicittraceform9}
\eeqa
where we have used the relations
(\ref{temperaturedependentFandD})
and
(\ref{deltaRandK}).

\newpage

~~~Finally,
by adding
(\ref{deltaexplicittraceform9}) 
to
(\ref{deltaexplicittraceform4}),
the variation of the
$F[{\cal Z}]_{\mbox{{\scriptsize Res}}}$ is made
as follows:
\beqa
\begin{array}{c}
\delta
F[{\cal Z}]_{\mbox{{\scriptsize Res}}}
=
{\displaystyle \frac{1}{2}}
\sum _{r\!=\!1}^n
\sum _{i_r \!=\!1}^N \\
\\[-10pt]
\!\times\!
\left[ {}^{^{^{^{^{^{.}}}}}} \!\!\!\!
F_{\mbox{{\scriptsize Res}}:rr;\beta \alpha }
^{\mbox{{\scriptsize thermal}} } \!\!
\left\{ \!
a^{\alpha \star }_{~i_r} \!
\Delta w_{r i_r} \!\!
a^{\beta }_{~i_r}
\!-\!
b_{\alpha i_r}
\Delta w_{r i_r} \!\!
b^\star_{\beta i_r} \!\!
\right\}
\!+\!
F_{\mbox{{\scriptsize Res}}:rr;\beta \alpha }
^{\mbox{{\scriptsize thermal}} \star } \!
\left\{ \!
a^{\alpha }_{~i_r} \!
\Delta w_{r i_r} \!\!
a^{\beta \star }_{~i_r}
\!-\!
b^\star_{\alpha i_r}
\Delta w_{r i_r} \!\!
b_{\beta i_r} \!\!
\right\}
\right. \\
\\[-4pt]
-3 D_{\mbox{{\scriptsize Res}}:rr;\beta \alpha }
^{\mbox{{\scriptsize thermal}} } \!\!
\left\{ \!
a^{\alpha }_{~i_r} \!
\Delta w_{r i_r} \!
b^\star _{\beta i_r} \!
\!-\!
b^\star_{\alpha i_r}
\Delta w_{r i_r} \!
a^{\beta }_{~i_r} \!\!
\right\} \!
\!-\!
3 D_{\mbox{{\scriptsize Res}}:rr;\beta \alpha }
^{\mbox{{\scriptsize thermal}} \star } \!\!
\left\{ \!
a^{\alpha \star }_{~i_r} \!
\Delta w_{r i_r} \!
b_{\beta i_r} \!
\!-\!
b_{\alpha i_r}
\Delta w_{r i_r} \!
a^{\beta \star }_{~i_r} \!\!
\right\}  \\
\\[-4pt]
\left.
-
{\displaystyle \frac{2}{\beta }}
\ln 
\left[
{\displaystyle
\frac{
f_r
\!-\!
w _{r i_r}
}{
w _{r i_r}
}
}
\right] \!
\Delta w_{r i_r}
\right] \!
\delta \! 
\left( \! \beta \epsilon _{r i_r } \! \right) \\
\\[-4pt]
\!=\!
{\displaystyle \frac{1}{2}} \!
\left\{ \!
[a^\star _{i_r}, b^\star _{i_r}] \!\!
\left[ \!\!\!\!
\begin{array}{cc}
F^{\mbox{{\scriptsize thermal}} \star }
_{\mbox{{\scriptsize Res}}:rr}&\!\!
3 D^{\mbox{{\scriptsize thermal}} \star }
_{\mbox{{\scriptsize Res}}:rr} \\ 
\\[-4pt]
-3 D^{\mbox{{\scriptsize thermal}} }
_{\mbox{{\scriptsize Res}}:rr}
&\!\!-F^{\mbox{{\scriptsize thermal}} }
_{\mbox{{\scriptsize Res}}:rr}
\end{array} \!\!\!\!
\right] \!\!
\Delta w_{r i_r} \!\!\!
\left[ \!\!\!
\begin{array}{c}
a _{i_r} \\
\\
b _{i_r}
\end{array} \!\!\!\!
\right] \!
-
{\displaystyle \frac{1}{\beta }}
\ln 
\left[ \!
{\displaystyle
\frac{
f_r
\!\!-\!\!
w _{r i_r}
}{
w _{r i_r}
}
} \!
\right] \!
\Delta w_{r i_r}
\right. \\
\\
\left.
~~~~+\!
[a _{i_r}, b _{i_r}] \!\!
\left[ \!\!\!\!
\begin{array}{cc}
F^{\mbox{{\scriptsize thermal}} }
_{\mbox{{\scriptsize Res}}:rr}&\!\!
3 D^{\mbox{{\scriptsize thermal}} }
_{\mbox{{\scriptsize Res}}:rr} \\
\\ 
-3 D^{\mbox{{\scriptsize thermal}} \star }
_{\mbox{{\scriptsize Res}}:rr}
&\!\!-F^{\mbox{{\scriptsize thermal}}  \star }
_{\mbox{{\scriptsize Res}}:rr}
\end{array} \!\!\!\!
\right] \!\!
\Delta w_{r i_r} \!\!\!
\left[ \!\!\!
\begin{array}{c}
a^\star  _{i_r}\\
\\
b^\star _{i_r}
\end{array} \!\!\!
\right] \!
-
{\displaystyle \frac{1}{\beta }} 
\ln 
\left[ \!
{\displaystyle
\frac{
f_r
\!\!-\!\!
w _{r i_r}
}{
w _{r i_r}
}
} \!
\right] \!
\Delta w_{r i_r} \!
\right\} \!
\delta \! 
\left( \! \beta \epsilon _{r i_r } \! \right)  \! .
\end{array}
\label{deltaResHBfreeenergy}
\eeqa
Then the variation equation of the Res-HB free energy
$\!F[{\cal Z}]_{\mbox{{\scriptsize Res}}}$
is given by
\beqa
\begin{array}{c}
\delta
F[{\cal Z}]_{\mbox{{\scriptsize Res}}}
=
0 ,
\end{array}
\label{variationequationResHBfreeenergy}
\eeqa
from which,
for $r \!=\! 1, \cdots , n$,
we have
\beqa
\!\!\!\!\!\!
\begin{array}{c}
\left[ \!\!
\begin{array}{cc}
F^{\mbox{{\scriptsize thermal}} \star }
_{\mbox{{\scriptsize Res}}:rr}
\!-\!
{\displaystyle \frac{1}{\beta }}
\ln 
\left[ \!
{\displaystyle
\frac{
f_r
\!\!-\!\!
w _{r i_r}
}{
w _{r i_r}
}
} \!
\right] \!
\!\cdot\!
I_{N}
&
3 D^{\mbox{{\scriptsize thermal}} \star }
_{\mbox{{\scriptsize Res}}:rr} \\ 
\\[-10pt]
-3 D^{\mbox{{\scriptsize thermal}} }
_{\mbox{{\scriptsize Res}}:rr}
&
-F^{\mbox{{\scriptsize thermal}} }
_{\mbox{{\scriptsize Res}}:rr}
\!\!-\!
{\displaystyle \frac{1}{\beta }}
\ln 
\left[ \!
{\displaystyle
\frac{
f_r
\!\!-\!\!
w _{r i_r}
}{
w _{r i_r}
}
} \!
\right] \!
\!\cdot\!
I_{N}
\end{array} \!\!
\right] \!
\Delta w_{r i_r} \!\!\!
\left[ \!\!
\begin{array}{c}
a _{i_r} \\
\\ \\
b _{i_r}
\end{array} \!\!\!
\right] \!
\!\!=\!\!
\left[ \!\!
\begin{array}{c}
a _{i_r} \\
\\ \\
b _{i_r}
\end{array} \!\!\!
\right] \!\!
E _{r i_r}^{\mbox{{\scriptsize thermal}}} \! .
\end{array}
\label{thermalResHBeigenvalueequation}
\eeqa
The
$E _{r i_r}^{\mbox{{\scriptsize thermal}}}$
is a new quasi-particle energy different from the one
arisen in
$w_{r i_r} \!
\left( \!
\!\equiv\! 
{\displaystyle \frac{1}{1 \!\!+\!\! e^{ \beta \epsilon _{r i_r }}}} \!\!\!
\right) \!
$.
We write
(\ref{thermalResHBeigenvalueequation})
as
${\cal H}_{\mbox{{\scriptsize Res}}:r}
^{\mbox{{\scriptsize thermal}}}u_{r i_r}
\!\!=\!\!
E ^{\mbox{{\scriptsize thermal}}}_ru_{r i_r}
(r \!\!=\!\! 1, \!\cdots\!, n)$.
In it there exist factors
$f_r$ defined in
(\ref{explicittraceform})
which contains $S_{rs}$ directly
and
$\Delta w_{r i_r}\!$ defined in
(\ref{variationalformula})
which has a diagonal element of
inverse matrix of overlap integral
$(S^{-1})_{rr}
(=\! \sum_{k\!=\!1}^n \! c_r^{(k)*} \! c_r^{(k)};$
See II.).
This means that
the equation
${\cal H}_{\mbox{{\scriptsize Res}}:r}
^{\mbox{{\scriptsize thermal}}}u_{r i_r}
\!\!=\!\!
E ^{\mbox{{\scriptsize thermal}}}_ru_{r i_r}
$
forces us to couple the HB WF $|g_r\rangle$ with
the other $|g_s\rangle$ and $\cdots$, 
though no direct couplings
among the mixing coefficients manifestly exist.
Thus we can reach our ultimate goal, i.e.,
the thermal Res-HB coupled eigenvalue equation
within the present approximation of
partition function and free energy.
This thermal Res-HB coupled eigenvalue equation, however,
has the very different form from that of
the previous thermal Res-HB coupled one
proposed in II
\cite{NishiProviOhnishi.13},
for $r \!\!=\!\! 1, \cdots, n$,
which is described as
\beqa
\left.
\!\!\!\!\!\!\!\!\!\!\!\!
\begin{array}{rl}
&
[{\cal F}_{\mbox{{\scriptsize Res}}:r}
^{\mbox{{\scriptsize thermal}}}u_r]_i
\!=\!
\epsilon_{ri}^{\mbox{{\scriptsize thermal}}}u_{ri},~~
\epsilon_{ri}^{\mbox{{\scriptsize thermal}}}
\!\equiv\! 
\widetilde{\epsilon}_{ri}^{\mbox{{\scriptsize thermal}}}
\!-\!
\sum_{k =1}^n \!
\left\{ \!
H[W_{\mbox{{\scriptsize Res}}:rr}
^{\mbox{{\scriptsize thermal}}}] \!-\! E^{(k)} \!
\right\} \!
|c_r ^{(k)}|^2 , \\
\\[-2pt]
&\!\!
{\cal F}_{\mbox{{\scriptsize Res}}:r}
^{\mbox{{\scriptsize thermal}}}
\!\equiv\! 
{\cal F}[W_{\mbox{{\scriptsize Res}}:rr}
^{\mbox{{\scriptsize thermal}}}] \!
\sum_{k =1}^n \! |c_r ^{(k)} |^2
\!+\!\!
\sum_{k =1}^n \!
\sum_{s=1}^{\prime~n} \!
\left\{ \!
{\cal K}_{\mbox{{\scriptsize Res}}:rs}
^{\mbox{{\scriptsize thermal}}(k)} \!
c_r ^{(k)*} c_s^{(k)} \!
\!+\! 
{\cal K}_{\mbox{{\scriptsize Res}}:rs}
^{\mbox{{\scriptsize thermal}}(k)\dag } \!
c_r ^{(k)} c_s^{(k)*} \!
\right\} , \!\!\!\!
\end{array}
\right\}
\label{thermalRes-HBeigenvalueequation0}
\eeqa
where the quantity
${\cal K}_{\mbox{{\scriptsize Res}}:rs}
^{\mbox{{\scriptsize thermal}}(k)}$
is defined by
(\ref{thermalRes-HBequation}).
This is the reason
why we have started from
the quadratic Res-HB Hamiltonian
(\ref{quadraticHBHamiltonian2})
and why we also have used
the solved set of the Res-HB eigenvalue equations
$[{\cal F}_r u_r]_i \!=\! \epsilon _{ri} u_{ri}
~(r \!=\! 1, \cdots , n)$,
(\ref{generalRes-HBeigenvalueequation}).

\newpage

%%%%%%%%%%%%%%%%%%%%%%%
%                                                               %
%  6  Summary and further perspectives   %
%                                                               %
%%%%%%%%%%%%%%%%%%%%%%%

\def\thesection{\arabic{section}}
\setcounter{equation}{0}
\renewcommand{\theequation}{\arabic{section}.\arabic{equation}}
\section{Summary and further perspectives}

~~~
In this paper
we have given a rigorous thermal Res-HBT and
Res-MF approximation,
to describe a superconducting fermion system.
We have used a Res-HB subspace spanned by Res-HB
ground and excited states.
Using the projection operator $P$ to the Res-HB subspace,
the partition function in the Res-HB subspace is given as
$
\mbox{Tr} ( \! Pe^{- \beta H} \!)
\!=\!
\sum _{r,s=1}^n
\langle g _r |
e^{- \beta H}
|g _s \rangle (S^{-1})_{sr}
$.
In principle this trace formula can be computed
within the Res-HB subspace
by using the Laplace transform of
$e^{-\beta H}$
and the projection method
\cite{Naka.58,Zwan.60,Mori.65,Fulde.93},
however,
whose computation by the IMCF is cumbersome.
A group action on a HB-Hamiltonian and -density matrix
at finite temperature are defined.
The variation of the Res-HB free energy
is made parallel to the usual thermal BCS theory
\cite{BCS.57,Bogo.59,KA.59,AGD.65}. %,Abrikosov.88}.
It leads to
the thermal Res-HB CI equation and
the thermal Res-HB equation
which is equivalent with
the thermal Res-HB coupled eigenvalue equations
${\cal F}_{\mbox{{\scriptsize Res}}:r}
^{\mbox{{\scriptsize thermal}}}u_r
\!=\!
\epsilon ^{\mbox{{\scriptsize thermal}}}_ru_r$
for the thermal Res-FB operator
${\cal F}_{\mbox{{\scriptsize Res}}:r}
^{\mbox{{\scriptsize thermal}}}$.
The variation of the Res-MF free energy
brings us the $r$th thermal HB density matrix
$W_{\mbox{{\scriptsize Res}}:rr}
^{\mbox{{\scriptsize thermal}}}$ 
expressed in terms of
the $r$th thermal Res-FB operator
${\cal F}_{\mbox{{\scriptsize Res}}:r}
^{\mbox{{\scriptsize thermal}}}$
as
$
W_{\mbox{{\scriptsize Res}}:rr}
^{\mbox{{\scriptsize thermal}}}
\!=\!
$
$
\{  1_{\!2N} \!+\! \exp \!
( \beta {\cal F} _{\mbox{{\scriptsize Res}}:r}
^{\mbox{{\scriptsize thermal}}} )
\}^{\!-1}
$.
Instead of the above formal theory,
we here have proposed a more practical way to the Res-HB theory
by approximation of partition function and free energy.

Recently,
to demonstrate the predominance of the Res-HB MF theory for
superconducting fermion systems with large quantum fluctuations,
in II we have applied it to a naive BCS Hamiltonian for singlet pairing
\cite{NishiProviOhnishi.10}.
A state with large quantum fluctuations
is approximated by superposition of two HB WFs
which are non-orthogonal CS reps with different correlation structures.
We have optimized directly the Res-MF energy functionals
by variations of the Res-MF ground-state energy
with respect to the Res-MF parameters,
the so-called {\it energy-gaps}.
The Res-MF ground and excited states
generated with the two HB WFs
explain most of the two energy-gaps in $\mbox{MgB}_2$.
Both the large {\em energy-gap} and the small one
have a significant physical meaning
because electron systems,
composed of condensed electron pairs,
have now strong correlations among fermions.
We also have treated the special case of
{\it equal energy-gaps}
and obtained interesting analytic solutions
\cite{NishiProviOhnishi.13}.

A time-reversed single-particle state $\bar{\alpha }$
is obtained from $\alpha$
and
a phase factor is used
$s_{\alpha }$
in the time reversal of physical quantities.
For the naive singlet-pairing interaction
$
[\alpha \gamma| \beta \delta ]
=
- g s_{\alpha }
\delta_{\alpha \bar{\beta}} s_{\gamma }
\delta_{\gamma \bar{\delta}},~(g:\mbox{force strength})
$,
the thermal pairing potential
$D_{\mbox{{\scriptsize Res}}:rr;\alpha \beta }
^{\mbox{{\scriptsize thermal}}}$
is expressed as
\begin{eqnarray}
D_{\mbox{{\scriptsize Res}}:rr;\alpha \beta }
^{\mbox{{\scriptsize thermal}}}
=
- s_{\alpha } \delta_{\alpha \bar{\beta }}
\Delta_{\mbox{{\scriptsize Res}}:rr}
^{\mbox{{\scriptsize thermal}}} ,
~~
\Delta_{\mbox{{\scriptsize Res}}:rr}
^{\mbox{{\scriptsize thermal}}}
=
{\displaystyle \frac{g}{2}} s_{\delta } 
K_{\mbox{{\scriptsize Res}}:rr; \delta \bar{\delta }}
^{\mbox{{\scriptsize thermal}}} .~(r = 1,\cdots, n)
\label{pair potential}
\end{eqnarray}
Combining 
(\ref{pair potential})
with
${\cal H}_{\mbox{{\scriptsize Res}}:r}
^{\mbox{{\scriptsize thermal}}}u_{r i_r}
\!=\!
E ^{\mbox{{\scriptsize thermal}}}_ru_{r i_r}
~(r = 1, \!\cdots\! ,  n)$,
i.e.,
(\ref{thermalResHBeigenvalueequation}),
we reach our ultimate goal of  the coupled thermal Res-HB multi-gap equations.
It may be expected to open a new research area
in the vigorous pursuit by the radical spirit of the Res-HBT
to develop a theoretical framework
appropriate for exploring the problem of high-$T_c$ superconductors.

Finally, it should be emphasized that
we may also provide a thermal Res-HF approximation.
We already have an expression for partition function in
a $U(N)$ CS rep $| u \rangle$
\cite{Perelomov.72},
$\mbox{Tr}(e^{-\beta H})
\!\!=\!\!
{}_N C_n \! \int \! \langle u |e^{-\beta H}| u \rangle du,~
{}_N C_n
\!\!=\!\!
N!/n!(N \!-\! n)!$
($n$: Number of occupied orbitals)
where
the integration is the group integration on the group $U(N)$.
Following Fukutome
\cite{Fuku.88},
using the projection operator $P$ to the Res-HF subspace,
the partition function in the Res-HF subspace can also be computed.
The variation of the Res-HF free energy
is made in the same way as the present thermal Res-HBT
and
it may be applied to a 1-D half-filled Hubbard model
\cite{Hubbard.63}
and
a simple LMG nuclear-model
\cite{LMG.65}.
These works will appear elsewhere.

\newpage

%%%%%%%%%%%
%                           %
%     Appendix      %
%                           %
%%%%%%%%%%%

\leftline{\large{\bf Appendix}}
\appendix

\vspace{-0.5cm}

\def\thesection{\Alph{section}}
\setcounter{equation}{0}
\renewcommand{\theequation}{\Alph{section}.\arabic{equation}}
\section{Calculation of the first few matrices
$\mathbb{S}_{rs}^{(l)}$
and 
$\mathbb{L}_{ur}^{(l)}$}

\vspace{-0.2cm}

~~~~
 In
(\ref{projectionoperator1})
by inserting the explicit expression for the projection operators
$P^{(1)}$
into
$Q^{(1)}$,\\[-14pt]
\beqa
\begin{array}{c}
Q^{(1)}
\equiv
1
-
\sum _{u,v=1}^n
|g _u^{(1)} \rangle (\mathbb{S}^{(1)-1})_{uv} \langle g _v^{(1)} |
=
1
-
\sum _{u,v=1}^n 
QH|g _u \rangle (\mathbb{S}^{(1)-1})_{uv} \langle g _v |HQ ,
\end{array}
\label{projectionoperatorQ1}
\eeqa\\[-12pt]
and denoting
$\langle g _r | H^m | g _s \rangle$
as
$(\mathbb{H}^{m})_{rs}$,
first few matrices
$\mathbb{S}_{rs}^{(l)}$
and 
$\mathbb{L}_{ur}^{(l)}$
can be calculated as
\beqa
\!\!\!\!\!\!\!\!\!
\left.
\begin{array}{ll}
&\!\!\!
\mathbb{S}_{rs}^{(1)}
=
\langle g _r^{(1)} | g _s^{(1)} \rangle
=
\langle g _r |
HQH
|g _s \rangle 
=
\left(
\mathbb{H}^{2}
\right)_{rs}
-
\left(
\mathbb{H}\mathbb{S}^{-1}\mathbb{H}
\right)_{rs} ,\\
\\[-10pt]
&\!\!\!
\left(
\mathbb{H}^{2}
\right)_{rs}
\!=\!
{\displaystyle \frac{1}{n}}
\sum _{t,t^\prime=1}^n
(\mathbb{H})_{rt}
\sum _{k=1}^n
c_t ^{(k)} c_{t^\prime } ^{(k)*}
(\mathbb{H})_{t^\prime s}
\!=\!
{\displaystyle \frac{1}{n}}
\left(
\mathbb{H}\mathbb{S}^{-1}\mathbb{H}
\right)_{rs} ,
\end{array}
\right\}
\label{S1form}
\eeqa
\vspace{-0.2cm}
\beqa
\!\!\!\!\!\!\!\!\!\!\!\!\!\!\!\!\!\!\!\!\!\!\!\!\!\!\!\!\!\!\!\!\!\!\!\!\!\!\!\!\!\!\!\!\!\!\!\!\!\!\!\!\!\!\!\!\!\!\!\!\!\!\!
\left.
\begin{array}{ll}
&\!\!\!
\mathbb{L}_{rs}^{(1)}
=
-
\langle g _r^{(1)} | H^{(1)} | g _s^{(1)} \rangle
=
-
\langle g _r |
HQHQH
|g _s \rangle \\
\\
&\!\!\!
~~~~~
=
-
\left(
\mathbb{H}^{3}
\right)_{rs}
+
\left(
\mathbb{H}\mathbb{S}^{-1}\mathbb{H}^{2}
+
\mathbb{H}^{2}\mathbb{S}^{-1}\mathbb{H}
\right)_{rs}
-
\left(
\mathbb{H}\mathbb{S}^{-1}\mathbb{H}\mathbb{S}^{-1}\mathbb{H}
\right)_{rs} , \\
\\[-10pt]
&\!\!\!
\left(
\mathbb{H}^{3}
\right)_{rs}
=
{\displaystyle \frac{1}{n}}
\sum _{t,t^\prime=1}^n
(\mathbb{H})_{rt}
\sum _{k=1}^n
c_t ^{(k)} c_{t^\prime } ^{(k)*}
(\mathbb{H}^2 )_{t^\prime s}
=
{\displaystyle \frac{1}{n}}
\left(
\mathbb{H}\mathbb{S}^{-1}\mathbb{H}^2
\right)_{rs} ,
\end{array}
\right\}
\label{L1form}
\eeqa
%and
\beqa
\!\!\!\!\!\!\!\!\!\!\!\!\!\!\!\!\!\!\!\!\!\!\!\!\!\!\!\!\!\!\!\!\!\!\!\!\!\!\!\!\!\!\!\!\!\!\!\!\!\!\!\!\!\!\!\!\!\!\!\!\!\!\!
\!\!\!\!\!\!\!\!\!\!\!
\begin{array}{ll}
&\!\!\!
\mathbb{S}_{rs}^{(2)}
=
\langle g _r^{(1)} |
H^{(1)}Q^{(1)}H^{(1)}
| g _s^{(1)} \rangle
=
\langle g _r |
HQHQQ^{(1)}QHQH
|g _s \rangle \\
\\
&\!\!\!
~~~~~=
\langle g _r |
HQHQHQH
|g _s \rangle
-
\left(
\mathbb{L}^{(1)}\mathbb{S}^{(1)-1}\mathbb{L}^{(1)}
\right)_{rs} ,
\end{array}
\label{S2form}
\eeqa
where
\beqa
\begin{array}{ll}
&\!\!\!
\langle g _r |
HQHQHQH
|g _s \rangle 
=
\left(
\mathbb{H}^{4}
\right)_{rs}
-
\left(
\mathbb{H}\mathbb{S}^{-1}\mathbb{H}^{3} 
+
\mathbb{H}^{3}\mathbb{S}^{-1}\mathbb{H}
+
\mathbb{H}^{2}\mathbb{S}^{-1}\mathbb{H}^{2}
\right)_{rs}\\
\\
&\!\!\!
+
\left(
\mathbb{H}\mathbb{S}^{-1}\mathbb{H}\mathbb{S}^{-1}\mathbb{H}^{2}
+
\mathbb{H}^{2}\mathbb{S}^{-1}\mathbb{H}\mathbb{S}^{-1}\mathbb{H}
+
\mathbb{H}\mathbb{S}^{-1}\mathbb{H}^{2}\mathbb{S}^{-1}\mathbb{H}
\right)_{rs}
-
\left(
\mathbb{H}\mathbb{S}^{-1}\mathbb{H}\mathbb{S}^{-1}
\mathbb{H}\mathbb{S}^{-1}\mathbb{H}
\right)_{rs} ,
\end{array}
\label{S2form2}
\eeqa
%and
\beqa
\!\!\!\!\!\!\!\!\!\!\!\!\!\!\!\!\!\!\!
\begin{array}{ll}
\mathbb{L}_{rs}^{(2)}
&\!\!\!
=
-
\langle g _r^{(2)} | H^{(2)} | g _s^{(2)} \rangle 
=
-
\langle g _r |
HQHQQ^{(1)}QHQQ^{(1)}QHQH
|g _s \rangle \\
\\
&\!\!\!
=
-
\langle g _r |
HQHQHQHQH
|g _s \rangle \\
\\
&\!\!\!
~~-
\left\{
\left(
\mathbb{S}^{(2)}
+
\mathbb{L}^{(1)}\mathbb{S}^{(1)-1}\mathbb{L}^{(1)}
\right)
\mathbb{S}^{(1)-1}\mathbb{L}^{(1)}
+
\mathbb{L}^{(1)}\mathbb{S}^{(1)-1}
\left(
\mathbb{S}^{(2)}
+
\mathbb{L}^{(1)}\mathbb{S}^{(1)-1}\mathbb{L}^{(1)}
\right)
\right\}_{rs}\\
\\
&\!\!\!
~~-
\left(
\mathbb{L}^{(1)}\mathbb{S}^{(1)-1}
\mathbb{L}^{(1)}
\mathbb{S}^{(1)-1}\mathbb{L}^{(1)}
\right)_{rs} ,
\end{array}
\label{L2form}
\eeqa
where
\beqa
\!\!\!\!\!
\begin{array}{ll}
&\!\!\!
\langle g _r |
HQHQHQHQH
|g _s \rangle
=
\left(
\mathbb{H}^{5}
\right)_{rs}
-
\left(
\mathbb{H}\mathbb{S}^{-1}\mathbb{H}^{4}
+
\mathbb{H}^{2}\mathbb{S}^{-1}\mathbb{H}^{3}
+
\mathbb{H}^{3}\mathbb{S}^{-1}\mathbb{H}^{2}
+
\mathbb{H}^{4}\mathbb{S}^{-1}\mathbb{H}
\right)_{rs}\\
\\
&\!\!\!
+
\left(
\mathbb{H}\mathbb{S}^{-1}\mathbb{H}\mathbb{S}^{-1}\mathbb{H}^{3}
+
\mathbb{H}\mathbb{S}^{-1}\mathbb{H}^{3}\mathbb{S}^{-1}\mathbb{H}
+
\mathbb{H}^{3}\mathbb{S}^{-1}\mathbb{H}\mathbb{S}^{-1}\mathbb{H}
\right)_{rs}\\
\\
&\!\!\!
+
\left(
\mathbb{H}\mathbb{S}^{-1}\mathbb{H}^{2}\mathbb{S}^{-1}\mathbb{H}^{2}
+
\mathbb{H}^{2}\mathbb{S}^{-1}\mathbb{H}\mathbb{S}^{-1}\mathbb{H}^{2}
+
\mathbb{H}^{2}\mathbb{S}^{-1}\mathbb{H}^{2}\mathbb{S}^{-1}\mathbb{H}
\right)_{rs}\\
\\
&\!\!\!
-
\left(
\mathbb{H}\mathbb{S}^{-1}\mathbb{H}\mathbb{S}^{-1}
\mathbb{H}\mathbb{S}^{-1}\mathbb{H}^{2}
\!+\!
\mathbb{H}\mathbb{S}^{-1}\mathbb{H}\mathbb{S}^{-1}
\mathbb{H}^{2}\mathbb{S}^{-1}\mathbb{H}
\!+\!
\mathbb{H}\mathbb{S}^{-1}\mathbb{H}^{2}\mathbb{S}^{-1}
\mathbb{H}\mathbb{S}^{-1}\mathbb{H}
\!+\!
\mathbb{H}^{2}\mathbb{S}^{-1}\mathbb{H}\mathbb{S}^{-1}
\mathbb{H}\mathbb{S}^{-1}\mathbb{H}
\right)_{rs}\\
\\
&\!\!\!
+
\left(
\mathbb{H}\mathbb{S}^{-1}\mathbb{H}\mathbb{S}^{-1}
\mathbb{H}\mathbb{S}^{-1}\mathbb{H}\mathbb{S}^{-1}\mathbb{H}
\right)_{rs}.
\end{array}
\label{L2form2}
\eeqa
The higher-order matrices become more complex
and tangled to calculate.
They have an intimate relation with the {\em cumulants} 
of the Hamiltonian or the {\em connected} diagrams
\cite{Kubb.62,Goldstone.57}.

\newpage

\def\thesection{\Alph{section}}
\setcounter{equation}{0}
\renewcommand{\theequation}{\Alph{section}.\arabic{equation}}
\section{Proof of relations
(\ref{modificationofWFW1}) 
and 
(\ref{modificationofWFW2})}

~~
The two Res-HB equations
(\ref{generalRes-HBeigenvalueequation}) 
and 
(\ref{generalRes-HBequation})
are equivalent.
The two thermal Res-HB ones
derived in II,
whose respective forms are the same as those of two Res-HB equations,
also turn out to be equivalent.
Here, following I, 
we prove this equivalence.
From now let us denote
$W_{\mbox{{\scriptsize Res}}:rs}^{\mbox{{\scriptsize thermal}}},
{\cal F}_{\mbox{{\scriptsize Res}}:r}^{\mbox{{\scriptsize thermal}}},
{\cal K}_{\mbox{{\scriptsize Res}}:rs}^{(k) \mbox{{\scriptsize thermal}}}$
and
${\cal F} [W_{\mbox{{\scriptsize Res}}:rs}^{\mbox{{\scriptsize thermal}}}]$
simply as
$W_{rs}, {\cal F}_r , {\cal K}_{rs}^{(k)}$
and
${\cal F} [W_{rs}]$.
First, due to idempotent-like product properties
$W_{rs}W_{rr} \!=\! W_{rs}$ and $W_{sr}W_{rr} \!=\!W_{rr}$,
we have important relations
${\cal K}_{rs}^{(k)}W_{rr}
\! = \!
{\cal K}_{rs}^{(k)}$
and
${\cal K}_{rs}^{(k)\dagger } W_{rr}
\! = \!
\left\{
H[W_{sr}] \! - \! E^{(k)}
\right\}
W_{rr}
\!\cdot\!
[\det z_{sr}]^{\frac{1}{2}}$.
Next multiplication of the second equation in
(\ref{generalRes-HBeigenvalueequation})
by $W_{rr}$ from the right yields
\vspace{-0.1cm}
\beqa
\!\!\!
\begin{array}{rl}
{\cal F}_r W_{rr}
\!=\! &\!\!\!
{\cal F}[W_{rr}] W_{rr} \sum_{k =1}^n |c_r ^{(k)}|^2 \\
\\[-2pt]
& ~~~~~+
\sum_{k =1}^n \sum_{s = 1} ^{\prime~n} \!
\left[
{\cal K}_{rs}^{(k)} c_r ^{(k)*} c_s^{(k)}
\!+\!
\left\{H[W_{sr}] \!-\! E^{(k)}\right\} W_{rr} 
\!\cdot \!
[\det z_{sr}]^{\frac{1}{2}} c_r ^{(k)} c_s^{(k)*}
\right] \\
\\[-2pt]
\!=\! &\!\!\!
{\cal F}[W_{rr}] W_{rr}
\sum_{k =1}^n |c_r ^{(k)}|^2
\!-\!
\sum_{k =1}^n {\cal K}_{rr}^{(k)} |c_r ^{(k)}| ^2
\!-\!
\sum_{k =1}^n
\left\{H[W_{rr}] \!-\! E^{(k)}\right\} W_{rr} |c_r ^{(k)}| ^2 \\
\\[-2pt]
& ~~~+
\sum_{k =1}^n \sum_{s = 1} ^n \!
\left[
{\cal K}_{rs}^{(k)} c_r ^{(k)*} c_s^{(k)}
\!+\! 
\left\{H[W_{sr}] \!-\! E^{(k)}\right\}
\!\cdot\! 
[\det z_{sr}]^{\frac{1}{2}}
c_s^{(k)*}
\!\cdot\!
W_{rr} c_r ^{(k)}
\right] \! .
\end{array}
\label{modificationofFW}
\eeqa
Using (\ref{generalRes-HBCIequation}),
the second term in the last line of R. H. S. of
(\ref{modificationofFW}) is vanished.
Substituting to ${\cal K}_{rr}^{(k)}$
whose explicit form is obtained from
(\ref{generalRes-HBequation}),
thus, ${\cal F}_r W_{rr}$ is cast into
\beqa
\begin{array}{rl}
&\!\!\!\!\!\!\!\!
{\cal F}_r W_{rr}
\!=\! 
{\cal F}[W_{rr}] W_{rr} \sum_{k =1}^n |c_r ^{(k)}|^2 \\
\\[-2pt]
&\!\!\!\!
-
\sum_{k =1}^n
\left[
(1_{2N} \! - \! W_{rr}) {\cal F} [W_{rr}]
\! + \!
2\left\{H[W_{rr}] \! - \! E^{(k)}\right\}
\right]
\cdot W_{rr} |c_r ^{(k)}|^2
\! + \!
\sum_{k =1}^n \sum_{s = 1} ^n
{\cal K}_{rs}^{(k)} c_r ^{(k)*} c_s^{(k)} \\
\\[-2pt]
&\!\!\!\!\!\!\!\!
=\!
W_{rr} {\cal F}[W_{rr}] W_{rr} \!
\sum_{k =1}^n \! |c_r ^{(k)}|^2
\! - \!
\sum_{k =1}^n \!
2 
\left\{ \! H[W_{rr}] \! - \! E^{(k)} \! \right\} \! 
W_{rr} |c_r ^{(k)}| ^2 
\! + \!
\sum_{k =1}^n \! \sum_{s = 1} ^n \!
{\cal K}_{rs}^{(k)} c_r ^{(k)*} c_s^{(k)} \!.
\end{array}
\label{modificationofFW2}
\eeqa
Further taking the hermitian conjugate of both sides of
(\ref{modificationofFW2}),
we have
\beq
{\cal F}_r W_{rr}
\!=\!
W_{rr}{\cal F}_r .
\label{commutabilityofFandW}
\eeq
This means the two hermitian matrices
${\cal F}_r$
and
$W_{rr}$
have common eigenvectors to diagonalize them
so that it leads to
(\ref{generalRes-HBeigenvalueequation}).
Therefore,
(\ref{generalRes-HBeigenvalueequation})
and
(\ref{commutabilityofFandW})
are equivalent.
Due to the idempotency relation
$
W_{rr}^2
\!=\!
W_{rr}
$,
(\ref{commutabilityofFandW})
is equivalent to
\beq
{\cal F}_r W_{rr} \!-\! W_{rr}{\cal F}_r W_{rr}
\!=\!
0 .
\label{FWequalWFW}
\eeq
Thus the relations
(\ref{modificationofWFW1}) 
and 
(\ref{modificationofWFW2})
are proved.
Further mutiplying
(\ref{modificationofFW2})
by $W_{rr}$ from the left
and using the explicit form of ${\cal K}_{rs}^{(k)}$,
we obtain
\beqa
\begin{array}{rl}
&\!\!\!\!
W_{rr} {\cal F}_r W_{rr}
\!=\!
W_{rr} ^2 {\cal F}[W_{rr}] W_{rr} \sum_{k =1}^n |c_r ^{(k)}|^2
\!-\!
\sum_{k =1}^n 
2 \left\{H[W_{rr}] \!-\! E^{(k)}\right\} W_{rr} ^2 |c_r ^{(k)}|^2 \\
\\[-2pt]
&~~+
\sum_{k =1}^n \sum_{s = 1} ^n
W_{rr}
\left\{
(1_{2N} \!-\! W_{rs}) {\cal F} [W_{rs}] \!+\! H[W_{rs}] - E^{(k)}
\right\}
\!\cdot\! W_{rs} \!\cdot\! 
[\det z_{rs}]^{\frac{1}{2}} c_r ^{(k)*} c_s^{(k)} \\
\\[-2pt]
&\!\!\!\!
= \!
W_{rr} {\cal F}[W_{rr}] W_{rr} \sum_{k =1}^n |c_r ^{(k)}|^2
\!-\!
\sum_{k =1}^n 
2 \left\{H[W_{rr}] \!-\! E^{(k)}\right\} W_{rr} |c_r ^{(k)}| ^2 .
\end{array}
\label{modificationofFW3}
\eeqa
Subtracting
(\ref{modificationofFW3})
from
(\ref{modificationofFW2}),
it is easy to derive an equivalence relation
\beqa
\begin{array}{c}
\sum_{k =1}^n \sum_{s = 1} ^n
{\cal K}_{rs}^{(k)} c_r ^{(k)*} c_s^{(k)}
\!=\!
{\cal F}_r W_{rr} \!-\! W_{rr} {\cal F}_r W_{rr} .
\end{array}
\label{modificationofWFW}
\eeqa
Thus,
the equivalence of
two types of the thermal Res-HB equations
in II is proved.
The equivalent relation
(\ref{modificationofWFW})
also makes a crucial role in the variation of Res-HB free energy.

\newpage

%%%%%%%%%%%%%%%
%                                       %
%   Acknowledgements     %
%                                       %
%%%%%%%%%%%%%%%

\vskip1.5cm
\begin{center}
{\bf Acknowledgements}
\end{center}
~~~~S. N. would like to
express his sincere thanks to 
Prof. Manuel Fiolhais for kind and
warm hospitality extended to
him at the Centro de F\'\i sica Computacional,
Universidade de Coimbra, Portugal.
This work was supported by FCT (Portugal) under the project
CERN/FP/83505/2008.
The authors are indebted to Professor Emeritus
M. Ozaki of Kochi University and
Prof. N. Tomita of Yamagata University
for their invaluable discussions and useful comments.
The authors thank the Yukawa Institute for Theoretical Physics
at Kyoto University. 
Discussions during the YITP workshop
YITP-W-06-07 and YITP-W-13-13 on 
Thermal Quantum Field Theories and Their Applications
were useful to complete this work.

\newpage

%%%%%%%%%%%
%                           %
%  References        %
%                           %
%%%%%%%%%%%

\end{document}